\documentclass[reqno,11pt]{amsart}
\usepackage{enumitem}
\usepackage{graphicx}
\usepackage{amscd}
\usepackage{slashed}
\usepackage{amssymb}
\usepackage[mathscr]{eucal}

\numberwithin{equation}{section}
\allowdisplaybreaks[1]

\title[Spinors on Singular Spaces]{Spinors on Singular Spaces and
the Topology of Causal Fermion Systems}

\author[F.\ Finster]{Felix Finster}
\address{Fakult\"at f\"ur Mathematik \\ Universit\"at Regensburg \\ D-93040 Regensburg \\ Germany}
\email{finster@ur.de}

\author[N.\ Kamran]{Niky Kamran \\ \\ September 2016}
\address{Department of Mathematics and Statistics \\ McGill University \\ Montr{\'e}al \\ Canada}
\email{nkamran@math.mcgill.ca}

\newtheorem{Def}{Definition}[section]
\newtheorem{Thm}[Def]{Theorem}
\newtheorem{Prp}[Def]{Proposition}
\newtheorem{Lemma}[Def]{Lemma}

\newtheorem{Example}[Def]{Example}

\newcommand{\Thanks}{\vspace*{.5em} \noindent \thanks}
\newcommand{\beq}{\begin{equation}}
\newcommand{\eeq}{\end{equation}}
\newcommand{\Proof}{\begin{proof}}
\newcommand{\QED}{\end{proof} \noindent}
\newcommand{\QEDrem}{\hspace*{0.1em} \ \hfill $\Diamond$}
\newcommand{\la}{\langle}
\newcommand{\ra}{\rangle}
\newcommand{\bra}{\mathopen{<}}
\newcommand{\ket}{\mathclose{>}}
\newcommand{\Sl}{\mathopen{\prec}}
\newcommand{\Sr}{\mathclose{\succ}}

\newcommand{\C}{\mathbb{C}}
\newcommand{\R}{\mathbb{R}}
\newcommand{\1}{\mbox{\rm 1 \hspace{-1.05 em} 1}}
\newcommand{\Z}{\mathbb{Z}}
\newcommand{\N}{\mathbb{N}}

\renewcommand{\H}{\mathscr{H}}

\newcommand{\U}{{\rm{U}}}
\newcommand{\A}{{\rm{U}^\pm}}
\newcommand{\GL}{{\rm{GL}}}
\newcommand{\SU}{{\rm{SU}}}
\newcommand{\SO}{{\rm{SO}}}

\newcommand{\G}{\mathscr{G}}
\newcommand{\e}{{\mathfrak{e}}}
\newcommand{\f}{{\mathfrak{f}}}

\newcommand{\bep}{\begin{pmatrix}}
\newcommand{\enp}{\end{pmatrix}}

\renewcommand{\O}{\mathscr{O}}

\newcommand{\F}{{\mathscr{F}}}
\newcommand{\Dir}{{\mathcal{D}}}
\newcommand{\D}{{\mathscr{D}}}
\newcommand{\K}{{\mathcal{K}}}
\renewcommand{\P}{{\mathcal{P}}}

\newcommand{\B}{{\mathcal{B}}} 
\renewcommand{\O}{{\mathscr{O}}}

\newcommand{\Lin}{\text{\rm{L}}}

\newcommand{\Cl}{{\mathscr{C}}\ell}
\newcommand{\con}{\text{\rm{con}}}
\newcommand{\Pin}{\text{\rm{Pin}}}
\newcommand{\Spin}{\text{\rm{Spin}}}
\newcommand{\scrM}{\myscr{M}}

\DeclareFontFamily{OT1}{rsfso}{}
\DeclareFontShape{OT1}{rsfso}{m}{n}{ <-7> rsfso5 <7-10> rsfso7 <10-> rsfso10}{}
\DeclareMathAlphabet{\myscr}{OT1}{rsfso}{m}{n}

\DeclareMathOperator{\Tr}{Tr}
\DeclareMathOperator{\tr}{tr}
\DeclareMathOperator{\Symm}{\mbox{\rm{Symm}}}
\newcommand{\AC}{{\mathfrak{A}}}

\DeclareMathOperator{\supp}{supp}
\DeclareMathOperator{\rg}{rg}
\DeclareMathOperator{\dist}{d}

\newcommand{\p}{\mathfrak{p}}
\newcommand{\q}{\mathfrak{q}}

\newcommand{\dsf}{\mathsf{d}}
\newcommand{\pseudo}{\Gamma}

\begin{document}

\begin{abstract}
Causal fermion systems and Riemannian fermion systems are proposed
as a framework for describing non-smooth geometries.
In particular, this framework provides a setting for spinors on singular spaces.
The underlying topological structures are introduced and analyzed. The connection
to the spin condition in differential topology is worked out.
The constructions are illustrated by many simple examples like the
Euclidean plane, the two-dimensional Minkowski space, a conical singularity,
a lattice system as well as the curvature singularity of the Schwarzschild space-time.
As further examples, it is shown how complex and K\"ahler structures
can be encoded in Riemannian fermion systems.
\end{abstract}

\maketitle
\tableofcontents

\section{Introduction}
Causal fermion systems arise in the context of relativistic quantum theory
(see~\cite{cfs} or the survey articles~\cite{srev, dice2014}).
From the mathematical point of view, they provide a framework
for describing generalized space-times (so-called ``quantum space-times'')
which do not need to have the structure of a Lorentzian manifold.
Nevertheless, many structures of Lorentzian spin geometry (like causality, a 
distinguished time direction, spinors, connection and curvature)
have a generalized meaning in these space-times
(cf.~\cite{lqg}, the survey article~\cite{rrev} or the introduction~\cite{nrstg}).
The present work is the first paper in which the topology of causal fermion systems
is analyzed. Moreover, we extend the framework to the Riemannian setting
by introducing so-called {\em{topological fermion systems}}.
Topological fermion systems are of interest from a purely mathematical
perspective as a general framework for describing non-smooth geometries.
In particular, they provide a setting for introducing spinors on singular spaces.

A central idea behind topological fermion systems is to encode the geometry of
space (or space-time) in a collection of linear operators on a Hilbert space~$\H$.
In the smooth setting in which space (or space-time) is a differentiable manifold,
one chooses~$\H$ to be the span of certain
spinorial wave functions on the manifold, typically formed of solutions of the Dirac equation.
The spatial (or space-time) dependence of the wave functions is then encoded in the so-called
{\em{local correlation operators}}, which are bounded linear operators on~$\H$.
Identifying the points of the manifold with the corresponding local correlation
operators, we describe space (or space-time) by a subset of~$\Lin(\H)$.
Finally, taking the push-forward of the volume measure gives rise to a measure on~$\Lin(\H)$,
the so-called {\em{universal measure}}.
This leads to the general setting of topological fermion systems
that will be introduced in Definition~\ref{defparticle} below.

Topological fermion systems also allow for the description of non-smooth geometries in which
the underlying space (or space-time) is {\em{not}} a differentiable manifold.
This can be understood from the fact that the solutions of a partial differential equation
often have better regularity than the coefficients of the PDE they solve. 
As a consequence, in many situations the wave functions, which are solutions of a
geometric PDE, remain continuous even when
the metric or curvature develop singularities. Since we derive all relevant objects
(like the spinor bundle and Clifford structures) from the local correlation operators, our framework
remains well-defined even in such singular situations.
Moreover, we can describe discrete spaces (like lattices) or other highly singular spaces,
and our methods endow such spaces with non-trivial topological data.

Our framework is very flexible because there is a lot of freedom
in choosing the wave functions in~$\H$. This has the advantage that one can describe
many different geometric situations by tailoring~$\H$ in regard to the specific application.
It is a main purpose of the present paper to explain in various examples how this can be done.
We remark that the framework becomes much more rigid if one assumes that the configurations of
wave functions are minimizers of causal variational principles (see~\cite{continuum})
or corresponding Riemannian analogs. The analysis of such variational principles
is a separate subject which we cannot enter here.
Instead, we refer the interested reader to~\cite{support, lagrange, cauchy, noether, jet, sphere}
and the references therein.

The paper is organized as follows. In Section~\ref{secex} we give the general definition
of topological fermion systems and explain in various examples how such systems can be constructed.
In Section~\ref{secstruct} we introduce the basic structures inherent to topological fermion systems,
starting from the most general singular situation and then specializing in several steps until
we end up in the smooth setting.
In Section~\ref{sectopspin} we define so-called topological spinor bundles on a topological manifold
and work out the connection to the structures on a usual spin manifold.
In Section~\ref{secaddex} it is explained how almost complex and complex structures
as well as K\"ahler structures can be encoded in topological fermion systems.
In Section~\ref{sectangential} we address the question of whether a topological fermion system determines
a distinguished Clifford structure.
In Section~\ref{secdiscrete} we present methods for getting topological information
on topological fermion systems for which the underlying space (or space-time) does not even
have the structure of a topological manifold.
In Section~\ref{secexamples} we illustrate our constructions by the examples of
the Euclidean plane and two-dimensional Minkowski space.
Section~\ref{secsing} is devoted to examples for spinors on singular spaces:
In Section~\ref{secconformal} we consider singularities of the curvature tensor which can
be removed by a conformal transformation. In this case, a rescaling of the spinorial
wave functions makes it possible to eliminate the singularity.
Section~\ref{secgenuine} treats curvature singularities which cannot be removed by
a conformal transformation. In Section~\ref{secschwarzschild} we describe the curvature
singularity of the Schwarzschild black hole. Finally, in Section~\ref{seclattice} we illustrate the topology
of singular spaces in the example of a two-dimensional lattice.

We finally point out that all our constructions are meant to be {\em{topological
but not differential geometric}} in the following sense: Starting from a Lorentzian manifold, getting into
the framework of causal fermion systems makes it necessary to introduce a regularization
on a microscopic length scale~$\varepsilon$
(for details see~\cite[Section~4]{finite}, where this is referred to as ``ultraviolet regularization'').
This means that the system must be ``smeared out''
on the microscopic scale. As a consequence, the macroscopic geometry of space-time can be seen only
on scales which are larger than the regularization length~$\varepsilon$.
This subtle point is taken care of in the constructions in~\cite{lqg} by working with the notions
of ``generically timelike separation'' and ``spin-connectable space-time points.'' Moreover, 
the spin connection in~\cite{lqg} gives a parallel transport along a discrete ``chain'' of points, and the
correspondence to the spinorial Levi-Civita connection is obtained by first taking the
limit~$\varepsilon \searrow 0$ and then letting the number of points of the chain tend to infinity
(see~\cite[Theorem~5.12]{lqg}). Similarly, the Euclidean sign operator (which will be introduced
after~\eqref{spinsplit} below) depends in an essential way on the regularization,
so that in the constructions in~\cite{lqg} it is handled with care.
On the other hand, the ultraviolet regularization can be regarded as a continuous deformation
of the geometry for small distances, having no influence on the topology.
With this in mind, we here take the point of view that for analyzing topological questions,
one can make use of the local behavior of the causal fermion system in an arbitrarily small neighborhood
of a given space-time point. This leads to different types of constructions which we will explore here.
In this way, the topological constructions given in this paper complement the differential
geometric constructions in~\cite{lqg} and give a different viewpoint on causal fermion systems.

\section{Basic Definitions and Simple Examples} \label{secex}
Causal fermion system were first introduced in~\cite{rrev}.
Here we give a slightly more general definition and explain it afterwards in a few examples.
\begin{Def} \label{defparticle} {\em{
Given a complex Hilbert space~$(\H, \la .|. \ra_\H)$ 
and parameters~$\p,\q \in \N_0$ with~$\p \leq \q$, we let~$\F \subset \Lin(\H)$ be the set of all
self-adjoint operators on~$\H$ of finite rank, which (counting with multiplicities) have
at most~$\p$ positive and at most~$\q$ negative eigenvalues. On~$\F$ we are given
a positive measure~$\rho$ (defined on a $\sigma$-algebra of subsets of~$\F$), the so-called
{\em{universal measure}}. We refer to~$(\H, \F, \rho)$ as a {\em{topological fermion system}} of spin
signature~$(\p,\q)$.

In the case~$\p=\q$, we call~$(\H, \F, \rho)$ a {\em{causal fermion system}} of spin dimension~$n:=\p$.
If~$\p=0$, we call~$(\H, \F, \rho)$ a {\em{Riemannian fermion system}} of spin dimension~$n:=\q$.}}
\end{Def} \noindent
It should be noted that the assumption~$\p \leq \q$ merely is a convention, because otherwise one may always arrange to replace~$\F$ by~$-\F$.

A basic feature of topological fermion systems is that the geometry and topology are encoded in
terms of linear operators on a Hilbert space. The support of the universal measure~$\rho$, defined by
\[ \supp \rho = \left\{ x \in \F \:|\: \rho(U) > 0 \text{ for every open neighborhood~$U$ of~$x$} \right\} \subset \F \:, \]
takes the role of the base space, usually referred to as ``space'' or ``space-time.''
This concept is illustrated by the following examples.

\begin{Example} (Dirac spheres) \label{exsphere} {\em{
\begin{itemize}[leftmargin=2em]
\item[(i)] We choose~$\H=\C^2$ with the canonical scalar product. Moreover,
let~$\scrM=S^2 \subset \R^3$ and~$d\mu$ the Lebesgue measure on~$\scrM$.
Consider the mapping
\beq \label{Fexdef}
F : \scrM \rightarrow \Lin(\H)\:, \qquad F(p) = 2 \sum_{\alpha=1}^3 p^\alpha \sigma^\alpha + \1 \:,
\eeq
where~$\sigma^\alpha$ are the three Pauli matrices
\beq \label{Pauli}
\sigma^1 = \begin{pmatrix} 0 & 1 \\ 1 & 0 \end{pmatrix} , \qquad
\sigma^2 = \begin{pmatrix} 0 & -i \\ i & 0 \end{pmatrix} , \qquad
\sigma^3 = \begin{pmatrix} 1 & 0 \\ 0 & -1 \end{pmatrix} .
\eeq
Using the identities for the Pauli matrices
\[ \Tr \big(\sigma^\alpha \big) = 0 \qquad \text{and} \qquad
\Tr \big( \sigma^\alpha \sigma^\beta \big) = 2\: \delta^{\alpha \beta} \]
(where~$\delta^{\alpha \beta}$ is the Kronecker delta and~$\alpha, \beta \in \{1, 2,3\}$),
one readily finds that for any~$p \in S^2$,
\[ \tr \big( F(p) \big) = 2 \:,\qquad \tr \big( F(p)^2 \big) = 10 \:. \]
As a consequence, the eigenvalues of~$F(p)$ are equal to~$1 \pm 2$.
In particular, one eigenvalue is positive and one eigenvalue is negative, so that~$F(p) \in \F$
if we chose~$\p=\q=1$. We introduce the universal measure
as the push-forward measure~$\rho = F_* \mu$ (i.e.\ $\rho(\Omega) := \mu(F^{-1}(\Omega))$).
Then~$(\H, \F, \rho)$ is a causal fermion system of spin dimension one.
The support of~$\rho$ is homeomorphic to~$S^2$. We refer to this example as
a {\em{Dirac sphere}}.
\item[(ii)] We again choose~$\H=\C^2$ with the canonical scalar product. 
Taking two different parameters~$\tau_\pm>1$, we introduce the mappings
\[ F^\pm : \scrM \rightarrow \Lin(\H)\:, \qquad F(p) = \tau_\pm \sum_{\alpha=1}^3 p^\alpha \sigma^\alpha + \1 \:, \]
and define the universal measure as the sum of the 
corresponding push-forward measures,
\beq \label{sumpush}
\rho = F^+_* \mu + F^-_* \mu \:.
\eeq
Then~$(\H, \F, \rho)$ is again a causal fermion system of spin dimension one.
The support of~$\rho$ is homeomorphic to the disjoint union of two spheres.

As an alternative to taking the sum of the push-forward measures~\eqref{sumpush}, one can also
realize~$\rho$ as the push-forward of the Lebesgue measure on the disjoint union of two
spheres. To this end, we choose~$\scrM = S^2 \dot{\cup} S^2$ as the disjoint
union of two spheres and define the mapping
\beq \label{Fident}
F \::\: \scrM = S^2 \dot{\cup} S^2 \rightarrow \Lin(\H)
\eeq
by the condition that~$F(p)=F^+(p)$ and~$F^-(p)$ if~$p$ lies on the first and second sphere, respectively.
Taking the push-forward measure~$\rho = F_* \mu$ (where~$\mu$ is the Lebesgue measure
on~$S^2 \dot{\cup} S^2$), we again obtain the causal fermion system~$(\H, \F, \rho)$.
\item[(iii)] We consider the mappings
\[ F^\pm : \scrM \rightarrow \Lin(\H)\:, \qquad F^\pm(p) = 2 \sum_{\alpha=1}^3 p^\alpha \sigma^\alpha + \1
\pm \sigma^3 \]
and introduce the universal measure again as the sum of the 
corresponding push-forward measures~\eqref{sumpush}.
Then~$(\H, \F, \rho)$ is again a causal fermion system of spin dimension one.
The support of~$\rho$ is homeomorphic to two spheres glued together along 
circles of latitude (see Figure~\ref{figcircle}).
\begin{figure} %
\begin{picture}(0,0)%
\includegraphics{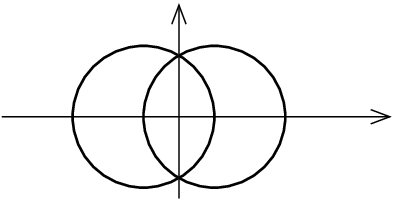}%
\end{picture}%
\setlength{\unitlength}{2486sp}%
\begingroup\makeatletter\ifx\SetFigFont\undefined%
\gdef\SetFigFont#1#2#3#4#5{%
  \reset@font\fontsize{#1}{#2pt}%
  \fontfamily{#3}\fontseries{#4}\fontshape{#5}%
  \selectfont}%
\fi\endgroup%
\begin{picture}(4994,2528)(1779,-6608)
\put(2082,-4875){\makebox(0,0)[lb]{\smash{{\SetFigFont{11}{13.2}{\familydefault}{\mddefault}{\updefault}$\supp \rho$}}}}
\put(4271,-4311){\makebox(0,0)[lb]{\smash{{\SetFigFont{11}{13.2}{\familydefault}{\mddefault}{\updefault}$\frac{1}{2} \Tr(x \sigma^{1\!/\!2})$}}}}
\put(6196,-6006){\makebox(0,0)[lb]{\smash{{\SetFigFont{11}{13.2}{\familydefault}{\mddefault}{\updefault}$\frac{1}{2} \Tr(x \sigma^3)$}}}}
\put(1801,-4291){\makebox(0,0)[lb]{\smash{{\SetFigFont{11}{13.2}{\familydefault}{\mddefault}{\updefault}$x \in \F$}}}}
\put(4531,-5816){\makebox(0,0)[lb]{\smash{{\SetFigFont{11}{13.2}{\familydefault}{\mddefault}{\updefault}$1$}}}}
\put(5441,-5811){\makebox(0,0)[lb]{\smash{{\SetFigFont{11}{13.2}{\familydefault}{\mddefault}{\updefault}$3$}}}}
\end{picture}%
\caption{Two intersecting Dirac spheres.}
\label{figcircle}
\end{figure}
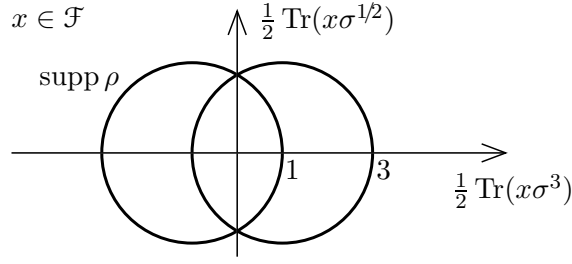
We refer to this example as {\em{two intersecting Dirac spheres}}.

Instead of taking the sum of the push-forward measures, one can again realize~$\rho$
as the push-forward of the Lebesgue measure on the disjoint union of two spheres,
\beq \label{Fident2}
\rho = F_* \mu \:,\qquad F \::\: \scrM := S^2 \dot{\cup} S^2 \rightarrow \Lin(\H)\:,
\eeq
where again~$F(p)=F^+(p)$ and~$F^-(p)$ if~$p$ lies on the first and second sphere, respectively.
\QEDrem
\end{itemize}
}}
\end{Example} \noindent
As already becomes clear in these simple examples, there are usually many ways to realize a topological
space as the support of a universal measure. The reason is that a topological fermion system encodes
more structures than just the topology, so that prescribing only the topology leaves a lot of freedom
to modify all the additional structures.

A particular structure of a topological fermion system is the Hilbert space~$\H$.
The vectors in~$\H$ have the interpretation as
the wave functions corresponding to the quantum states of the physical system.
More generally, a basic underlying concept is to encode
the geometry and topology in a certain family of functions
defined on space or in space-time. This is illustrated in the next examples.
\begin{Example} (Scalar and vector fields on a closed Riemannian manifold) \label{exscalar} {\em{
\begin{itemize}[leftmargin=2em]
\item[(i)] Let~$(\scrM,g)$ be a smooth compact Riemannian manifold without boundary and~$\Delta$ the Laplace-Beltrami operator
acting on complex-valued scalar functions on~$\scrM$. Then the operator~$-\Delta$
with domain~$C^\infty(\scrM)$ is an essentially self-adjoint operator on~$L^2(\scrM)$.
It has a purely discrete spectrum lying on the positive real axis. For a given parameter~$L>0$
we let~$\H$ be the span of all eigenfunctions corresponding to eigenvalues at most~$L$, i.e.
\[ \H = \rg \chi_{[0,L]}(-\Delta) \subset L^2(\scrM)\:, \]
where~$\chi_{[0,L]}(-\Delta)$ is defined via the functional calculus for the characteristic function~$\chi_{[0,L]}$,
and~$\rg(A):= A(L^2(\scrM))$ denotes the range of an operator~$A$.
Then~$\H$ is a finite-dimensional Hilbert space which, by elliptic regularity theory,
consists of smooth functions. Hence, for every~$p \in \scrM$ the 
bilinear form~$(\psi, \phi) \mapsto -\overline{\psi(p)} \phi(p)$ is well-defined and continuous on~$\H \times \H$.
By the Fr{\'e}chet-Riesz theorem, there is a unique linear operator~$F(p)$ with the property that
\beq \label{scalar}
-\overline{\psi(p)} \phi(p) = \la \psi | F(p) \phi \ra_{L^2(\scrM)} \qquad \text{for all~$\psi, \phi \in \H$} \:.
\eeq
This operator is obviously negative semi-definite. Moreover, since the left side of~\eqref{scalar} only
involves~$\phi$ evaluated at~$p$, it follows by linearity that
\[ F(p)\, \phi = \phi(p)\, \big( F(p) \,1_\scrM \big) \:, \]
where~$1_{\scrM}$ is the constant function one on~$\scrM$. This shows that the operator~$F(p)$
has rank at most one. Varying~$p$, we thus obtain a mapping~$F : \scrM \rightarrow \F$ if we
choose~$\p=0$ and~$\q=1$.
Finally, we define~$\rho = F_* \mu$ as the push-forward measure of the volume measure
(i.e.\ $\rho(\Omega) := \mu(F^{-1}(\Omega))$.
Then~$(\H, \F, \rho)$ is a Riemannian fermion system of spin dimension one.
\item[(ii)] Let~$(\scrM,g)$ be a smooth compact Riemannian manifold of dimension~$k$ and~$\Delta$
the covariant Laplacian on smooth vector fields. Complexifying the vector fields and taking the
$L^2$-scalar product
\beq \label{L2sp}
\la u | v \ra_{L^2} = \int_{\scrM} g_{jk}\, \overline{u^j} \,v^k \: d\mu_{\scrM} \:,
\eeq
the operator~$-\Delta$ is essentially self-adjoint and has smooth eigenfunctions.
We again set~$\H = \rg \chi_{[0,L]}(-\Delta)$ and define the operator~$F(p) \in \Lin(\H)$ by
\beq \label{vector}
-g_{ij} \,\overline{u^i(p)} \,v^j(p) = \la u | F(p) v \ra_{L^2} \qquad \text{for all~$u,v \in \H$} \:.
\eeq
Now the operators~$F(p)$ are negative semi-definite and have rank at most~$k$.
We again introduce the universal measure by~$\rho = F_* \mu$.
Then~$(\H, \F, \rho)$ is a Riemannian fermion system of spin dimension~$k$.
\QEDrem
\end{itemize}
}}
\end{Example}

We now give an example involving holomorphic functions, which also gives a connection
to the well-known Bergman kernel.
\begin{Example} (Holomorphic functions and the Bergman kernel) \label{exbergman} {\em{
Let~$\Omega \subset \C$ be a bounded domain. We let~$\H$ be the holomorphic functions on~$\overline{\Omega}$
with the $L^2$-scalar product
\[ \la \psi | \phi \ra_\H = \int_\Omega \overline{\psi(z)}\: \phi(z)\: d\mu(z) \:, \]
where~$d\mu(z) = dx\, dy = (i/2)\, dz \wedge d\bar{z}$ is the two-dimensional Lebesgue measure.

As is obvious from the mean value property of holomorphic functions,
the $L^2$-limit of holomorphic functions is again holomorphic. Therefore, $\H$ is a Hilbert space and
\[ \H \subset L^2(\Omega, \C) \cap C^0(\Omega, \C) \:. \]
Since the functions in~$\H$ are continuous and pointwise bounded by their~$L^2$-norm,
for any~$z \in \Omega$ one can define the {\em{evaluation map}} $e_z \in \H^*$ by
\[ e_z(f) = f(z) \:. \]
We define the operator~$F(z)$ by
\[ F(z) = -e_z^* e_z \::\: \H \rightarrow \H \:. \]
A direct computation shows that, in analogy to~\eqref{scalar},
this operator satisfies the identity
\beq \label{corrberg}
-\overline{\psi(z)} \phi(z) = \la \psi | F(z) \phi \ra_\H \qquad \text{for all~$\psi, \phi \in \H$} \:.
\eeq
Thus, taking again the push-forward measure~$\rho = F_* \mu$, one obtains a Riemannian fermion system
of spin dimension one.

There is a close connection to the Bergman kernel, as we now explain.
Representing the linear functional~$e_z \in \H^*$ as a vector~$\psi_z \in \H$,
\[ e_z(f) = \int_\Omega \overline{\psi_z(\zeta)}\: f(\zeta)\: d\mu(\zeta) \:, \]
one obtains the integral representation
\beq \label{bergman}
f(z) = \int_\Omega K(z,\zeta)\: f(\zeta)\: d\mu(\zeta) \:,
\eeq
where~$K(z, \zeta) := \overline{\psi_z(\zeta)}$ is referred to as the {\em{Bergman kernel}}
(see for example~\cite[Section~1.4]{krantz}).
Using this integral representation on the left side of~\eqref{corrberg},
\begin{align*}
-\overline{\psi(z)} \phi(z) &= - \left( \int_\Omega \overline{K(z,\zeta')\: \psi(\zeta')}\: d\mu(\zeta') \right)
\left( \int_\Omega K(z,\zeta)\: \phi(\zeta)\: d\mu(\zeta) \right) \\
&= -\int_\Omega \overline{\psi(\zeta')} \left( \int_\Omega \overline{K(z,\zeta')} \:K(z,\zeta)\: \phi(\zeta)\: d\mu(\zeta) 
\right) d\mu(\zeta')\:,
\end{align*}
one recovers the operator~$F(z)$ as an integral operator whose kernel is the product of two
Bergman kernels, i.e.
\beq \label{FK2}
\big(F(z) \:\phi \big)(\zeta') =-\int_\Omega \overline{K(z,\zeta')} \,K(z,\zeta)\: \phi(\zeta)\: d\mu(\zeta) \:.
\eeq
Therefore, the operators~$F(z)$ encode all the information on the Bergman kernel, except
for the local phase described the transformation
\[ K(z,\zeta) \rightarrow e^{-i \varphi(z)}\: K(z,\zeta) \qquad \text{with} \qquad
\varphi \in C^\infty(\Omega, \R) \:, \]
which obviously leaves~\eqref{FK2} invariant. Using that the Bergman kernel is holomorphic
in~$z$, this local phase freedom reduces to a global phase change~$K(z,\zeta) \rightarrow e^{i \varphi} K(z,\zeta)$
with~$\varphi \in \R$.

Since the operators~$F(z)$ are encoded in the above topological fermion system, one sees
that, up to the above phase freedom, all the information of the Bergman kernel is contained in
our topological fermion system.
\QEDrem
}}
\end{Example}

Except for the above examples, in all applications worked out in the literature, the functions on space or space-time
are {\em{spinors}}.
In order to get the connection to topological fermion systems, we let~$(\scrM,g)$ be a spin manifold (Riemannian or Lorentzian)
and denote the corresponding spinor bundle by~$S\scrM$. Then the spinor space~$S_p\scrM$ at any
point~$p \in \scrM$ is endowed with an inner product, which we denote by
\beq \label{ssprod}
\Sl .|. \Sr_p \::\: S_p \scrM \times S_p \scrM \rightarrow \C
\eeq
and refer to as the {\em{spin scalar product}}. Next, we
choose~$\H \subset \Gamma(\scrM, S\scrM)$ as a subspace of
the continuous sections on~$\scrM$, together with a scalar product~$\la .|. \ra_\H$
(the choice of the scalar product depends on the signature of the metric and the particular application being considered).
Then for every~$p \in \scrM$, we can express the scalar product at a point~$p$ in terms of the
Hilbert space scalar product,
\beq \label{loccorr}
-\Sl \psi | \phi \Sr_p = \la \psi | F(p) \phi \ra_\H \qquad \text{for all~$\psi, \phi \in \H$} \:.
\eeq
According to the Riesz representation theorem, this determines a unique
linear operator~$F(p) \in \Lin(\H)$.
\begin{Def} \label{defloccor}
{\em{ The operator~$F(p) \in \Lin(\H)$ satisfying~\eqref{loccorr} is
referred to as the {\em{local correlation operator}} at the point~$p \in \scrM$. }}
\end{Def} \noindent
By construction, the operator~$F(p)$ has finite rank (indeed, its rank is at most
the dimension of~$S_p$), and its maximal number of positive and negative eigenvalues
is determined by the signature of the spin scalar product.
Therefore, we can regard~$F(p)$ as an element of~$\F \subset \Lin(\H)$
(for a suitable choice of the spin signature).
Varying~$p$, we obtain a mapping~$F : \scrM \rightarrow \F$.
Introducing the universal measure as the push-forward of the
volume measure~$d\mu$ on~$\scrM$, i.e.
\[ \rho(\Omega) := \mu(F^{-1}(\Omega)) \:, \]
we obtain a topological fermion system~$(\H, \F, \rho)$.
The concept behind taking the push-forward measure is that we want to identify the
point~$p \in \scrM$ with its local correlation operator~$F(p) \in \F$.
Likewise, the manifold~$\scrM$ should be identified with the subset~$F(\scrM)$ of~$\F$.
With this in mind, we do not want to work with objects on~$\scrM$, but instead with
corresponding objects on~$\F$.
Apart from giving a different point of view, this procedure makes it possible to extend
the notion of the manifold~$\scrM$ as well as the objects thereon to a more general setting.

To avoid confusion, we remark that the above-mentioned identification of~$p$ with~$F(p)$
clearly fails if the mapping~$F$ is not injective. For this reason, one usually chooses~$\H$
in such a way that~$F$ becomes injective. In certain applications, however, it is indeed preferable to work with
a mapping~$F$ which is not injective. In this case, all points of~$\scrM$ with the same image are identified
when forming the topological fermion system.
We already saw this effect in the example of the interacting Dirac spheres, where
the mapping~\eqref{Fident2} was not injective at the intersection points of the two spheres.
We will come back to this point in Example~\ref{ex2driemann} below.

The simplest setting in which the above construction of topological fermion systems using
spinors can be made precise is to choose~$\scrM$ as a closed Riemannian manifold:
\begin{Example}  (Spinors on a closed Riemannian manifold) \label{exspinor} {\em{
Let~$(\scrM, g)$ be a compact Riemannian spin manifold of dimension~$k \geq 1$.
The spinor bundle~$S\scrM$ is a vector bundle with fibre~$S_p\scrM \simeq \C^n$
with~$n=2^{[k/2]}$ (see for example~\cite{lawson+michelsohn, friedrich}).
The spin scalar product~\eqref{ssprod} is positive definite.
On the smooth sections~$\Gamma(S\scrM)$ of the spinor bundle we can thus introduce the scalar product
\beq \label{sprod-H}
\la \psi | \phi \ra = \int_{\scrM} \Sl \psi | \phi \Sr_p\: d\mu(p) \:,
\eeq
where~$d\mu = \sqrt{\det g}\, d^k x$ is the volume measure on~$\scrM$.
Forming the completion gives the Hilbert space~$L^2(\scrM, S\scrM)$.
The Dirac operator~$\Dir$ with domain of definition~$\Gamma(S\scrM)$
is an essentially self-adjoint operator on~$L^2(\scrM, S\scrM)$. It has a purely discrete spectrum
and finite-dimensional eigenspaces (for details see for example~\cite{ginoux}).
For a given parameter~$L>0$, we let~$\H$ be the space spanned by
all eigenvectors whose eigenvalues lie in the interval~$[-L, 0]$,
\[ \H = \rg \chi_{[-L,0]}(\Dir) \subset L^2(\scrM, S\scrM)\:. \]
Denoting the restriction of the $L^2$-scalar product to~$\H$ by~$\la .|. \ra_\H$,
we obtain a finite-dimensional Hilbert space~$(\H, \la .|. \ra_\H)$.
By elliptic regularity theory, the functions in~$\H$ are all smooth.

For every~$p \in \scrM$ we introduce the local correlation operator by~\eqref{loccorr}.
This operator is negative semi-definite and has rank at most~$n$.
Hence~$F(p)$ is an element of~$\F$ according to Definition~\ref{defparticle} if we choose~$\p=0$
and~$\q=n$. Varying~$p$, we obtain a mapping~$F : \scrM \rightarrow \F$.
Finally, we define~$\rho = F_* \mu$ as the push-forward measure of the volume measure.
Then~$(\H, \F, \rho)$ is a Riemannian fermion system of spin dimension~$n$.

This example can readily be extended to an infinite-dimensional Hilbert space
by choosing a function~$f \in C^0(\R)$ and by modifying the above construction to
\begin{align*}
\H &= \rg f^2(\Dir) \subset L^2(\scrM, S\scrM) \\
-\Sl f(\Dir) \psi | f(\Dir) \phi \Sr_p &= \la \psi | F(p) \phi \ra_\H \qquad \text{for all~$\psi, \phi \in \H$} \:.
\end{align*}
If~$f$ has suitable decay properties at infinity, then the operator~$f(\Dir)$ maps~$\H$ to the continuous
functions, so that~$F(p) \in \Lin(\H)$ is well-defined. We omit the details for brevity. \QEDrem
}} 
\end{Example}

Examples of causal fermion systems can be obtained similarly starting from a Lorentzian
manifold (see the introduction~\cite{nrstg}
a well as~\cite[Section~1.2]{cfs}, \cite[Section~1.1]{rrev}, \cite[Section~4 and~5]{lqg} or~\cite[Section~4]{finite}).
In this case, the space~$\H$ has the physical interpretation as describing all the
occupied Dirac quantum states of the system, including the so-called Dirac sea
(for the connection to physics see~\cite{srev} or the textbook~\cite{cfs}).
The scalar product on~$\H$ is typically deduced from the spatial integral over the
Dirac current and is thus closely related to the probabilistic interpretation of the
Dirac wave function. The fact that Dirac particles are fermions explains
the name {\em{fermion}} system.
The notion {{``\em{causal''}} in a causal fermion system can be understood as follows:
Taking the product of two operators~$x,y \in \F$,
we obtain an operator of rank at most~$\p+\q$.
This operator is in general no longer symmetric (because~$(x y)^* = y x$,
and thus~$xy$ is symmetric if and only if~$x$ and~$y$ commute).
Nevertheless, we can consider its characteristic polynomial. We denote its non-trivial zeros
(counted with algebraic multiplicities) by~$\lambda^{xy}_1, \ldots, \lambda^{xy}_{\mathfrak{p+q}}$.
For a Riemannian fermion system, these zeros are all real and non-negative.
Namely, in this case the operator~$-y$ is positive semi-definite, so that its square root~$\sqrt{-y}$
is well-defined as a positive semi-definite operator. Using that the spectrum
is invariant under cyclic permutations, it follows that
\beq \label{xyRiem}
xy = -x \:\sqrt{-y}\, \sqrt{-y} \qquad \text{is isospectral to} \qquad \sqrt{-y} \,(-x)\, \sqrt{-y} \:,
\eeq
and the last operator product is obviously positive semi-definite.
In the case~$\p>0$, the operator~$\sqrt{-y}$ no longer exists as a symmetric
operator, so that the argument~\eqref{xyRiem} breaks down.
It turns out that the~$\lambda^{xy}_j$ will in general be complex, giving rise to
the following notion of causality:

\begin{Def} (causal structure) \label{def2}
{\em{ Two points~$x,y \in \F$ are called {\em{timelike}} separated
if the non-trivial zeros~$\lambda^{xy}_1, \ldots, \lambda^{xy}_{\p+\q}$ of the
characteristic polynomial of the operator product~$xy$ are are all real and not all equal.
The points~$x$ and~$y$ are said to be {\em{spacelike}} separated if all the~$\lambda^{xy}_j$
have the same absolute value.
In all other cases, the points~$x$ and~$y$ are said to be {\em{lightlike}} separated. }}
\end{Def} \noindent
The relation between the causal structure of this definition and the usual
notion of causality in Minkowski space or on a Lorentzian manifold
is explained in~\cite[\S1.2.5]{cfs}, \cite[Sections~4 and~5]{lqg} or in non-technical
terms in~\cite[Section~4]{dice2014} and~\cite[Section~1.1]{rrev}.
Since this relation is not relevant to the main ideas of the present paper,
we here omit the details.

\section{Topological Structures} \label{secstruct}
In this section we work out the topological structures underlying a topological fermion system.
We begin with the most general
structures and then specialize the setting in several steps.
We first recall a few basic notions from~\cite{rrev} (see also~\cite[Section~1.1]{cfs}).
Let~$(\H, \F, \rho)$ be a topological fermion system.
On~$\F$ we consider the topology induced by the operator norm
\[ \|A\| := \sup \big\{ \|A u \|_\H \text{ with } \| u \|_\H = 1 \big\} \:. \]
The {\em{base space}}~$M$ (often referred to as ``space'' or ``space-time'')
is defined as the support of the universal measure, $M := \text{supp}\, \rho$.
On~$M$ we consider the topology induced by~$\F$.
For every~$x \in M$ we define the {\em{spin space}}~$S_x$ by~$S_x = x(\H)$; it is a subspace of~$\H$
of dimension at most~${\mathfrak{p+q}}$.
On~$S_x$ we introduce the {\em{spin scalar product}} $\Sl .|. \Sr_x$ by
\beq \label{ssp}
\Sl \psi | \phi \Sr_x = -\la \psi | x \phi \ra_\H \qquad \text{for all $\psi, \phi \in S_x$}\:;
\eeq
it is an indefinite inner product of signature~$(q_x, p_x)$ with~$p _x\leq \p$
and~$q_x \leq \q$ (the minus sign in~\eqref{ssp} is needed in order to be consistent
with the usual sign conventions for Dirac spinors in Minkowski space;
for details see~\cite{rrev}).

\subsection{A Sheaf}
The most general setting in which the topology of a topological fermion system of signature~$(\mathfrak{p,q})$
can be encoded is a that of a sheaf $S$ on $M$ whose stalks $(S_x, \Sl .|. \Sr_x)$ are indefinite inner product spaces of signature~$(q_x, p_x)$. 
Although this setting is too general for most of our constructions (we will mainly work with topological
vector bundles as will be described in Section~\ref{secbundle} below), we briefly explain how to get the
connection to sheaf theory.

For any~$x \in M$, we denote the orthogonal projection in~$\H$ to the spin space~$S_x$ by~$\pi_x$,
\beq \label{pidef}
\pi_x \::\: \H \rightarrow S_x\:.
\eeq
Projecting a given vector~$u \in \H$ to the spin spaces gives the mapping
\[ \psi_u \::\: M \rightarrow \H \:,\quad x \mapsto \pi_x u \in S_x \:. \]
We refer to~$\psi_u$ as the {\em{wave function of the occupied state}}~$u$.
For any open subset~$U \subset M$, we obtain a corresponding wave function by restriction,
\beq \label{psirestrict}
\psi_u|_U \::\: U \rightarrow \H \qquad \text{with} \qquad \psi(x) \in S_x \quad \text{for all~$x \in U$}\:.
\eeq
We denote the vector space of such wave functions on~$U$ by~$S_U$. 
We let~$S$ be the mapping which to an open set~$U$ assigns the vector space~$S_U$.
Moreover, for an open subset~$V \subset U$ we introduce the restriction map as the linear mapping
\beq \label{rUV}
r^U_V \::\: S_U \rightarrow S_V\:,\qquad \psi \mapsto \psi|_V\:.
\eeq
Obviously, these mappings have the following properties:
\begin{itemize}
\item[(I)] If~$U$ is the empty set, then~$S_U=\{0\}$.
\item[(II)] The linear map~$r^U_U$ is the identity. If~$W \subset V \subset U$, then~$r^U_W = r^V_W r^U_V$.
\end{itemize}
This gives the structure of a {\em{presheaf}} of complex vector spaces
over the topological space~$M$ (see~\cite{bredon} or~\cite[\S I.1.2]{hirzebruch}).
Introducing the corresponding {\em{sheaf}} by taking the direct limits of the vector spaces~$S_U$
(as outlined for example in~\cite[\S I.1.2]{hirzebruch}) gives the following structure: We
define~$S$ as the disjoint union of all spin spaces and~$\pi$ as the projection to the base point,
\[ S := \dot{\bigcup}_{x \in M} \:S_x \qquad \text{and} \qquad \pi \::\: S \rightarrow M \:,\;\; S_x \mapsto x \:.\]
Every~$\psi \in S_U$ defines the subset~$\cup_{x \in U} \psi(x) \subset S$.
On~$S$ we introduce the topology generated by all these subsets.
Then the triple~$(S, \pi, M)$ is a sheaf. The stalks~$(S_x, \Sl .|. \Sr_x)$ are indefinite inner product spaces
of signature~$(q_x,p_x)$.

Now the cohomology groups~$H^r(M, S)$, $r \geq 0$, with coefficients in a sheaf
(as defined for example in~\cite[\S I.2.6]{hirzebruch}) give topological information on the topological
fermion system. For example, globally defined continuous sections are naturally isomorphic to elements of
the cohomology group~$H^0(M,S)$ (see~\cite[Theorem~I.2.6.2]{hirzebruch}),
\[ \Gamma(S)\simeq H^{0}(M,S)\:. \]

\subsection{A Topological Vector Bundle} \label{secbundle}
We now show that under a certain regularity assumption (see Definition~\ref{defreg}),
topological fermion systems naturally give rise to topological vector bundles.
In preparation, we briefly recall the definition of a topological vector bundle (see~\cite{milnor+stasheff, steenrod}) and set up some notation.
Let~$\B$ and~$M$ be topological spaces and~$\pi : \B \rightarrow M$ a continuous surjective map.
Moreover, let~$Y$ be a complex vector space and~$G \subset \GL(Y)$ a group acting on~$Y$.
Then~$\B$ is a topological vector bundle with fibre~$Y$ and structure group~$G$
if every point~$x \in M$ has an open neighborhood~$U$ equipped with a
homeomorphism~$\phi_{U}: \pi^{-1}(U)\to U\times Y$, called a local trivialization or a bundle chart, such
that the diagram
\beq \label{commdiag}
\begin{array}[c]{cccc}
\pi^{-1}(U)&\stackrel{\phi_{U}}\longrightarrow&U\times Y\\
&\searrow&\downarrow \\
&&U
\end{array}
\eeq
commutes, where the projection maps are $\pi$ and the projection onto the first factor,
respectively. Furthermore, on overlaps $U\cap V$, we have
\beq \label{changechart}
\phi_{U}\circ \phi_{V}^{-1} \big|_{\{x\}\times Y}= g_{UV}(x) \qquad \text{for all~$x \in U \cap V$}\:,
\eeq
where $g_{UV}: U\cap V\to G$ is a continuous transition function.

Again setting~$M = \supp \rho$, we want to construct a topological vector bundle
having the spin space~$S_x$ as the fibre at the point~$x \in M$.
To this end, all the spin spaces must have the same dimension and signature,
making it necessary to impose the following condition:
\begin{Def} \label{defreg} {\em{ The topological fermion system is called {\em{regular}} if for all~$x \in M$,
the operator~$x$ has the maximal possible rank~${\mathfrak{p+q}}$. }}
\end{Def} \noindent
Clearly, the topological fermion systems of Example~\ref{exsphere} are all regular.
The topological fermion systems in Examples~\ref{exscalar} and~\ref{exspinor} are regular if and only
if for every~$p \in M$, the vectors~$\psi(p)$ with~$\psi \in \H$ span the fibre at~$p$.
We note that most of our constructions can be extended to non-regular topological fermion systems by
decomposing~$M$ into subsets on which~$x$ has fixed rank and a fixed number of positive and
negative eigenvalues. For clarity, we postpone this decomposition to Section~\ref{secdiscrete}
and for now restrict attention to regular topological fermion systems.

We define~$\B$ as the set of pairs
\[ \B = \{ (x, \psi) \:|\: x \in M,\: \psi \in S_x \} \]
and let~$\pi$ be the projection onto the first component.
Moreover, we let~$(Y, \Sl .|. \Sr)$ be an indefinite inner product space of signature~$({\mathfrak{q,p}})$,
and choose~$G = \U({\mathfrak{q,p}})$ as the group of unitary transformations on~$Y$.
In order to construct the bundle charts, for any given~$x \in M$ we choose a unitary
mapping~$\sigma : S_x \rightarrow Y$. By restricting the projection~$\pi_x$ in~\eqref{pidef}
to~$S_y$, we obtain the mapping
\[ \pi_x|_{S_y} \::\: S_y \rightarrow S_x \:. \]
In order to compute its adjoint with respect to the spin scalar product~\eqref{ssp},
for~$\psi \in S_x$ and~$\phi \in S_y$ we make the computation
\begin{align*}
\Sl \psi \,|\, \pi_x|_{S_y} \,\phi \Sr_x &= -\la \psi | x \phi \ra_\H
= -\la x \psi | \phi \ra_\H = -\la \pi_y \,x \,\psi | \phi \ra_\H
= -\la y \, (y|_{S_y})^{-1}\,\pi_y \,x \,\psi | \phi \ra_\H \\
&= -\big\la (y|_{S_y})^{-1}\, \pi_y \,x\, \psi \,\big|\, y \phi \big\ra_\H
= \Sl (y|_{S_y})^{-1}\, \pi_y \,x\, \psi \,|\, \phi \Sr_y \:.
\end{align*}
Hence
\[ \big( \pi_x|_{S_y} \big)^* =  (y|_{S_y})^{-1}\, \pi_y \,x |_{S_x} \:. \]
We now introduce the operator
\[ T_{xy} = \big( \pi_x|_{S_y} \big) \big( \pi_x|_{S_y} \big)^*
= \pi_x \,(y|_{S_y})^{-1}\, \pi_y \,x|_{S_x} \::\: S_x \rightarrow S_x \:. \]
By construction, this operator is symmetric and~$T_{xx} =\1$.
We now form the polar decomposition of~$T_{xy}$ to obtain a unitary operator~$U_{xy}$:
By continuity, there is a neighborhood~$U$ of~$x$ such that for all~$y \in U$,
the operator~$T_{xy}$ is invertible and has a unique square root~$\rho_{xy}$
(defined for example by the power series~$\sqrt{T_{xy}} = \sqrt{\1 + (T_{xy}-\1)}
= \1 + \frac{1}{2}\: (T_{xy} - \1) + \cdots$).
Introducing the mapping
\[ U_{x,y} = \rho_{xy}^{-1}\: \pi_x|_{S_y} : S_y \rightarrow S_x \:, \]
the calculation
\[ U_{x,y} \,U_{x,y}^* = \rho_{xy}^{-1} \:\pi_x|_{S_y}  \big( \pi_x|_{S_y} \big)^* \rho_{xy}^{-1} 
= \rho_{xy}^{-1} \:T_{xy}\: \rho_{xy}^{-1} = \1_{S_x} \]
shows that the mapping~$U_{xy}$ is unitary. Moreover, it clearly depends continuously on~$y \in U$.

We define the bundle chart~$\phi_U$ by
\[ \phi_U(y,v) = \big( y,  (\sigma \circ U_{x,y})(v) \big) \:. \]
The commutativity of the diagram~\eqref{commdiag} is clear by construction.
Moreover, the transition functions~$g_{UV}$ in~\eqref{changechart} are in~$G$ because
we are working with unitary mappings of the fibres throughout.
We choose the topology on~$\B$ such that all the bundle charts are homeomorphisms.

\begin{Def} \label{defassoc} The topological bundle~$\B \rightarrow M$ is referred to as
the vector bundle associated to the regular topological fermion system~$(\H, \F, \rho)$,
or simply the {\bf{associated vector bundle}}.
\end{Def}

\subsection{A Bundle over a Topological Manifold} \label{secbuntop}
In many applications, $M$ has a manifold structure.
This motivates us to specialize our setting by assuming that~$M$ is a {\em{topological manifold}}
of dimension~$k \geq 1$ (with or without boundary).
From the topological point of view, this is a major simplification which excludes many
examples (like the intersecting spheres in Example~\ref{exsphere}~(iii)).
The main benefit of this simplifying assumption  is that
one can choose local coordinates and work with partitions of unity.
As is made precise in the following theorem, these properties ensure that {\em{every}} such
bundle can be realized by a topological fermion system.
The proof illustrates our concept of encoding the topology of the bundle
in a suitable family of sections.

\begin{Thm} Let~$X \rightarrow \scrM$ be a vector bundle over a $k$-dimensional
topological manifold~$\scrM$, whose fibres are isomorphic to
an indefinite inner product space of signature~$(\mathfrak{q, p})$.
Then there is a regular topological fermion system~$(\H, \F, \rho)$ of signature~$(\mathfrak{p,q})$
such that the associated vector bundle (see Definition~\ref{defassoc})
is isomorphic to~$X$. If~$\scrM$ is compact, the Hilbert space~$\H$ can be chosen to
be finite-dimensional.
\end{Thm}
\Proof Let~$\{(x_\alpha, U_\alpha)\}$ be an at most countable atlas of~$\scrM$
such that the bundle has a trivialization on every~$U_\alpha$.
Thus we can choose continuous sections~$e_\alpha^1, \ldots, e_\alpha^{\mathfrak{p+q}}$
on~$U_\alpha$ which at every point~$p \in U_\alpha$ form a pseudo-orthonormal basis of
the fibre, which we denote by~$(S_p, \Sl .|. \Sr_p)$.
Next, we let~$(Z_i)_{i \in I}$ be an at most countable, locally finite covering of~$\scrM$
by relatively compact open subsets, which is subordinate to the atlas~$\{(x_\alpha, U_\alpha)\}$
(meaning that for every~$i \in I$, there is~$\alpha=\alpha(i)$ with~$Z_i \subset U_{\alpha(i)}$).
Starting from the sets~$(Z_i)_{i \in I}$, we now want to construct non-empty
open sets~$\Omega_i$, $V_i$ and~$W_i$
with the following properties:
\begin{itemize}
\item[(i)] The~$V_i$ are relatively compact and~$\overline{\Omega_i} \subset V_i \subset
\subset W_i \subset U_{\alpha(i)}$ for all~$i \in I$
(where~$V \subset \subset W$ stands for~$V \subset \overline{V} \subset W$).
\item[(ii)] The family~$(V_i)_{i \in I}$ is a locally finite covering of~$\scrM$.
\item[(iii)] $\overline{\Omega_i} \cap W_j = \varnothing$ for all~$i \neq j$.
\end{itemize}
To this end, we proceed inductively in the index~$i$. If the index set~$I$ is finite,
we denote it by~$I=\{1, \ldots, K\}$ with~$K \in \N$. If~$I$ is infinite, we set~$I=\N$, $K=\infty$
and use the notation~$Z_1 \cup \cdots \cup Z_K \equiv \cup_{i \in \N} Z_i$.
We let~$W_1=Z_1$ and set~$A_1 = \complement  (Z_2 \cup \cdots \cup Z_K)$. Then~$A_1$ is a
closed (possibly empty) set contained in~$Z_1$. We choose non-empty open sets~$\Omega_1$
and~$V_1$ such that~$A_1 \subset \Omega_1 \subset \subset V_1 \subset \subset W_1$.
For the induction step, assume that~$\Omega_i$, $V_i$ and~$W_i$ have already been constructed
for some~$i$. We set
\begin{align}
W_{i+1} &= Z_{i+1} \setminus (\overline{\Omega_1} \cup \cdots \cup \overline{\Omega_i}) \\
A_{i+1} &= \complement \big(V_1 \cup \cdots \cup V_i \cup Z_{i+2} \cup \cdots \cup Z_K \big)\:. \label{Aidef}
\end{align}
Then~$A_{i+1}$ is a closed subset of~$W_{i+1}$.
If~$W_{i+1}$ is empty, we skip this step and increase~$i$. Otherwise,
we choose non-empty open sets~$\Omega_{i+1}$
and~$V_{i+1}$ such that
\beq A_{i+1} \subset \Omega_{i+1} \subset \subset V_{i+1} \subset \subset W_{i+1} \:. \label{Videf}
\eeq
Let us verify that the resulting sets~$\Omega_i$, $V_i$ and~$W_i$ really have the above properties~(i)--(iii).
The properties~(i) and~(iii) are obvious by construction. To prove~(ii), let~$p \in \scrM$.
Since~$(Z_i)_{i \in I}$ is a locally finite covering, there is an index~$i \in I$ such that~$p \not \in Z_j$ for
all~$j>i+2$. But then one sees from~\eqref{Aidef} and~\eqref{Videf}
that~$p \in V_1 \cup \cdots \cup V_{i+1}$. We conclude that~(i)--(iii) hold.

Next, as in the usual construction of the partition of unity, we choose
non-negative continuous functions~$\eta_i \in C^0(\scrM, \R^+_0)$ with~$\eta_i|_{V_i} \equiv 1$,
$\eta_i|_{W_i \setminus \overline{V_i}} < 1$ and $\supp \eta_i \subset W_i$.
We consider the family of compactly supported continuous sections
\beq \label{family}
\eta_i\, e_{\alpha(i)}^\ell \qquad \text{and} \qquad \eta_i\,  x_{\alpha(i)}^j \, e_{\alpha(i)}^\ell \:,
\eeq
where~$i \in I$, $\ell=1,\ldots, \mathfrak{p+q}$ and~$j=1,\ldots, k$.
Let us verify that these sections are linearly independent. Thus suppose that a linear combination
of these functions vanishes. Restricting the functions to~$\Omega_i$, by property~(iii)
all the functions with~$j \neq i$ drop out. Thus it remains to show that for any fixed~$i$,
the sections in~\eqref{family} restricted to~$\Omega_i$ are linearly independent.
But this follows immediately from the fact that the~$e_{\alpha(i)}^\ell$ are linearly independent
at every~$p \in \Omega_i$, and that the coordinate functions of the chart are linearly independent
and not locally constant.

We let~$\H_0$ be the vector space spanned by the family of sections~\eqref{family}.
In order to introduce a scalar product, we represent a function~$\phi \in \H_0$
locally in components,
\[ \phi(p) = \sum_{\ell=1}^{\p+\q} \phi_\ell \big( x_\alpha(p) \big) \:e_{\alpha(i)}^\ell \qquad \text{for~$p \in U_\alpha$}\:, \]
and introduce the $L^2$-scalar product
\beq \label{sproddef}
\la \phi | \psi \ra_\H = \sum_{i \in I} \int_{x_{\alpha(i)}(U_{\alpha(i)})} \eta_i(x)\:
\sum_{\ell=1}^{\p+\q} \overline{\phi_\ell(x)}\, \psi_\ell(x) \: d^k x
\eeq
(thus we define the~$e_{\alpha(i)}^\ell$ to be orthonormal and take the measure
as a weighted sum of the Lebesgue measures in the charts). 
This scalar product is well-defined and finite because the functions in~$\H_0$ all have compact
support and because the sum in~\eqref{sproddef} is locally finite.
Moreover, it is clear from our construction that~$\H_0$ is {\em{locally finite-dimensional}} in the sense that
for any compact~$K \subset \scrM$, the function space
\[ \H|_K := \{ \psi|_K \text{ with } \psi \in \H_0 \} \]
is finite-dimensional.

Let us analyze whether~$(\H_0, \la .|. \ra_\H)$ is complete. Thus let~$\phi_n \in \H_0$ be a Cauchy sequence.
Then for every compact~$K \subset \scrM$, the functions~$\phi_n|_K$ are also a $L^2$-Cauchy sequence.
Since the space~$\H|_K$ is finite-dimensional, it follows immediately that the sequence~$\phi_n|_K$
converges in~$\H|_K$. However, the functions~$\phi_n$ need not converge globally in~$\H_0$.
For this reason, we introduce~$\H$ as the completion of~$\H_0$,
\[ \H := \overline{\H_0}^{\la .|. \ra_\H} \:. \]
Clearly, $(\H, \la .|. \ra_\H)$ is a Hilbert space. Moreover, the functions in~$\H$ are again locally
finite and~$\H|_K = \H_0|_K$. In particular, the functions in~$\H$ are all continuous.
If~$\scrM$ is compact, then~$\H$ is obviously finite-dimensional.

For any~$p \in \scrM$, we define the local correlation operator~$F(p)$ again by~\eqref{loccorr}.
Since the functions~\eqref{family} restricted to any point~$p$ span the fibre, the operator~$F(p)$
has maximal rank~${\mathfrak{p+q}}$. Similar as in~\eqref{sproddef}, we choose on~$\scrM$
the measure~$d\mu =  \sum_{i \in I} \eta_i(x)\: d^k x$.
Introducing the universal measure as the push-forward measure~$\rho = F_*(\mu)$,
we obtain a regular topological fermion system~$(\H, \F, \rho)$ of spin signature~$(\mathfrak{p,q})$.

It remains to prove that~$\scrM$ is homeomorphic to~$M:=\supp \rho \subset \F$,
and that the bundle~$\B \rightarrow M$ is homeomorphic to the bundle~$X \rightarrow \scrM$.
First, since~$\H$ is locally finite-dimensional and the functions in~$\H$ are all continuous,
the mapping~$F : \scrM \rightarrow \F$ is continuous. As a consequence, the pre-image
of any open neighborhood of a point~$q \in F(\scrM)$ is a non-zero open subset of~$\scrM$,
and thus it has non-zero $\mu$-measure. This implies that~$F(\scrM) \subset \supp \rho$.
Thus it suffices to show that the mapping
\beq \label{Fdef}
F : \scrM \rightarrow \supp \rho \subset \F \qquad \text{is a homeomorphism}\:.
\eeq
In order to show that~$F$ is injective, let~$p, q \in \scrM$ with~$F(p)=F(q)$.
Then in view of~\eqref{loccorr}, we know that
\beq \label{injcond}
\Sl \phi(p) | \psi(p) \Sr_p = \Sl \phi(q) | \psi(q) \Sr_q \qquad \text{for all~$\phi, \psi \in \H$} \:.
\eeq
Evaluating these relations for the functions~$\eta_i e^\ell_{\alpha(i)}$
in~\eqref{family}, we conclude that~$\eta_i(p) = \eta_i(q)$ for all~$i$.
As a consequence, we can choose the index~$i$ such that~$p, q \in \overline{V_i}$.
Next, evaluating~\eqref{injcond} for~$\phi = \eta_i e_{\alpha(i)}^\ell$
and~$\psi = \eta_i x_{\alpha(i)}^j e_{\alpha(i)}^\ell$, we conclude that~$x_{\alpha(i)}^j(p)=x_{\alpha(i)}^j$
for all~$j=1,\ldots, k$, implying that~$p=q$.

We next prove that the mapping~\eqref{Fdef} is open: For given~$p \in \scrM$ we
choose~$i$ such that~$p \in V_i$. We let~$\H_i \subset \H$ be the subspace spanned by the
functions in~\eqref{family}, and denote the orthogonal projection to~$\H_i$
by~$\pi_{\H_i}$. Then for any~$q \in \scrM$,
\begin{align*}
\| & F(p) - F(q)\|_{\Lin(\H)} = \sup_{u \in \H, \|u\|=1} \big| \la u | (F(p) - F(q)) u \ra \big| \\
&\geq \sup_{u \in \H_i, \|u\|=1} \big| \la u | (F(p) - F(q)) u \ra \big| = \| \pi_{\H_i} (F(p) - F(q)) \,\pi_{\H_i} \|_{\Lin(\H)} \\
&\geq c \:\Big| \big\la \eta_i e_{\alpha(i)}^\ell \,\big|\, (F(p) - F(q)) \, \eta_i e_{\alpha(i)}^\ell \big\ra \Big|
+ c \sum_{j=1}^k \Big| \big\la \eta_i e_{\alpha(i)}^\ell \,\big|\, (F(p) - F(q)) \, \eta_i x_{\alpha(i)}^j e_{\alpha(i)}^\ell \big\ra \Big| \\
&= c \: \big| \eta_i(p)^2 - \eta_i(q)^2 \big|
+ c \sum_{j=1}^k \big| \eta_i(p)^2  \,x_{\alpha(i)}^j(p) - \eta_i(q)^2  \,x_{\alpha(i)}^j(q) \big| \:,
\end{align*}
where the constant~$c=c(i)>0$ depends on the scalar products of the vectors in~$\H_i$
(here we use that on a finite-dimensional vector space all norms are equivalent).
If the left side of this inequality tends to zero, then~$\eta(q) \rightarrow \eta(p)=1$,
so that~$q$ lies in~$W_i$. Moreover, $x_i^j(q) \rightarrow x^j(p)$, implying that~$q \rightarrow p$.
Hence~$F^{-1}$ is continuous. We conclude that~$F : \scrM \rightarrow M$ is a homeomorphism.

Finally, we show that the corresponding bundles~$X \rightarrow \scrM$ and~$\B \rightarrow M$
are homeomorphic. First, any function~$\psi$ in~\eqref{family} is a continuous section of the
bundle~$X \rightarrow M$.
Identifying~$\psi$ with a vector in~$\H$, the mapping~$x \mapsto \pi_x \psi$ defines a continuous
section in~$\B \rightarrow M$. This identification of sections can be used to construct a
homeomorphism of bundles:
For any~$u \in X_p$ we choose a function~$\psi \in \H_0$ with~$\psi(p)=u$.
Setting~$x=F(p)$, we obtain the vector~$\pi_x \psi \in S_x$.
As is obvious by construction, the mapping~$u \mapsto \pi_x \psi$ does not depend on the
choice of~$\psi$ and thus defines a mapping~$X \rightarrow \B$.
This mapping is compatible with the projections. Moreover, it is
a bijection of the fibres which depends continuously on the base point.
Thus it defines a homeomorphism of the bundles.
\QED

In the setting of a topological manifold, it would be natural to assume
that~$\rho$ should be a {\em{continuous measure}} in the sense that~$\rho(\{x\})=0$ for all~$x \in M$.
However, this assumption will not be needed in this paper.

\subsection{A Bundle over a Differentiable Manifold} \label{secbundiff}
In many situations, $M$ is even a {\em{differentiable manifold}} (again of dimension~$k \geq 1$).
We will assume that~$M$ is differentiable whenever the tangent space or the tangent bundle are needed
in our constructions (more precisely, in Section~\ref{secspinstruct}, Section~\ref{secconstruct}
and some of our examples).
In the differentiable setting, it is natural to assume that the universal measure is compatible with
the differentiable structure in the following way: First, we always assume that the injection
\beq \label{Frechet}
M \hookrightarrow \F \subset \Lin(\H) \qquad \text{is Fr{\'e}chet differentiable}\:.
\eeq
This assumption makes it possible to identify tangent vectors on~$M$ with tangent vectors in~$\F$.
Moreover, it is a reasonable assumption that the measure~$\rho$ should be {\em{absolutely continuous}}
with respect to the Lebesgue measure in a chart.

\section{Topological Spinor Bundles} \label{sectopspin}
The vector bundle associated to a regular topological fermion system (see Definition~\ref{defassoc})
is reminiscent of a spinor bundle. In particular, a section~$\psi$ of this bundle takes values in the
spin spaces, $\psi(x) \in S_x$, and can thus be regarded as a spinorial wave function.
However, important structures of a spinor bundle like Clifford multiplication are still missing.
We shall now introduce these additional structures and analyze the topological obstructions
to their existence.
This will make it possible to interpret the vector bundles constructed in Sections~\ref{secbundle}
and~\ref{secbuntop} above as true spinor bundles corresponding in the classical sense of the term to bundles
of Clifford algebras and representations of the spin group.  We shall see that the topological conditions which the
topological fermion system will be required to satisfy (see Theorem~\ref{Steen} below) are independent of the
standard topological condition for the existence of a spin structure, as expressed by the vanishing of the
second Stiefel-Whitney class (see Section~\ref{secspinstruct}). Thus there are conditions 
for the existence of spin structures that are specific to topological fermion systems.

\subsection{Clifford Sections} \label{seccliff}
In this section we only assume that the topological fermion system is {\em{regular}} (see Definition~\ref{defreg}).
Then for any~$x \in M$, the operator~$(-x)$ on~$\H$ has $\q$~positive
and $\p$~negative eigenvalues. We denote its positive and negative spectral subspaces
by~$S_x^+$ and~$S_x^-$, respectively. In view of~\eqref{ssp}, these subspaces are also orthogonal
with respect to the spin scalar product,
\beq \label{spinsplit}
S_x = S_x^+ \oplus S_x^- \:.
\eeq
Moreover, we introduce the {\em{Euclidean sign operator}}~$s_x$ as a symmetric operator on~$S_x$
whose eigenspaces corresponding to the eigenvalues~$\pm 1$ are the spaces~$S_x^+$
and~$S_x^-$, respectively. Clearly, for a Riemannian fermion system the Euclidean sign operator
is the identity on~$S_x$.

In order to get a connection to the usual Clifford multiplication, we
need the notions of a Clifford subspace and a Clifford extension.
These notions were first introduced in~\cite{lqg} for causal fermion systems of spin dimension two.
We now extend them to general spin dimension.
We denote the space of symmetric linear operators on~$S_x$ by~$\Symm(S_x) \subset \Lin(S_x)$.

\begin{Def} \label{defcliffsubspace} 
A subspace~$K \subset \Symm(S_x)$ is called
a {\bf{Clifford subspace}} of signature~$(r,s)$ at the point~$x$ (with~$r,s \in \N_0$)
if the following conditions hold:
\begin{itemize}[leftmargin=2em]
\item[(i)] For any~$u, v \in K$, the anti-commutator~$\{ u,v \} \equiv u v + v u$ is a multiple
of the identity on~$S_x$.
\item[(ii)] The bilinear form~$\la .,. \ra$ on~$K$ defined by
\beq \label{anticommute2}
\frac{1}{2} \left\{ u,v \right\} = \la u,v \ra \, \1 \qquad {\text{for all~$u,v \in K$}}
\eeq
is non-degenerate and has signature~$(r,s)$.
\end{itemize}
We denote the set of Clifford subspaces of signature~$(r,s)$ at~$x$ by~${\mathcal{K}}^{(r,s)}_x$.
\end{Def} \noindent
For Riemannian fermion systems, the bilinear form~$\la .,. \ra$ defined by~\eqref{anticommute2}
is always a scalar product:
\begin{Lemma} \label{lemmariemsig}
On a Riemannian fermion system, a Clifford subspace of dimension~$m$
has signature~$(m,0)$.
\end{Lemma} 
\Proof Since~$S_x$ is positive definite, every~$u \in K$ is a symmetric operator on
a Hilbert space. As a consequence, $u^2$ is positive semi-definite, and thus~$0 \leq u^2 = \la u,u \ra \1$.
It follows that~$\la u,u \ra \geq 0$.
\QED

In the setting~${\mathfrak{p=q}}$ of causal fermion systems, a
useful method for constructing Clifford subspaces is to begin with the Euclidean sign operator~$s_x$
and to add operators which anti-commute with it. This gives rise to the so-called Clifford extensions.

\begin{Def} \label{defclifford} On a causal fermion system,
a Clifford subspace~$K$ which contains the Euclidean sign operator is
referred to as a {\bf{Clifford extension}} (of the Euclidean sign operator~$s_x$).
\end{Def} \noindent
Analogous as stated in Lemma~\ref{lemmariemsig} for Riemannian fermion systems,
on a causal fermion system the signature of Clifford extensions is determined by their dimension.
\begin{Lemma} A Clifford extension of dimension~$m$ has Lorentzian signature~$(1, m-1$).
\end{Lemma}
\Proof We choose an orthonormal basis~$(s_x, e_1, \ldots, e_{m-1})$ of the Clifford extension~$K$.
Choosing a pseudo-orthonormal eigenvector basis of~$S_x$,
we can represent the Euclidean sign operator by the matrix
\[ s_x = \begin{pmatrix} \1 & 0 \\ 0 & -\1 \end{pmatrix} . \]
Due to the anti-commutation relations, in this basis the operators~$e_j$ have the matrix representations
\[ e_j = \begin{pmatrix} 0 & a_j^* \\ a_j & 0 \end{pmatrix} \qquad \text{and} \qquad
a_j^* = - a_j^\dagger \]
(where the dagger denotes transposition and complex conjugation). Hence
\[ (e_j^2) = -  \begin{pmatrix} a_j^\dagger a_j & 0 \\ 0 & a_j a_j^\dagger \end{pmatrix} \:. \]
Noting that this matrix is negative definite, we obtain the claim.
\QED
We denote the Clifford extensions of signature~$(1, r)$ by~${\mathcal{K}}^{s_x, r}_x$.

In order to introduce global sections of the Clifford algebra, we let~${\mathfrak{S}}_m$ be the set of $m$-dimensional subspaces of~$\Lin(\H)$ endowed with the topology coming from the metric
\[ \dist(K,L) = \sup_{u \in K,\: \|u\|=1} \:\inf_{v \in L,\: \|v\|=1} \|u-v\| \:+\:
\sup_{v \in L,\: \|v\|=1} \:\inf_{u \in K,\: \|u\|=1} \|u-v\| . \]
Moreover, we let~$\Cl M$ be a continuous mapping which to every space-time point associates
a Clifford subspace,
\beq \label{Cliffsec}
\Cl M \::\: M \rightarrow {\mathfrak{S}}_{r+s} \:,\quad x \mapsto \Cl_x \in {\mathcal{K}}^{(r,s)}_x \:.
\eeq
We refer to~$\Cl M$ as a {\bf{Clifford section}} of signature~$(r,s)$.
It is a {\bf{section of Clifford extensions}} if~$\Cl_x \in {\mathcal{K}}^{s_x, r}_x$ for
all~$x \in M$.

Choosing a Clifford section is the main step for getting into the setting
in which the elements of~$S_x$ can be interpreted as spinors.
In the remainder of Section~\ref{sectopspin}, we shall work out this setting in more detail.
More precisely, in Section~\ref{secobstruct} we shall analyze topological obstructions for the existence of
Clifford sections. In Section~\ref{SpinGroup} we will construct the analogs of the usual $\Pin$ and $\Spin$
groups. In Section~\ref{BundleCharts} spinors, spinor bases and bundle charts for spinor bundles will be defined. Finally, in Section~\ref{secspinstruct} we will introduce spin structures which
associate a tangent vector on a differentiable manifold to a vector in the corresponding Clifford
subspace. Altogether, these constructions extend the usual structures of spin geometry to the
framework of topological fermion systems.

Before entering the detailed analysis, we now give a brief overview of the different situations of interest.
First, recall that regular topological fermion systems~$(\H, \F, \rho)$ on a topological manifold~$M=\supp \rho$
are distinguished by their spin signature~$(\p, \q)$ and the dimension~$k$ of~$M$.
Modifying Example~\ref{exscalar} by taking a Hilbert space~$\H$ formed of mappings from a manifold~$\scrM^k$ 
to~$\C^{\q,\p}$, one sees that~$k$ and~$(\p, \q)$ can be chosen arbitrarily.
When choosing Clifford subspaces or Clifford sections,
the signature~$(r,s)$ of the Clifford subspace gives additional parameters.
For Riemannian fermion systems, the signature of the spin scalar product is~$(0,n)$,
and the signature of the Clifford subspace is~$(m,0)$.
Similarly, if one considers causal fermion systems and Clifford extensions, the signature of the
spin scalar product is~$(n,n)$ and the signature of the Clifford subspace is~$(1, m-1)$.
Thus both for Riemannian and for causal fermion systems we are left with
two parameters~$n$ and~$m$ to describe the signatures.
In usual spin geometry, the dimension of the Clifford subspace always coincides with the
dimension of the manifold, i.e.
\beq \label{spincoind}
k = \left\{ \begin{array}{cl}
r+s & \text{in general} \\
m & \text{for Riemannian fermion systems or Clifford extensions}\:.
\end{array} \right.
\eeq
In our setting, these relations are needed if we want to introduce Clifford multiplication with tangent vectors,
because then one needs to identify the tangent space~$T_xM$ with the corresponding Clifford
subspace~$\Cl_x$. This construction will be given in Section~\ref{secspinstruct} below.
At the present stage, there is no need to impose~\eqref{spincoind}. On the contrary, 
for the sake of having more flexibility it is preferable to carefully distinguish the dimension of
the manifold from the dimension of the Clifford subspace and to treat these dimensions as
independent parameters.

More specifically, in the case of spin signature~$(2,2)$ and Clifford extensions of dimension~$m=4$,
one can get the correspondence to Dirac spinors on a Lorentzian manifold.
In this case, the corresponding geometric structures are worked out in~\cite{lqg}.
Indeed, most of the constructions in~\cite{lqg} apply just as well to the case~$m=5$.
In order to model the interactions of the standard model, one should increase the spin signature
to~$(2 \ell, 2 \ell)$ with~$\ell=2$ (to get the weak interaction and gravity~\cite[Chapter~4]{cfs})
or~$\ell=8$ (to also include the strong interaction~\cite[Chapter~5]{cfs}).
These cases of higher spin signature have not yet been studied systematically.
If the spin signature is decreased to~$(1,1)$, the spin spaces are two-dimensional,
making it impossible to represent Dirac spinors. But one can describe
Pauli-like spinors in dimensions $m=1$, $2$ or~$3$.
In these lower-dimensional situations, the geometric constructions of~\cite{lqg} simplify considerably.
This gives hope that it should be possible to relate the geometric notions to the
methods and notions arising in the analysis of causal variational principles
(see~\cite{continuum, support, lagrange, cauchy, noether, jet, sphere}).
We consider this to be a promising starting
point for future research. The study of Riemannian fermion systems is also a project for the future.
The different cases are summarized in Table~\ref{figcases}.
\begin{table}
\begin{tabular}{|c|c|}
\hline
\text{spin signature~$(2,2)$} & \parbox[t]{10cm}{ This
case is relevant for the description of Dirac spinors (as is explained in~\cite{rrev, finite}). The 
corresponding Lorentzian quantum geometry is developed in~\cite{lqg}.
\begin{itemize}
\item[$m=5$:] A Lorentzian Clifford subspace of signature~$(1,4)$. Appears naturally in the setting of
the Lorentzian quantum geometry in~\cite{lqg}. Is a bit easier to handle than
the four-dimensional case.
\item[$m=4$:] A Lorentzian Clifford subspace with the ``physical'' signature~$(1,3)$.
\end{itemize} } \\
\hline
\text{spin signature~$(1,1)$} & \parbox[t]{10cm}{ Is a mathematical simplification.
Most analytic work has been done in this case (see~\cite{small, support, sphere}).
\begin{itemize}
\item[$m=3$:] A Lorentzian Clifford subspace of signature~$(1,2)$. Appears naturally in the
quantum geometry setting, similar to the case~$k=5$ in spin signature~$(2,2)$.
\item[$m=2$:] A Lorentzian Clifford subspace of signature~$(1,1)$. Seems a good starting point for
connecting the Lorentzian quantum geometry with an analysis of the causal action principle.
\end{itemize} } \\
\hline
\text{spin signature~$(0,n)$} & \parbox[t]{10cm}{ 
Riemannian fermion systems of dimension~$m \geq 2$ have not yet been studied.
But it seems an interesting future project to work out the resulting Riemannian quantum
geometries. \\[-0.8em] } \\
\hline
\end{tabular} 
\vspace*{1em}
\caption{Different cases for the spin signature and the dimension of the
Clifford subspaces.}
\label{figcases}
\end{table}

\subsection{Topological Obstructions} \label{secobstruct}
The goal of this section is to determine topological obstructions to the existence of Clifford sections.
We shall see that there is an interesting interplay between the conditions that need to be satisfied for the
existence of Clifford sections and the usual obstructions to the existence of spin structures
on a differentiable manifold as expressed by the vanishing of the second Stiefel-Whitney class.
The connection between these different conditions will become clear in Section~\ref{secspinstruct} below.
For the moment, the question whether a Clifford section exists is independent of the
usual topological spin condition. This is illustrated by the following example which shows
that a smooth manifold may fail to admit 
a section of Clifford extensions even if its tangent bundle is spin. 

\begin{Example} (Non-existence of Clifford sections on the Dirac sphere) \label{hairy}
{\em{ We return to the Dirac sphere of Example~\ref{exsphere}~(i).
At a point~$x=F(p) \in \supp \rho$, the spin scalar product~\eqref{ssp} takes the form
\[ \Sl .|. \Sr_x = -\la ., F(p) . \ra_{\C^2}\:. \]
By definition, the Euclidean sign operator has the same eigenspaces as the operator~$F(p)$ in~\eqref{Fexdef},
but corresponding to the eigenvalues~$\mp 1$, so that
\[ s_x = -p \!\cdot\! \sigma \]
(where the dot is a short notation for the sum over the products of components).

We now choose a convenient parametrization of the space~$\Symm(S_x)$.
Let~$|F(p)|$ be the absolute value of the operator~$F(p)$, i.e.\ by~\eqref{Fexdef}
\[ |F(p)| = \sum_{\alpha=1}^3 p^\alpha \sigma^\alpha + 2 \,\1 \equiv p \!\cdot\! \sigma + 2 \,\1 \:. \]
Writing a linear operator in the form~$A = |F(p)|^{-\frac{1}{2}} B |F(p)|^\frac{1}{2}$,
the computation
\begin{align*}
\Sl \psi | A \phi \Sr_p &= -\la \psi | F(p)\, A \phi \ra_{\C^2}
= \big\la \psi \,\big|\, s_x\, |F(p)|\: A \phi \big\ra_{\C^2}
= \big\la \psi \,\big|\, s_x\, |F(p)|^{\frac{1}{2}} \,B\, |F(p)|^\frac{1}{2} \phi \big\ra_{\C^2} \\
&= \big\la \psi \,\big|\, |F(p)|^{\frac{1}{2}} \,s_x\, B\, |F(p)|^\frac{1}{2} \phi \big\ra_{\C^2}
= \big\la |F(p)|^{\frac{1}{2}} \psi \,\big|\, s_x\, B\, |F(p)|^\frac{1}{2} \phi \big\ra_{\C^2}
\end{align*}
shows that~$A$ is symmetric with respect to the inner product~$\Sl .|. \Sr_p$ if and only if~$B$ is 
symmetric with respect to the inner product~$\la ., s_x . \ra_{\C^2}$. A short computation yields
\beq \label{symmform}
\begin{split}
\Symm(S_x) = \big\{ |F(p)|^{-\frac{1}{2}} & (\alpha\, \1 + \beta \,p \!\cdot\! \sigma
+ i \,u \!\cdot\! \sigma )\: |F(p)|^{\frac{1}{2}} \\
& \text{with} \quad
\alpha, \beta \in \R,\; u \in \R^3 \text{ and } u \perp p \big\} .
\end{split}
\eeq

In order to obtain a two-dimensional Clifford extension at a point~$p$, we need to
choose an operator in~$\Symm(S_x)$ which anti-commutes with~$s_x$ and
whose square equals~$-\1$. A short computation using~\eqref{symmform} yields
\[ {\mathcal{K}}^{s_x, 1}_x = \Big\{ 
\text{span} \big( s_x, i \, |F(p)|^{-\frac{1}{2}}\, u \!\cdot\! \sigma\, |F(p)|^{\frac{1}{2}} \big) \:\Big|\: u \in S^2 \subset \R^3 \text{ with } u \perp p \Big\}\:. \]
We conclude that a {\em{two-dimensional section of Clifford extensions}}
amounts to finding a tangent unit vector field on the sphere~$S^2$. However, such a vector field
does not exist by the well-known ``hairy ball theorem''.

The existence of general Clifford sections (which may not necessarily be Clifford extensions)
depends on the dimension and signature. A {\em{three-dimensional Clifford section}}
necessarily has signature~$(1,2)$. There is the unique Clifford section
\[ \Cl_x =  \big\{
|F(p)|^{-\frac{1}{2}} (\beta \,p \!\cdot\! \sigma
+ i \,u \!\cdot\! \sigma ) |F(p)|^{\frac{1}{2}} \text{ with } \beta \in \R,\; u \in \R^3 \text{ and } u \perp p \big\}\:, \]
which is also a section of Clifford extensions.
{\em{Two-dimensional Clifford sections}} exist in signature~$(0,2)$, like for example
\[ \Cl_x = \big\{ |F(p)|^{-\frac{1}{2}} (i \,u \!\cdot\! \sigma ) |F(p)|^{\frac{1}{2}}
\text{ with } u \in \R^3 \text{ and } u \perp p \big\} \:. \]
By applying a unitary transformation~$\Cl_x \rightarrow U(p) \Cl_x U(p)^{-1}$, where~$U$
is a continuous family of unitary transformations~$U(p) \in \U(S_x)$, one can construct many
other Clifford sections of signature~$(0,2)$.
In signature~$(1,1)$, however, {\em{no Clifford section exists}}, as the following argument shows:
Suppose that~$\Cl_x$ were a Clifford section of signature~$(1,1)$. Choosing a positive
definite vector unit vector~$v$ in~$\Cl_x$, it is of the form
\[ v = |F(p)|^{-\frac{1}{2}} \big( \pm \cosh(\alpha) \:p \!\cdot\! \sigma
+ i \sinh(\alpha) \:u \!\cdot\! \sigma \big) |F(p)|^{\frac{1}{2}} \]
with~$\alpha \in \R$ and~$u \perp p$. Varying~$p$, by continuity we can always choose the plus sign
to obtain a global section~$v(p)$. A short computation shows that
a unit vector~$w$ in the orthogonal complement of~$v(p)$ in~$\Cl_x$ must be of the form
\[ w = |F(p)|^{-\frac{1}{2}} \big( \beta \:p \!\cdot\! \sigma
+ i \:t \!\cdot\! \sigma \big) |F(p)|^{\frac{1}{2}} \]
with~$\beta \in \R$, where the vector~$t$ is tangential to~$S^2$ and has length at least one.
In this way, the orthogonal complement of~$v(p)$ determines a continuous family of
one-dimensional subspaces of the tangent space of~$S^2$, in contradiction to the ``hairy ball theorem.''
}} \QEDrem
\end{Example}

This example illustrates that sections of Clifford extensions do not exist on
all topological manifolds. Obstructions to the existence of Clifford sections can be derived for a fairly general class of topological manifolds from classical results on obstructions to the existence of continuous sections
of topological fiber bundles over cell complexes~\cite{steenrod}. In order to apply these results, we will need to assume that our topological manifold $M$ is homeomorphic to a
{\em{finite cell complex}}. This will be the case for example if $M$ is a compact topological manifold~\cite{kirby+siebenmann}. Let us recall the following result from~\cite[Corollary~34.4]{steenrod} which gives the topological obstructions to the existence of continuous cross sections:

\begin{Thm} \label{Steen} Let $\B$ be a bundle over a finite cell complex $K$, such that for all $1\leq q \leq
\dim K$, the fibre $Y$ is $(q-1)$-simple. If 
\begin{equation}\label{obstruct}
H^{q} \big( K, \B(\pi_{q-1}(Y)) \big) =0\qquad \forall \,\, 1\leq q \leq \dim K \:,
\end{equation}
then $\B$ admits a continuous section.
\end{Thm} \noindent
To explain the notions in this theorem, we first recall that a path-connected space~$Y$ is~$q$-simple
if the action of $\pi_{1}(Y,y_{0})$ on the $q^\text{th}$ homotopy group~$\pi_{q}(Y,y_{0})$ is trivial for some base
point~$y_{0} \in y$ (and therefore all base points).
The assumption that the fibre~$Y$ is $(q-1)$-simple implies that the homotopy group~$\pi_{q-1}(Y)$
is defined independent of a base point. Finally, $\B(\pi_{q-1}(Y))$ denotes the
bundle over~$K$ whose fibre at a point~$x \in K$ is 
the homotopy group~$\pi_{q-1}$ of the corresponding fibre~$\B(x) \simeq Y$.

For the Clifford extensions that we will consider in this paper, the fibres $Y$ are isomorphic to Lie subgroups of~$\SU(1,1)$, $\SU(2,2)$
or~$\SU(n)$, and are therefore $q$-simple, since topological groups are $q$-simple for
all $q$ (see~\cite[Theorem~7.3.9]{spanier} or~\cite[page~281, Section 3C]{hatcher}).
Therefore, the only obstructions to the existence of Clifford extensions are given by the cohomological conditions~(\ref{obstruct}).

We now formulate these conditions explicitly for causal fermion systems
in the cases in which $M$ is a topological manifold $M^k$ of dimension $k$ and the fibre $Y$ is the set
of Clifford extensions of dimension~$m$ according to Table~\ref{figcases}.
In the case~$m=5$, the fibre $Y$ in spin dimension $n=2$ is isomorphic to the circle group $S^{1}$ which acts simply transitively on the set of Clifford extensions
(see~\cite[Corollary~3.7 and~Lemma~3.8]{lqg}). We thus obtain
\begin{equation}
\pi_{0}(Y)=0 \:,\qquad\pi_{1}(Y)=\mathbb{Z} \:,\qquad \pi_{n}(Y)=0 \quad \forall\: n\geq 2 \:,
\end{equation}
and the obstructions~(\ref{obstruct}) reduce to
\begin{equation}
H^{2}(M^k,\mathbb{Z})=0 \:.
\end{equation}
Next, in the case $m=4$ of a four-dimensional Clifford extension, the fibre $Y$ (again in spin dimension $n=2$) is isomorphic to the product $S^{1}\times SO(4,\mathbb{R})$, so that 
\begin{equation}
\pi_{0}(Y)=0,\,\,\pi_{1}(Y)=\mathbb{Z}\times \mathbb{Z}_{2},\,\,\pi_{2}(Y)=0,\,\,\pi_{3}(Y)=\mathbb{Z}\times \mathbb{Z}.
\end{equation}
The cohomological obstructions~(\ref{obstruct}) to the existence of a Clifford extension are therefore given by
\begin{equation}
H^{2}(M^k, \mathbb{Z}\times \mathbb{Z}_{2})=0 \qquad \text{and} \qquad H^{4}(M^k, \mathbb{Z}\times \mathbb{Z})=0 \:.
\end{equation} 
The remaining cases of interest are $m=2$ and~$3$, with the spin dimension $n$ taken to be equal to one.
For~$m=3$, the fibre $Y$ reduces to a point, giving no topological obstructions to the existence of a Clifford extension.  In the case $m=2$, on the other hand, the fibre $Y$ is again a circle, and the
obstructions~\eqref{obstruct} are given by
\begin{equation}
H^{2}(M^k, \mathbb{Z})=0.
\end{equation}

We close with two remarks. We first go back to Example~\ref{hairy} in which we tried to construct Clifford extensions on the Dirac sphere. In this example, the topological obstruction $H^{2}(M^{2}, \mathbb{Z})=0$ is violated. This confirms the conclusion that we came to by explicit calculation that Clifford extensions do not exist on the Dirac sphere.

Next, we further illustrate the independence of the usual spin condition from the
topological conditions needed for the existence of a Clifford extension by discussing two examples.
The first is $P_{3}(\mathbb{R})$, which is a spin manifold since $w_{2}(TP_{3}(\mathbb{R}))=0$, but which does not admit a Clifford extension since $H^{2}(M^k, \mathbb{Z})={\mathbb{Z}}_{2}\neq 0$. This example shows that the existence of a spin structure does not imply that of a Clifford extension. Conversely, one can construct by a surgery argument an example of a manifold $N^k$ of dimension $k\geq 6$ with is not spin, {\it i.e.} such that $w_{2}(TN^{k})\neq 0$, but which satisfies the condition~$H^{2}(N^k, \mathbb{Z}) = 0$, thereby showing that the two conditions are indeed independent. We outline the main steps of the construction of such a manifold $N^{k}$ by surgery\footnote{We are grateful to Steven Boyer for providing us with this argument.}. We start with any closed, connected, orientable smooth manifold $M^k$ of dimension $k\geq 6$ with $w_{2}(TM^k)\neq 0$, and let $\nu M^k$ denote the stable normal bundle of $M^k$. By definition, $\nu M^k$ is such that $TM^k \oplus \nu M^{k}$ is a trivial bundle (the existence of a stable normal bundle, which is well-defined up to the addition of trivial bundles, follows from the Whitney embedding theorem). Since $w_{1}(TM^k)=0$ by our assumption of orientability, it follows that $w_{1}(\nu M^k)=0$ and $w_{2}(\nu M^k)=w_{2}(TM^k)$. As an oriented bundle over $M^k$, the stable normal bundle $\nu M^k$ is classified by the homotopy class of a map $f: M^k \to \text{BSO}$, where $\text{BSO}$ is the classifying space of $\text{SO}=\cup_{n}\text{SO}(n)$
(see~\cite[Theorem 3(7.2)]{husemoller}).
Now, the homotopy groups of $\text{BSO}$ are given by $\pi_{r}(\text{BSO})\cong \pi_{r-1}(\text{SO})$ and we have $\pi_{r}(\text{SO})=0$ for  $r=0,1$, while $\pi_{2}(\text{SO})={\mathbb Z}_2$. 
It then follows from the hypotheses on $M^k$ and~\cite[Theorem 10.3]{ranicki} that for $k\geq 6$, one can apply surgery to $f$ and $M^k$ so as to construct a manifold $N^k$ and a map $g: N^k \to \text{BSO}$ such that $N^k$ is closed, orientable and smooth, $g$ is a classifying map for the stable normal bundle $\nu N^k$ and $g_{*}: \pi_{r}(N^k) \to \pi_{r}(\text{BSO})$ is an isomorphism for $r \leq 2$. Therefore, we have $H_1(N^k,\mathbb{Z})=0$ and $H_{2}(N^k, \mathbb{Z})={\mathbb{Z}_2}$ which implies that $H^{2}(N^k, \mathbb{Z})=0$. Now by definition of the Stiefel-Whitney classes, we have $w_{2}(TN^k)=g^{*}(w_{2}(\text{BSO}))$, where $w_{2}(\text{BSO})$ is the second Stiefel-Whitney class of the universal bundle over $\text{BSO}$, which is non-zero. Since $g_{*}$ is an isomorphism, it follows that $g^{*}: H^{2}(\text{BSO}, \mathbb{Z}_{2}) \to H^{2}(N^k, \mathbb{Z}_{2})$ is an isomorphism. Therefore, $w_{2}(TN^k)\neq 0$, as claimed.   

\subsection{The Spin Group}\label{SpinGroup}
In this section we will show
that the usual $\text{Pin}$ and $\text{Spin}$ groups arise naturally in the context of
topological fermion systems by looking at the subgroups of automorphisms of the spin spaces
that stabilize Clifford subspaces.
We again assume that the topological fermion system of signature~$({\mathfrak{p,q}})$
is regular (see Definition~\ref{defreg}).
We denote the group of unitary endomorphisms of~$S_x$ by~$\U(S_x)$;
it is isomorphic to the group~$\U({\mathfrak{q,p}})$.
Moreover, we consider endomorphisms~$U$ of~$S_x$ which are unitary up to a sign, i.e.
\beq \label{Adef}
U^* \,U = \1 \qquad \text{or} \qquad U^* \,U = -\1
\eeq
(where the star again denotes the adjoint with respect to the spin scalar product~$\Sl .|. \Sr_x$).
Clearly, in the case of a topological fermion system, when the spin scalar product
is positive definite, the minus sign in~\eqref{Adef} cannot occur.
But in the case of a causal fermion system, the operators satisfying~\eqref{Adef}
form a group which contains the unitary group~$\U(S_x)$ as a proper subgroup.
We denote this extension of the unitary group by~$\A(S_x)$.

Given a Clifford subspace~$K \in {\mathcal{K}}^{(r,s)}_x$,
we introduce the following {\em{stabilizer subgroup}} of~$\A(S_x)$:
\beq \label{defstable}
\G_x^K = \left\{ U \in \A(S_x) \text{ with } U K U^{-1} = K \right\} \:.
\eeq
Special mappings in~$\G_x^K$ can be constructed using Clifford multiplication,
as we now explain. Suppose that~$v \in K$ is a unit vector (which may be timelike or
spacelike, i.e.\ $v^2= \pm \1$).
Then~$v^* v = v^2 = \pm\1$, so that~$v \in \A(S_x)$.
Using the anti-commutation relations,
one verifies that~$v K v^{-1} = K$. Therefore, $v$ can be regarded as an operator in~$\G_x^K$,
\beq
v \in \G_x^K \qquad \text{if~$v \in K$, $v^2= \pm \1$\:.} \label{odd1}
\eeq

\begin{Def} The {\bf{pin group}} $\Pin_x$ is defined as the subgroup of~$\G_x^K$
generated by products of operators of the form~\eqref{odd1}.
The {\bf{spin group}} $\Spin_x$ is defined as the subgroup of~$\Pin_x$
generated by products of even numbers of the operators~\eqref{odd1}.
\end{Def}

We now explain how the groups $\Pin_x$ and $\Spin_x$ are related to the usual pin and
spin groups defined on a Clifford algebra (see for example~\cite{baum, lawson+michelsohn},
the concise summary in~\cite[Section~2]{baer+gauduchon}
or similarly~\cite{friedrich} in the Riemannian setting). To this end, we let~$\Cl(K, \la .,. \ra)$ be the
real Clifford algebra on the inner product space~$(K, \la.,. \ra)$ (thus it is the algebra generated by
vectors in~$K$ with the only relation~$v w + w v = 2 \la v, w \ra$; see~\cite[\S I.1]{lawson+michelsohn},
where for consistency we use a different sign convention).
In view of the anti-commutation relations~\eqref{anticommute2}, in our setting~$\Cl(K, \la .,. \ra)$
is not only an abstract Clifford algebra, but it comes with a representation by
operators on~$(S_x, \Sl .|. \Sr_x)$.
Likewise, the groups~$\Pin_x$ and~$\Spin_x$ are not only isomorphic to the usual pin and spin groups
(see for example~\cite[\S II.1]{lawson+michelsohn}),
but they come with a representation on the inner product space~$(S_x, \Sl .|. \Sr_x)$.
An element~$g \in \Spin_x$ also acts on~$K$ by conjugation, giving rise to the mapping
\beq \label{Odef}
{\mathcal{O}}(g) \::\: K \rightarrow K\:,\qquad {\mathcal{O}}(g)\, u := g u g^{-1} \:.
\eeq
Since conjugation does not change the anti-commutation relations, this operator is an
isometry of~$(K, \la .,. \ra)$. 
Moreover, conjugation by an operator~$v$ as in~\eqref{odd1} flips the sign 
of all vectors which are orthogonal to~$v$. Consequently, an even number of such conjugations
preserves the orientation of~$K$.
We thus obtain a mapping
\[ {\mathcal{O}} \::\: \Spin_x \rightarrow \SO(K, \la .,.\ra) \:. \]
This mapping is the projection realizing~$\Spin_x$ as the
two-fold cover of~$\SO(K, \la .,.\ra)$.
We remark that restricting to the subgroup of the spin group consisting of unitary operators,
\[ \Spin_x^\uparrow := \Spin_x \cap \U(S_x) \:, \]
we obtain the two-fold cover of the
orthochronous (i.e.\ time orientation preserving) and orientation preserving isometries of~$K$.

We finally decompose the representation of $\Cl(K, \la .,. \ra)$
on~$(S_x, \Sl .|. \Sr_x)$ into irreducible components.
Denoting this representation by~$\rho$, such a decomposition clearly exists
(see~\cite[Proposition~I.5.4]{lawson+michelsohn}).
Moreover, the number~$\nu$ of inequivalent irreducible representations is given by
(see~\cite[Theorem~I.5.7]{lawson+michelsohn})
\[ \nu = \left\{ \begin{array}{cl} 1 & \text{if~$r+s$ is even} \\
2 & \text{if~$r+s$ is odd} \end{array} \right. \]
(the fact that~$K \subset \Symm(S_x)$ consists of symmetric operators
may give constraints for the possible representations, but this is irrelevant
for the following argument). For clarity, we treat the two cases after each other.

If~$r+s$ is even, we denote the irreducible representation by~$\Delta$ acting on an inner
product space~$E$. We then obtain the decomposition
\[ S_x = E \otimes V \qquad \text{and} \qquad \rho(u) = \Delta(u) \otimes \1 \:, \]
where~$V$ is another inner product space.
Now we can read off the stabilizer group~$\G_x^K$. If a linear operator in~$\A(S_x)$
commutes with all~$\Delta(u)$, its eigenspaces are invariant under~$\Delta$.
Since~$\Delta$ is irreducible, the eigenspaces are either trivial or they coincide with~$E$.
We conclude that
\begin{align}
\G_x^K &= {\mathcal{F}} \otimes \A(V) \:,
\end{align}
where~${\mathcal{F}}$ are the transformations in~$\A(E)$ which leave~$K$ invariant.

If~$r+s$ is odd, the above formulas must be modified in a straightforward way by
considering the two direct summands. Denoting the
two irreducible representations by~$\Delta_\ell$ acting
on inner product spaces~$E_\ell$, we obtain
\[ S_x = E_1 \otimes V_1 \oplus E_2 \otimes V_2 \qquad \text{and} \qquad \rho(u) = \Delta_1(u) \otimes \1 \oplus \Delta_2 \otimes \1\:, \]
where~$V_1$ and~$V_2$ are inner product spaces which might be trivial.
Moreover,
\begin{align}
\G_x^K &= \big\{ \Delta_1(g) \otimes \A(V_1) \oplus \Delta_2(g) \otimes \A(V_2) \:|\: g \in \Pin(r,s) \big\} \:.
\end{align}

\subsection{Construction of Bundle Charts}\label{BundleCharts}
We now show how the choice of a Clifford section~$\Cl M$ of signature~$(r,s)$
gives rise to topological vector bundles with structure groups~$\SO(r,s)$
and~$\G^0 \times \Spin(r,s)$, where~$\G^0$ denotes the stabilizer group
\beq \label{G01}
\G^0 = \left\{ U \in \A(\C^{{\mathfrak{q,p}}}) \text{ with } U v U^{-1} = v \text{ for all~$v \in K$} \right\} \:.
\eeq
In order to choose bundle charts, we fix a matrix representation~$(e_1, \ldots, e_{p+q})$
of a Clifford algebra of signature~$(r,s)$ on~$\C^{{\mathfrak{q,p}}}$,
which is isomorphic to the Clifford subspaces~$\Cl_x$ coming from our Clifford section.
This gives rise to a corresponding Clifford subspace~$K \subset \Lin(\C^{\mathfrak{q,p}})$
as well as a matrix representation of the group~$\G^0$ on~$\C^{\mathfrak{q,p}}$
(see~\eqref{G01}).
For any~$x \in M$, we now choose a pseudo-orthonormal spinor basis~$(\f_1, \ldots, \f_{\mathfrak{p+q}})$
of~$S_x$, i.e.
\[ \Sl \f_\alpha | \f_\beta \Sr_x = s_\alpha \,\delta_{\alpha \beta} \:, \]
where~$s_1, \ldots, s_\q=1$ and~$s_{\q+1}, \ldots, s_{\mathfrak{p+q}}=-1$.
This allows us to write the spinors in~$x$ in components,
\beq \label{psicomp}
S_x \ni \psi = \sum_{\alpha=1}^{\mathfrak{p+q}} \psi^\alpha \,\f_\alpha\:.
\eeq
The spinor basis also gives rise to a matrix representation of~$\Cl_x$.
A pseudo-orthonormal spinor basis~$(\f_1, \ldots, \f_{\mathfrak{p+q}})$ is called {\em{Clifford compatible}}
if the resulting matrix representation of~$\Cl_x$ coincides with our fixed Clifford subspace~$K$.
A Clifford compatible spinor basis makes it possible to
represent the vectors in~$\Cl_x$ as vectors in~$\R^{r,s}$ by
\beq \label{veccomp}
\Cl_x \ni u = \sum_{i=1}^{r+s} u^i \,e_i \:.
\eeq
By construction, Clifford compatible spinor bases are related to each other by the action of the
spin group~$\G^0_x \times \Spin_x$, i.e.\ by transformations of the form
\[ \f_\alpha \rightarrow \tilde{\f}_\alpha = U^{-1} \f_\alpha \qquad \text{with} \qquad
U \in \G^0_x \times \Spin_x \:. \]
This transforms the components in~\eqref{psicomp} according to
\beq \label{spintrans}
\psi^\alpha \rightarrow \tilde{\psi}^\alpha = U^\alpha_\beta \psi_\beta \qquad
\text{with} \qquad U=(U^\alpha_\beta) \in \G^0 \times \Spin(r,s) \subset \U(\mathfrak{q,p}) \:.
\eeq
Likewise, the components of the vector in~\eqref{veccomp} transform to
\[ u^i \rightarrow \tilde{u}^i = O^i_j \,u^j \qquad \text{with} \qquad O \in \SO(r,s) \:, \]
where~$O$ is given by~$U e_j U^{-1} = O^i_j e_i$ (and~$U$ is the matrix in~\eqref{spintrans}).

At every point~$p_0 \in M$, we choose a neighborhood~$V$ such that
there is a continuous mapping which to every~$x \in V$ associates a
Clifford compatible spinor basis~$(\f_1(x),\ldots, \f_{\mathfrak{p+q}}(x))$
(such continuous families of spinor bases can be constructed similar as explained in Section~\ref{secbundle}
using projections and polar decompositions in~$\H$). Moreover, we choose a chart in~$V$.
This gives an atlas for the bundles.

\subsection{Spin Structures} \label{secspinstruct}
In this section we will explain that the usual concept of a spin structure on the tangent bundle of a manifold 
has a counterpart in the context of causal fermion systems, provided that the system satisfies
certain assumptions, which we now state. 
As in Section~\ref{seccliff}, we assume that our topological fermion system is regular
of signature~$(\p, \q)$. Moreover, we again assume that there is a Clifford section
of signature~$(r,s)$ (see~\eqref{Cliffsec}).
In addition, we now assume that the support~$M^k$ of the universal measure is a {\em{differentiable manifold}}
(see Section~\ref{secbundiff}) and that the dimension of the manifold coincides with that of
the Clifford subspaces~\eqref{spincoind}.
\begin{Def} \label{defspinstruct} A bundle isomorphism
\[ \gamma \::\: TM \rightarrow \Cl M \]
is referred to as a {\bf{spin structure}}.
\end{Def} \noindent
A spin structure gives rise to the usual Clifford multiplication,
\[ \gamma \::\: T_pM \rightarrow \Symm(S_p) \:, \]
satisfying the anti-commutation relations
\[ \gamma(u)\, \gamma(v) + \gamma(v)\, \gamma(u) = 2\, \la u,v \ra_x \:, \]
where~$\la .,. \ra_x$ is the bilinear form in~\eqref{anticommute2} of signature~$(r,s)$.
Denoting this bilinear form by~$g$, we obtain a {\em{pseudo-Riemannian manifold}}~$(M,g)$
of signature~$(r,s)$.
We point out that, in contrast to the usual construction on spin manifolds, we do not need to assume
an orientation neither of the manifold~$M$ nor of the the Clifford subspaces.

The existence of spin structures is subject to the usual topological obstruction:
\begin{Thm} \label{thmspin}
A spin structure exists only if the second Stiefel-Whitney class of~$TM$ vanishes.
\end{Thm} \noindent
This theorem follows from a more general result which applies to vector bundles over a topological manifold
whose fibres are modules for the action of the spin group $\Spin(r,s)$.
We now show for the sake of completeness that in this setting, the usual topological condition guaranteeing the existence of a spin structure through the vanishing of the second Stiefel-Whitney class of the bundle is satisfied
on a regular causal fermion systems admitting a Clifford section.
Theorem~\ref{thmspin} will follow as a corollary. We note that the topological arguments appearing in this section are not new. They are essentially an adaptation of the argument given in~\cite[page~83]{lawson+michelsohn}. (We refer to~\cite{milnor+stasheff} and~\cite{mccleary} for the construction of characteristic classes and the
Serre exact sequence.)

Thus we consider the topological vector bundle $\B$ with structure group $G=\rm{Spin}(r,s)$ determined by a
given Clifford section~$\Cl M$. To define a spin structure on $\B$, it is convenient to work with the principal
bundle $\pi_{P(\B)}: P(\B)\to M$ associated to $\B$. In this setting, a spin structure is a twofold cover
$$\begin{array}[c]{ccccc}
{\tilde P}({\B})&\stackrel{p}\rightarrow&P(\B)&\\
&\searrow&\downarrow&\\
&&M
\end{array}$$
such that the mapping~$p|_{{\tilde P}_{x}}:{\tilde P}_{x} \to  P_{x}$ (where ${\tilde P}_{x}=\pi_{\tilde{P}(\B)}^{-1}(\{x\})$ and $P_{x}=\pi_{P(\B)}^{-1}(\{x\})$ denote respectively the fibres of $\tilde{P}(\B)$ and $P(\B)$ over $x\in M$), is a copy of the universal covering map ${\tilde P}_{x}\simeq {\rm{Spin}}(r,s)\to P_{x}\simeq \SO(r,s)$. Note that, by the
constructions in Section~\ref{seccliff}, we know that a Clifford section~$\Cl M$ defines precisely such a twofold cover.  
We conclude by standard covering space theory that a spin structure on $\B$ is given by a cohomology class~$[\sigma]\in H^{1}(P(\B),{\mathbb{Z}}_{2})$. The restriction-induced homomorphism $H^{1}(P(\B),{\mathbb{Z}}_{2})\to H^{1}(P_{x},{\mathbb{Z}}_{2})$ 
maps the cohomology class~$[\sigma]$ 
to a generator of $H^{1}(P_{x},{\mathbb{Z}}_{2})\simeq {\mathbb{Z}}_{2}$. Therefore,
$\B$ admits a spin structure if and only if the sequence
\begin{equation}\label{exact1}
0\to H^{1}(M,{\mathbb{Z}}_{2}) \to H^{1}(P(\B),{\mathbb{Z}}_{2}) \to H^{1}(P_{x},{\mathbb{Z}}_{2}) \to 0
\end{equation}
is exact, and a spin structure on $\B$ is simply a splitting of this sequence. On the other hand, the Serre spectral sequence \cite{mccleary} implies that the sequence
\begin{equation}\label{exact2}
0\to H^{1}(M,{\mathbb{Z}}_{2}) \to H^{1}(P(\B),{\mathbb{Z}}_{2}) \to H^{1}(P_{x},{\mathbb{Z}}_{2}) \to H^{2}(M,{\mathbb{Z}}_{2})\to \cdots,
\end{equation}
is exact, with the image of the generator of $H^{1}(P_{x},{\mathbb{Z}}_{2})\simeq {\mathbb{Z}}_{2}$ in $H^{2}(M,{\mathbb{Z}}_{2})$ being the second Stiefel-Whitney class $w_{2}(\B)$.
We conclude that~$\B$ admits a spin structure if and only if $w_{2}(\B)=0$.

\section{Further Examples} \label{secaddex}
In this section we illustrate in various examples how spin structures as well as complex and K\"ahler
structures can be encoded in a topological fermion system.
The reader not interested in these examples may skip this section.

\subsection{Compact Riemannian Spin Manifolds}
We now specialize the setting of a spin manifold considered in Section~\ref{secex}
to the case that~$(\scrM, g)$ is compact and Riemannian.
We let~$\H \subset L^2(\scrM, S\scrM) \cap C^0(\scrM, S\scrM)$ be a closed subspace
of the Hilbert space of square-integrable spinors (with the scalar product~\eqref{sprod-H})
composed of continuous wave functions.
We again introduce the local correlation operators~$F(p)$ by~\eqref{loccorr}.
Taking the push-forward measure~$\rho = F_* \mu$ gives a Riemannian fermion system~$(\H, \F, \rho)$
whose spin dimension coincides with the fiber dimension of~$S\scrM$.

In order get a connection between the spinor space~$S_p\scrM$
and the corresponding spin space~$S_x$ (see before~\eqref{ssp}),
for any~$p \in \scrM$ we define the {\em{evaluation map}}~$e_p$ by
\[ e_p \::\: \H \rightarrow S_p\scrM \:,\qquad
e_p \,\psi = \psi(p)\:. \]
Its adjoint is defined as usual, taking into account the corresponding inner products on the
domain and the target space, i.e.
\[ \la (e_p)^* \chi \,|\, \psi \ra_\H = \Sl \chi \,|\,e_p \,\psi \Sr_x
\qquad \text{for all~$\chi \in S_p\scrM$ and~$\psi \in \H$}\:. \]
We denote this adjoint by~$\iota_p$,
\[ \iota_p := (e_p)^* \::\: S_p \scrM \rightarrow \H\:. \]
Multiplying~$e_p$ by~$\iota_p$ gives us
back the local correlation operator~$F(p)$. Namely,
\begin{align*}
\la \psi \,|\, F(p)\, \phi \ra_\H = 
- \overline{(\psi)(p)} (\phi)(p)
= -\overline{\big( e_p \psi \big)} \big(e_p \phi \big)
= - \la \psi \,|\, \iota_p e_p \,\phi \ra_\H
\end{align*}
and thus
\beq \label{Fepsdef}
F(p) = -\iota_p \,e_p
= -\iota_p \,\big(\iota_p)^* \::\: \H \rightarrow \H \:.
\eeq

The next proposition gives the desired connection between the spinor space~$S_p\scrM$
and the corresponding spin space~$S_x$..
\begin{Prp} \label{prpisometry}
For any~$p \in \scrM$ and~$x=F(p)$, the mapping
\[ e_p|_{S_x} \::\: S_x \rightarrow S_p \scrM \quad
\text{is an isometric embedding}\:. \]
If the point~$x$ is regular (see Definition~\ref{defreg}), the inverse is given by
\beq \label{invform}
\big(e_p|_{S_x}\big)^{-1} =  -\big( x|_{S_x} \big)^{-1}
\iota_p\::\: S_p\scrM \rightarrow S_x \:.
\eeq
\end{Prp}
\Proof Let~$\psi, \phi \in S_x$. Then
\begin{align*}
\overline{\big( e_p \psi \big)} \big(e_p \phi \big)
&= \la \psi \:|\: (e_p)^* \,e_p \,\phi \ra_\H
= \la \psi \:|\: \iota_p \,e_p \,\phi \ra_\H
\overset{\eqref{Fepsdef}}{=} -\la \psi \:|\: x \,\phi \ra_\H = \Sl \psi | \phi \Sr \:.
\end{align*}
Finally, if~$x$ is regular, then
\[ -\big( x|_{S_x} \big)^{-1} \iota_p \:e_p|_{S_x}
\overset{\eqref{Fepsdef}}{=} \big( x|_{S_x} \big)^{-1} \:x|_{S_x} = \1_{S_x} \:, \]
proving that the inverse of~$e_p|_{S_x}$ is indeed given by the expression in~\eqref{invform}.
\QED

This proposition makes it possible to identify the spin space~$S_x \subset \H$ 
endowed with the inner product~$\Sl .|. \Sr_x$
with a subspace of the spinor space~$S_p\scrM$ with the inner product~$\Sl .|. \Sr_p$.
If~$x$ is singular (see Definition~\ref{defreg}),
this is all we can expect, because in this case the spaces~$S_x$
and~$S_p\scrM$ have different dimensions and are clearly not isomorphic.
In most cases of interest, however, the points of~$M$ will be regular,
so that the mapping~$e_p|_{S_x}$ gives a canonical identification of~$S_x$ with~$S_p\scrM$.

\subsection{Almost-Complex Structures on Riemannian Manifolds} \label{secalmostcomplex}
Let~$(\scrM, g)$ be a $2n$-dimensional compact Riemannian manifold (the compactness assumption
could be relaxed by suitable decay conditions at infinity). Assume that~$\scrM$ has an
{\em{almost complex structure}}~$J$, i.e.~$J$ acts linearly on the fibres of the tangent bundle~$T\scrM$
with the property
\begin{itemize}
\item[(i)] $J_p^2 = -\1_{T_p\scrM}\:.$ 
\end{itemize}
Moreover, we assume that the almost-complex structure is compatible with the
Riemannian metric:
\begin{itemize}
\item[(ii)] $J_p$ is an isometry on the fibres,
\[ g_p\big( Ju, Jv \big) = g(u,v) \qquad \text{for all~$u,v \in T_p\scrM$}\:. \]
\end{itemize}
We let~$T^c\scrM = T\scrM \otimes \C$ be the complexified tangent bundle.
Moreover, let~$L^2(\scrM, T^c\scrM)$ be the square integrable sections of~$T^c\scrM$, again with the scalar
product~\eqref{L2sp}. We introduce~$\H$ as a $J$-invariant closed subspace of~$L^2(\scrM, T^c\scrM)$ 
formed of continuous functions.
We again introduce the local correlation operators~$F(p)$ by~\eqref{vector}.
Taking the push-forward measure~$\rho = F_* \mu$ gives a Riemannian fermion system~$(\H, \F, \rho)$
of spin dimension~$2n$.

By pointwise multiplication with~$J$, we obtain a unitary operator on~$\H$, which we again denote by~$J$.
Acting pointwise, it also gives rise to an operator on the restriction to an open subset~$U$
as considered in~\eqref{psirestrict}. Moreover, these operators are compatible with the restriction
maps~$r^U_V$ in~\eqref{rUV}, thereby giving rise to a well-defined mapping on the stalks
of the resulting presheaf,
\[ J_x \::\: S_x \rightarrow S_x \qquad \text{for} \qquad x \in M := \supp \rho \:. \]
The calculation
\begin{align*}
\la J_p u \,|\, F(p) J_p v \ra_\H = -g_{ij} \,\overline{\big(J_p u(p)\big)^i} \,\big(J_p v(p)\big)^j(p)  =
-g_{ij} \,\overline{u^i(p)} \,v^j(p) = \la u | F(p) v \ra_\H
\end{align*}
(valid for any~$u,v \in T^c_p\scrM$) shows that the operators~$J_x$ are indeed unitary
(with respect to the spin scalar product).
Repeating the last calculation for~$J_p^2$ and using the above property~(i), 
one sees that also the operators~$J_x$ acting on the spin spaces square to minus the identity.
We conclude that
\[ J_x^* = -J_x \qquad \text{and} \qquad J_x^2 = -\1_{S_x} \:. \]
In this way, the almost-complex structure has been encoded in the topological fermion system.

\subsection{Complex Structures on Riemannian Manifolds}
Let~$(\scrM, g)$ be a $2n$-dimen\-sio\-nal compact Riemannian manifold with a complex
structure~$J$ (i.e.\ an integrable almost-complex structure; see for example~\cite{chern}).
Thus in addition to the assumptions~(i) and~(ii) in the previous section, we can work with
holomorphic and anti-holomorphic functions on~$\scrM$. Their derivatives~$\partial f$
and~$\overline{\partial} f$ are sections in the complexified cotangent bundle~$(T^c)^*\scrM$.
We choose two subspaces of~$L^2(\scrM, (T^c)^*\scrM)$ generated 
by the following functions:
\begin{align*}
\H^\text{hol} &= \text{span} \big\{ \partial f \quad \text{with $f$ holomorphic} \big\} \\
\H^\text{anti-hol} &= \text{span} \big\{ \overline{\partial} f
\quad \text{with $f$ anti-holomorphic} \big\}
\end{align*}
(where~$J$ is now considered as acting on the cotangent bundle;
note that these subspaces are $J$-invariant).
Exactly as explained in Example~\ref{exbergman} for a complex domain,
these subspaces are closed and consist of continuous functions.
Taking their direct sum, we obtain a Hilbert space~$(\H, \la .|. \ra_\H)$
with scalar product~\eqref{L2sp}.
We again introduce the local correlation operators~$F(p)$ by~\eqref{vector}.
Taking the push-forward measure~$\rho = F_* \mu$ gives a Riemannian fermion system~$(\H, \F, \rho)$
of spin dimension~$2n$.

Similar as explained in the previous section, the complex structure gives rise to the operators
\[ J \::\: \H \rightarrow \H \quad \text{unitary} \:,\qquad J^2 =-\1_\H, \quad J^* = -J \]
and corresponding operators~$J_x$ acting on the spin spaces,
\[ J_x \::\: S_x \rightarrow S_x \quad \text{unitary} \:,\qquad J_x^2 = -\1_{S_x}, \quad J_x^* = -J_x  \:. \]
Moreover, the eigenspaces of~$J$ are now formed of holomorphic and anti-holomorphic
sections of~$(T^c)^*\scrM$, respectively.
Note that, in contrast to the operators~$J_x$ describing the almost-complex structure in 
Section~\ref{secalmostcomplex} above,
the operator~$J$ acts on the Hilbert space of sections of the complexified bundle,
thereby encoding the integrability condition in the Riemannian fermion system.

\subsection{K\"ahler Structures} Let~$(\scrM,g,J)$ be a compact K\"ahler manifold.
Thus it is a complex manifold with the additional property that the complex structure~$J$ is parallel
with respect to the Levi-Civita connection, i.e.
\beq \label{nablaJ}
\nabla J = 0 \:.
\eeq
The idea is to build in this condition by a suitable choice of~$\H$, as we now explain.
In order to give a somewhat different viewpoint, we consider all holomorphic sections.
But similarly to the previous section, one could include anti-holomorphic
sections and/or restrict attention to holomorphic sections of the form~$\partial f$
and~$J \partial f$.

We choose~$\H$ as
\begin{align*}
\text{span} \Big\{ (u, \alpha, &\nabla_u \alpha) \quad \text{with $u \in T^c\scrM$ holomorphic,
$\alpha \in (T^c)^*\scrM$ holomorphic} \big\} \\
& \subset L^2 \big(\scrM, T^c\scrM \big) \oplus L^2 \big(\scrM, (T^c)^*\scrM \big) \oplus
L^2 \big( \scrM, (T^c)^*\scrM \big) \:.
\end{align*}
Exactly as explained in Example~\ref{exbergman},
this subspace is closed and consists of continuous functions.
We again introduce the local correlation operators~$F(p)$ by~\eqref{vector}.
Taking the push-forward measure~$\rho = F_* \mu$ gives a Riemannian fermion system~$(\H, \F, \rho)$
of spin dimension~$6n$.

The condition~\eqref{nablaJ} gives rise to an action of~$J$ on~$\H$ by
\[ J (u, \alpha, \nabla_u \alpha) = (u, J \alpha, J \nabla_u \alpha) = \big( u, J \alpha, \nabla_u (J \alpha) \big) \:. \]
It gives rise to corresponding operators on~$\H$ and on the spin spaces,
\begin{align*}
J &\::\: \H \rightarrow \H \quad \text{unitary} \:,& J^* &= -J \:,\quad
\;\;\;\;\;\,\,J^2 (u,\alpha,\beta) = (u,-\alpha,-\beta) \\
J_x &\::\: S_x \rightarrow S_x \quad \text{unitary} \:,& J_x^* &= -J_x \:,\quad
J_x^2 (u,\alpha,\beta)(x) = (u,-\alpha,-\beta)(x) \:.
\end{align*}
In this way, the K\"ahler structure is encoded in the Riemannian fermion system.

\section{Tangent Cone Measures and the Tangential Clifford Section} \label{sectangential}
In the previous section, we saw that a Clifford section is essential for giving a
topological fermion system the additional structure of a topological spinor bundle.
We also worked out the topological obstructions for the existence of a Clifford section.
The important remaining question is how to choose a Clifford section. In particular, can the
universal measure be used to distinguish a specific Clifford section?
We shall now analyze this question, giving an affirmative answer in terms of
the so-called {\em{tangential Clifford section}}.
Before beginning, we point out that all our constructions will be {\em{local}} in the sense that
they will involve the universal measure only in an arbitrarily small neighborhood of a given point~$x \in M$.
Consequently, the assumptions needed for the constructions to work will also be local,
making it possible to easily verify them in concrete examples by direct computation.
Provided that these local conditions are fulfilled at every point~$x \in M$, we shall obtain the
tangential Clifford section, implying in particular that the topological obstructions
of Section~\ref{secobstruct} are fulfilled. In this way, we get a connection between local properties
of the topological fermion system and its global topological structure.

\subsection{The Tangent Cone Measure} \label{sectcm}
In this section we again assume that the topological
fermion system is regular (see Definition~\ref{defreg}). Moreover, we assume
that the universal measure~$\rho$ is {\em{locally bounded}}
in the sense that every~$x \in M$ has an open neighborhood~$U$ with~$\rho(U) < \infty$.
Finally, we assume that~$\rho$ is the {\em{completion of a Borel measure}},
meaning that every Borel
set in~$\F$ is $\rho$-measurable and that every subset of a set of measure zero is measurable
(and clearly also has measure zero). For basic definitions see~\cite[\S7 and~\S52]{halmosmt}.
Restricting attention to completions of Borel measures is no major restriction because
the universal measures obtained by minimizing the causal action are always of this form (see~\cite{continuum}).

We want to analyze the subset~$M \subset \F$ in a neighborhood of a given point~$x \in M$. To this end,
it is useful to consider a continuous mapping~${\mathcal{A}}$ from~$M$ to the symmetric operators on the spin
space at~$x$. We always assume that this mapping vanishes at~$x$, i.e.\
\beq \label{Afunct}
{\mathcal{A}} : M \rightarrow \Symm(S_x) \qquad \text{with} \qquad {\mathcal{A}}(x)=0 \:.
\eeq
There are different possible choices for~${\mathcal{A}}$. The simplest choice is
\beq \label{cchain}
{\mathcal{A}} \::\: y \mapsto \pi_x \,(y-x)\, x |_{S_x}\:.
\eeq
Here the factor~$x$ on the right is needed for the operator to be symmetric, because
\begin{align*}
\Sl \psi | {\mathcal{A}} \phi \Sr_x &\overset{\eqref{ssp}}{=} -\la \psi \,|\, x \, (\pi_x \,(y-x)\, x)\, \phi \ra_\H
= -\la \psi \,|\, x \,\,(y-x)\, x\, \phi \ra_\H \\
&\;\:= -\la \pi_x \,(y-x)\, x\, \psi \,|\, x \, \phi \ra_\H 
= \Sl {\mathcal{A}} \psi | \phi \Sr_x \:.
\end{align*}
Alternatively, one can consider mappings involving the operators~$s_y$ or~$\pi_y$, like
\begin{align}
{\mathcal{A}} &\::\: y \mapsto \pi_x \,(s_y-s_x)\, x |_{S_x}\label{As} \\
{\mathcal{A}} &\::\: y \mapsto \pi_x \,(\pi_y-\pi_x)\, x |_{S_x} \label{Ap}
\end{align}
(where~$\pi_x$ again denotes the orthogonal projection in~$\H$ on~$S_x$).
Moreover, one might be interested more specifically in the contributions which are block diagonal or
block off-diagonal in the decomposition~\eqref{spinsplit}, which we denote by
\beq \label{Adodef}
{\mathcal{A}}^\text{d} := \frac{1}{2} \Big( {\mathcal{A}} + s_x \,{\mathcal{A}}\, s_x \Big) \qquad \text{and} \qquad
{\mathcal{A}}^\text{o} := \frac{1}{2} \Big( {\mathcal{A}} - s_x \,{\mathcal{A}}\, s_x \Big) \:.
\eeq
In the applications, the results will depend sensitively on how the functional~${\mathcal{A}}$ is chosen.
However, in the following construction we do not need to specify~${\mathcal{A}}$.

Let~${\mathcal{A}}$ be any functional~\eqref{Afunct}.
In order to get finer information, one can introduce {\em{scaling parameters}}~$\alpha$ and~$\beta$
with~$0 \leq \alpha < \beta$. Defining the set
\[ \Lambda^{\alpha, \beta} = \left\{ y \in \F \:|\: \|y-x\|^\beta < \| {\mathcal{A}}(y)\| < \|y-x\|^\alpha \right\} \]
and multiplying by its characteristic function, we obtain the measure
\beq \label{rhoabdef}
\rho^{\alpha, \beta} := \chi_{\Lambda^{\alpha, \beta}} \, \rho \:.
\eeq
This construction is illustrated on the left of Figure~\ref{figcone}.
\begin{figure} %
\begin{picture}(0,0)%
\includegraphics{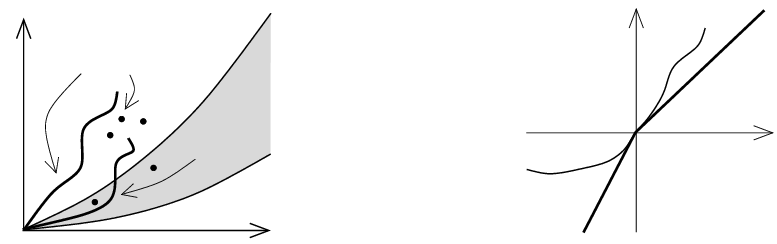}%
\end{picture}%
\setlength{\unitlength}{2486sp}%
\begingroup\makeatletter\ifx\SetFigFont\undefined%
\gdef\SetFigFont#1#2#3#4#5{%
  \reset@font\fontsize{#1}{#2pt}%
  \fontfamily{#3}\fontseries{#4}\fontshape{#5}%
  \selectfont}%
\fi\endgroup%
\begin{picture}(9857,3073)(-3084,-6963)
\put(-314,-5311){\makebox(0,0)[lb]{\smash{{\SetFigFont{11}{13.2}{\familydefault}{\mddefault}{\updefault}$\Lambda^{\alpha, \beta}$}}}}
\put(-3069,-6876){\makebox(0,0)[lb]{\smash{{\SetFigFont{11}{13.2}{\familydefault}{\mddefault}{\updefault}$x$}}}}
\put(-2614,-4241){\makebox(0,0)[lb]{\smash{{\SetFigFont{11}{13.2}{\familydefault}{\mddefault}{\updefault}$y \in \F$}}}}
\put(-2219,-4646){\makebox(0,0)[lb]{\smash{{\SetFigFont{11}{13.2}{\familydefault}{\mddefault}{\updefault}${\supp{\rho}}$}}}}
\put(-844,-5766){\makebox(0,0)[lb]{\smash{{\SetFigFont{11}{13.2}{\familydefault}{\mddefault}{\updefault}${\supp{\rho^{\alpha, \beta}}}$}}}}
\put(196,-6566){\makebox(0,0)[lb]{\smash{{\SetFigFont{11}{13.2}{\familydefault}{\mddefault}{\updefault}$\|x-y\|$}}}}
\put(481,-5841){\makebox(0,0)[lb]{\smash{{\SetFigFont{11}{13.2}{\familydefault}{\mddefault}{\updefault}$\|{\mathcal{A}}(y)\|=\|x-y\|^\beta$}}}}
\put(461,-4121){\makebox(0,0)[lb]{\smash{{\SetFigFont{11}{13.2}{\familydefault}{\mddefault}{\updefault}$\|{\mathcal{A}}(y)\|=\|x-y\|^\alpha$}}}}
\put(3000,-6341){\makebox(0,0)[lb]{\smash{{\SetFigFont{11}{13.2}{\familydefault}{\mddefault}{\updefault}$\supp(
\mathcal{A}_* \rho^\bullet)$}}}}
\put(3476,-4541){\makebox(0,0)[lb]{\smash{{\SetFigFont{11}{13.2}{\familydefault}{\mddefault}{\updefault}$\Symm(S_x)$}}}}
\put(5986,-4916){\makebox(0,0)[lb]{\smash{{\SetFigFont{11}{13.2}{\familydefault}{\mddefault}{\updefault}${\mathcal{C}}^\bullet_x$}}}}
\end{picture}%
\caption{The scaling parameters~$\alpha, \beta$ and the tangent cone~${\mathcal{C}}^\bullet_x$.}
\label{figcone}
\end{figure}
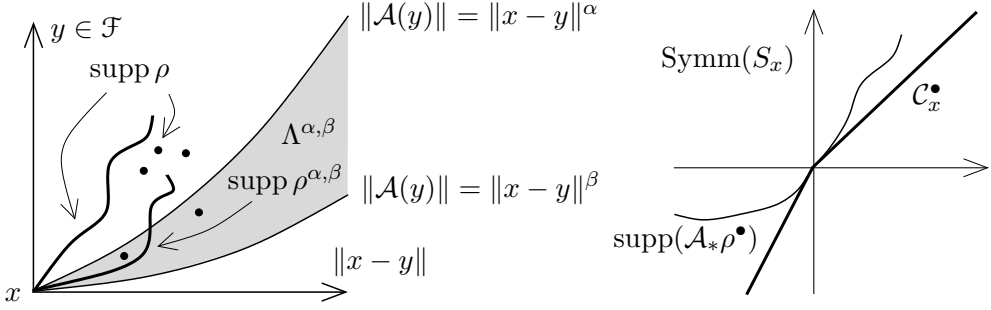 %
We now let~$\rho^\bullet$ be either the measure~$\rho$ or the measure~$\rho^{\alpha, \beta}$.
A {\em{conical set}} is a set of the form~$\R^+ A$ with~$A \subset \Symm(S_x)$.
We denote the conical sets whose preimages under~${\mathcal{A}}$
are both $\rho$- and~$\rho^\bullet$-measurable by~${\mathfrak{M}}$.
For a conical set~$A \subset \Symm(S_x)$ we consider countable coverings by
measurable conical sets,
\[ A \subset \bigcup_{k=1}^\infty A_k \qquad \text{with} \qquad
\text{$A_k \in {\mathfrak{M}}$}\:. \]
We denote the set of such coverings by~${\mathcal{P}}$. We define
\beq \label{defmus}
\mu^*_\con(A) = \inf_{\mathcal{P}} \;\sum_{k=1}^\infty \: \liminf_{\delta \searrow 0}\:
\frac{1}{\rho \big( B_\delta(x) \big)}\: 
\rho^\bullet \Big( {\mathcal{A}}^{-1} \big(  A_k \big) \cap B_\delta(x) \Big)
\eeq
(where~$B_\delta(x) \subset \Lin(\H)$ is the Banach space ball).
We remark for clarity that, since~$x \in M := \supp \rho$, it follows that the
measure~$\rho( B_\delta(x))$ is non-zero.

\begin{Lemma} $\mu^*_\con$ is an outer measure on the conical sets in~$\Symm(S_x)$.
\end{Lemma}
\Proof According to the definitions (see~\cite[\S7 and~\S10]{halmosmt}), we need to show that \\[-0.8em]
\begin{itemize}
\item[(i)] $\mu^*_\con(\varnothing)=0$.
\item[(ii)] $A \subset B \Rightarrow \mu^*_\con(A) \leq \mu^*_\con(B) \quad\:\,$ (monotonicity).
\item[(iii)] $\displaystyle \mu^*_\con \Big( \bigcup_{n=1}^\infty A_n \Big) \leq \sum_{n=1}^\infty
\mu^*_\con(A_n) \quad$ (countable subadditivity).
\end{itemize}
The properties~(i) and~(ii) are obvious from the definition of~$\mu^*_\con$.
In order to show~(iii), for any~$\varepsilon>0$ and~$n \in \N$, we choose coverings~$(A^{(n)}_k)_{k \in \N}$
of~$A_n$ by measurable sets such that
\[ \sum_{k=1}^\infty \: \liminf_{\delta \searrow 0}\:
\frac{1}{\rho \big( B_\delta(x) \big)}\: 
\rho^\bullet \Big( {\mathcal{A}}^{-1} \big( A^{(n)}_k \big) \cap B_\delta(x) \Big)
\leq \mu^*_\con(A_n) + \varepsilon\, 2^{-n} \:. \]
Then the sets~$(A^{(n)}_k)_{n,k \in \N}$ are a countable covering of~$A$ and thus
\[ \mu^*_\con(A) \leq \sum_{k,n=1}^\infty \: \liminf_{\delta \searrow 0}\:
\frac{1}{\rho \big( B_\delta(x) \big)}\: 
\rho^\bullet \Big( {\mathcal{A}}^{-1} \big( A^{(n)}_k \big) \cap B_\delta(x) \Big)
\leq \sum_{n=1}^\infty \mu^*_\con(A_n) + \varepsilon \:. \]
Since~$\varepsilon$ is arbitrary, the result follows.
\QED
Now we can apply the Carath{\'e}odory extension lemma to construct a corresponding measure.
To this end, we say that a conical set~$A$ is $\mu_\con${\bf{-measurable}} if
\beq \label{Caracond}
\mu^*_\con(Z) = \mu^*_\con(Z \cap A) + \mu^*_\con(Z \setminus A)
\qquad \text{for every conical set~$Z$}\:.
\eeq
By the Carath{\'e}odory extension lemma (see for example~\cite[\S12]{halmosmt}
or~\cite[Lemma~2.8]{broeckerana2}), the $\mu_\con${\bf{-measurable}} sets form a $\sigma$-algebra,
and~$\mu^*_\con$ is a measure on this $\sigma$-algebra.
We now show that this $\sigma$-algebra includes all conical Borel sets:
\begin{Lemma} \label{lemmameas}
The $\mu_\con${\bf{-measurable}} sets form a $\sigma$-algebra~${\mathfrak{M}}_\con$
which contains all conical Borel sets in~$\Symm(S_x)$. The mapping~$\mu_\con := \mu^*_\con|_{{\mathfrak{M}}_\con} :
{\mathfrak{M}}_\con \rightarrow [0, \infty]$ is a measure.
\end{Lemma}
\Proof It remains to show that~\eqref{Caracond} holds for any measurable conical
set~$A \in {\mathfrak{M}}$ and any conical set~$Z$. The inequality~$\mu^*_\con(Z) \leq \mu^*_\con(Z \cap A)
+ \mu^*_\con(Z \setminus A)$ follows from the countable subadditivity of~$\mu^*_\con$.
In order to prove the opposite inequality, for any given~$\varepsilon>0$ we choose
a covering~$(A_k)_{k \in \N}$ of~$Z$ by measurable conical sets such that
\[ \sum_{k=1}^\infty \liminf_{\delta \searrow 0}\:
\frac{1}{\rho(B_\delta(x))}\: \rho^\bullet \big( 
{\mathcal{A}}^{-1} (A_k) \cap B_\delta(x) \big)  \leq \mu^*_\con(Z) + \varepsilon \:. \]
Then the sets~$B_k := A_k \cap A$ and~$C_k := A_k \setminus A$ are again measurable,
and using the additivity of the measure~$\mu^\bullet$, we obtain
\[ \sum_{k=1}^\infty \liminf_{\delta \searrow 0}\:
\frac{\rho^\bullet \big( {\mathcal{A}}^{-1}(B_k) \cap B_\delta(x) \big)}{\rho(B_\delta(x))}
+ \sum_{k=1}^\infty \liminf_{\delta \searrow 0}\:
\frac{\rho^\bullet \big( {\mathcal{A}}^{-1}(C_k) \cap B_\delta(x) \big)}{\rho(B_\delta(x))}
\leq \mu^*_\con(Z) + \varepsilon \:. \]
Since the sets~$(B_k)$ and~$(C_k)$ cover~$Z \cap A$ respectively~$Z \setminus A$, it follows that
\[ \mu^*_\con(Z \cap A) + \mu^*_\con(Z \setminus A) \leq \mu^*_\con(Z) + \varepsilon \:. \]
Since~$\varepsilon$ is arbitrary, the result follows.
\QED

For convenience, we usually restrict the measures to the Borel sets, keeping in mind that we
can take the completion whenever needed. Then the result of Lemma~\ref{lemmameas}
makes it possible to introduce the following notions.
\begin{Def} We denote the conical Borel sets of~$\Symm(S_x)$ by~${\mathfrak{B}}_\con(x)$.
We denote the measure obtained by applying the above construction
with~${\mathcal{A}}$ according to~\eqref{Afunct} by
\[ \mu^\bullet_x \::\: {\mathfrak{B}}_\con(x) \rightarrow [0, \infty] \:. \]
It is referred to as the {\bf{tangent cone measure}} corresponding to~${\mathcal{A}}$.
The {\bf{tangent cone}}~${\mathcal{C}}^\bullet_x$ is defined as the support of the tangent cone measure,
\[ {\mathcal{C}}^\bullet_x := \supp \mu^\bullet_x \subset \Symm(S_x)\:. \]
If~$\rho^\bullet$ is chosen according to~\eqref{rhoabdef}, we also write
the tangent cone as~${\mathcal{C}}^{\alpha, \beta}_x$. Likewise, if~$\rho^\bullet=\rho$,
we simply omit the superscripts.
\end{Def} \noindent
This definition is local in the sense that it only depends on the behavior of the
topological fermion system in an arbitrarily small neighborhood of~$x$.
Also note that~$\mu_x$ is a normalized measure, whereas the measure~$\mu_x^\bullet$
has total volume at most one.
The tangent cone is illustrated on the right of Figure~\ref{figcone}.

The next lemma illustrates the usefulness of the scaling parameters.
\begin{Lemma} \label{lemmaac} We choose~${\mathcal{A}}$ according to~\eqref{As}
and consider the scaling parameters in the range
\[ 0 \leq \alpha < \beta < 2\:. \]
Then the vectors in~${\mathcal{C}}_x^{\alpha, \beta}$ anti-commute with~$s_x$,
\[ \{ s_x, v \} = 0 \qquad \text{for all~$v \in {\mathcal{C}}_x^{\alpha, \beta}$}\,. \]
\end{Lemma}
\Proof Setting~$\Delta s = s_y - s_x$, the functional~\eqref{As} becomes
\[ {\mathcal{A}} = \pi_x \,\Delta s\, x |_{S_x}\:. \]
Since the definition of~${\mathcal{C}}^{\alpha,\beta}$ involves~$y$
only in an arbitrarily small neighborhood of~$x$, we want to treat~$\Delta s$ perturbatively.
We know that for all~$y$, the operator~$s_y$ has the eigenvalues~$\pm 1$.
A standard perturbation argument yields that to first order, the operator~$\Delta s$
vanishes on the eigenspaces of~$s_x$, i.e.
\[ 0 = \{ s_x, \pi_x\, \Delta s\, \pi_x \} + \O \big( (\Delta s)^2 \big) \:. \]
Note that here only the projection of~$\Delta s$ to~$S_x$ enters.
The subtle point is that to second and higher order in perturbation theory, the
remaining contributions to~$\Delta s$, i.e. the operator
\[ \Delta s - \pi_x\, \Delta s\, \pi_x \]
also comes into play, and we must make sure that the resulting contribution to the eigenvalues
of~$s_y$ is  dominated by the first order in perturbation theory. 
Here is where the scaling parameter~$\beta$ enters: For example by expressing the signature
operators with contour integrals, one readily sees that for~$y$ in a small neighborhood of~$x$,
the inequality
\beq \label{Delupper}
\| \Delta s \| \leq c\, \|y-x\|
\eeq
holds (where the constant~$c$ depends on the eigenvalues of~$x$).
According to the definition of~$\Lambda^{\alpha, \beta}$, the measure~$\rho^{\alpha, \beta}$
is non-trivial only for points~$y$ for which
\[ \| {\mathcal{A}}(y) \| \geq \|y-x\|^\beta \gg \|\Delta s\|^2 \:, \]
where in the last step we used~\eqref{Delupper} and made~$y-x$ sufficiently small.
This shows that we may restrict attention to points~$y$ for which the first order in perturbation
theory is non-zero. Then the second and higher orders vanish in the limit~$\delta \searrow 0$.
\QED

\subsection{Construction of a Tangential Clifford Section}
We now address the question whether the geometry of the causal fermion system
in a neighborhood of a given point~$x \in M$ makes it possible to distinguish a specific
Clifford subspace at~$x$. We give a construction which achieves this goal under generic assumptions.

We introduce on~$\Symm(S_x)$ the symmetric bilinear form
\beq \label{symmbil}
\la .,. \ra \::\: \Symm(S_x) \times \Symm(S_x) \rightarrow \C \:,\qquad
\la u, v \ra = \frac{1}{\mathfrak{p+q}}\: \Tr(u v)
\eeq
(note that this bilinear form generalizes~\eqref{anticommute2} because for
two operators~$u, v$ in a Clifford subspace~$K$, the formulas agree).
In the Riemannian case, this  bilinear form~\eqref{symmbil} is positive definite
and coincides with the usual Hilbert-Schmidt scalar product.
For causal fermion systems, however, the inner product~$\la .,. \ra$ is indefinite.
This makes it necessary to treat these two situations separately. We first give the
abstract constructions. A more explicit analysis of the cases shown in Figure~\ref{figcases}
will be given afterwards.

We begin with {\em{Riemannian fermion systems}}, where the construction is somewhat simpler.
At a given point~$x \in M$, we consider a tangent cone measure~$\mu^\bullet_x$
(either without or for a suitable choice of the scaling parameters).
Then~$\mu^\bullet_x$ gives
rise to a Borel measure on the unit sphere~$S_1(0) \subset \Symm(S_x)$ simply by 
taking the measure of the cone~$\R^+ \Omega$ over the corresponding set~$\Omega \subset S_1(0)$.
Denoting the orthogonal projection onto a subspace~$U \subset \Symm(S_x)$ by~$\pi_U$,
the function~$L$ defined by
\beq \label{Ldef}
L(U) = \int_{S_1(0) \subset \text{Symm}(S_x)} \Tr_{\text{Symm}(S_x)}(\pi_U \,\pi_{\bra e \ket}) \: d\mu^\bullet_x(\R^+ e)
\eeq
(where~$\bra e \ket$ denotes the span of~$e$),
is non-negative and tells us about the position of the subspace~$U$ 
relative to the tangent cone~${\mathcal{C}}^\bullet_x$. For instance, if~${\mathcal{C}}^\bullet_x$ is
a linear subspace of~$\text{Symm}(S_x)$, the functional is maximal if~$U$ is contained in this
subspace. This motivates us to maximize over all Clifford subspaces,
\beq \label{max1}
\text{maximize $L(.)$ on~${\mathcal{K}}^{(m,0)}_x$}\:.
\eeq
\begin{Def} \label{defRnondeg} For a Riemannian fermion system,
the tangent cone measure~$\mu_x^\bullet$  is {\bf{non-degenerate}} if for all~$x \in M$,
the optimization problem~\eqref{max1} has a unique
maximizer~$K(x) \in {\mathcal{K}}^{(m,0)}_x$, which depends continuously on~$x$.
Setting~$\Cl_x = K(x)$ defines the {\bf{tangential Clifford section}}.
\end{Def}

In the case of {\em{causal fermion systems}}, we need to construct a distinguished
Clifford extension~$K \in {\mathcal{K}}^{s_x, r}_x$.
Thus our task is to find the vectors which extend~$s_x$ to the Clifford subspace~$K$.
Since these vectors all anti-commute with~$s_x$, it is useful to
introduce the set~$\AC\{s_x\}$ of all symmetric operators on~$S_x$ which
anti-commute with~$s_x$,
\[ \AC\{s_x\} = \big\{ u \in \Symm(S_x) \:|\: \{u, s_x\}=0 \big\} \subset \Symm(S_x) \:. \]

\begin{Lemma} The bilinear form~$\la .,. \ra|_{\AC\{s_x\} \times \AC\{s_x\}}$ is negative definite.
\end{Lemma}
\Proof In view of~\eqref{ssp}, for any~$v \in \Symm(S_x)$ the operator~$xv$
is a symmetric operator on~$\H$. Using that the Hilbert-Schmidt norm is positive, we conclude that
\[ \Tr (xv\,xv) >0 \qquad \text{for all~$v \in \Symm(S_x), v \neq 0$} \]
(here we used that~$x$ is invertible on~$S_x$ by definition of the spin space).
For any~$u \in \AC\{s_x\}$ with~$u \neq 0$, we
choose~$v=|x|^{-\frac{1}{2}} \,u\, |x|^{-\frac{1}{2}}$. Then, using that~$s_x$ has eigenvalues~$\pm1$
and anti-commutes with~$u$, we obtain
\[ \Tr(u u) = \Tr(s_x s_x u u) = - \Tr(s_x u s_x u) =  -\Tr(xvxv) < 0 \:, \]
concluding the proof.
\QED
We now consider a tangent cone measure~$\mu^\bullet_x$ on the block off-diagonal
functional~${\mathcal{A}}^\text{o}$ in~\eqref{Adodef}. Modifying the functional~\eqref{Ldef} to
\beq \label{Ldef2}
L(U) = \int_{S_1(0) \cap \AC\{s_x\}} \Tr_{\text{Symm}(S_x)}(\pi_U \,\pi_{\bra e \ket}) \: d\mu^\bullet_x(\R^+ e) \:,
\eeq
we can formulate the variational principle
\beq \label{max2}
\text{maximize $L(.)$ on~${\mathcal{K}}^{s_x, r}_x$}\:.
\eeq
We note for clarity that the trace in~\eqref{Ldef2} is taken effectively only over~$\AC\{s_x\} \subset \Symm(S_x)$,
because the operator~$\pi_{\bra e \ket}$ vanishes on~$\AC\{s_x\}^\perp$. Thus we could also have written the 
trace in~\eqref{Ldef2} as
\[ \Tr_{\AC\{s_x\}}( \pi_{\AC\{s_x\}} \,\pi_U\, \pi_{\AC\{s_x\}}\;\pi_{\bra e \ket}) \:, \]
but we prefer the more compact notation in~\eqref{Ldef2}.

\begin{Def} \label{defLnondeg} For a causal fermion system, the tangent cone measure~$\mu_x^\bullet$ 
is {\bf{non-degenerate}} if for all~$x \in M$, the
optimization problem~\eqref{max1} has a unique
maximizer~$K(x) \in {\mathcal{K}}^{s_x, r}_x$, which depends continuously on~$x$.
Setting~$\Cl_x = K(x)$ defines the {\bf{tangential Clifford section}}.
\end{Def}

Clearly, a tangential Clifford section can exist only if the general topological obstructions
of Section~\ref{secobstruct} are satisfied. Therefore, if we know that there is
a non-degenerate tangent cone measure (see Definitions~\ref{defRnondeg} and~\ref{defLnondeg}),
we can infer that there are no topological obstructions to the existence of Clifford sections
respectively Clifford extensions. Since the variational principles~\eqref{Ldef} and~\eqref{Ldef2}
are local (because the tangent cone measure~$d\mu_x^\bullet$ only depends on the universal measure
in an arbitrarily small neighborhood of~$x$), we thus obtain a method to deduce global topological properties
of a causal fermion system from its local behavior at every point~$x \in M$.

Examples for the construction of the tangential Clifford section will be given in Sections~\ref{secssEuklid}
and~\ref{secssMin} below. The example of the shifted Dirac sphere (Example~\ref{exsphere}~(ii))
is worked out in detail in the Master's thesis~\cite{fischer}.

\subsection{Construction of a Spin Structure} \label{secconstruct}
We now assume in addition that~$M^k$ is a {\em{differentiable manifold}}
and that the universal measure~$\rho$ is {\em{absolutely continuous}} with respect to the Lebesgue
measure (see Section~\ref{secspinstruct}). Moreover, we assume as in~\eqref{spincoind} that the
dimension of the manifold coincides with the dimension of the Clifford subspaces,
\[ r+s=k\:. \]
Our goal is to construct a spin structure (see Definition~\ref{defspinstruct}).

We first explain how the tangent cone simplifies under the differentiability and regularity
assumptions.
For simplicity, we work with the definition of the tangent space~$u \in T_xM$ as
equivalence classes of curves. For any~$u \in T_xM$ we choose
a representative~$\gamma \::\: (-\varepsilon, \varepsilon) \rightarrow M$
with~$\gamma(0)=x$ and~$\gamma'(0)=u$.
Then~$\gamma(t)$ is a one-parameter family of linear operators in~$\F$.
Composing with the operator~${\mathcal{A}}$, \eqref{Afunct}, we obtain a family of
operators in~$\Symm(S_x)$. We denote the directional derivative by
\beq \label{kappaform}
\dsf_u {\mathcal{A}} :=  \frac{d}{dt} {\mathcal{A}} \big( \gamma(t) \big) \Big|_{t=0} \:.
\eeq
Here the derivative exists in view of our assumption~\eqref{Frechet}.
Obviously, this definition is independent of the choice of the representative~$\gamma$.
Moreover, $\dsf_u {\mathcal{A}}$ is again a symmetric operator on~$S_x$. We thus obtain a mapping
\beq \label{kappadef}
\dsf {\mathcal{A}} \::\: T_xM \rightarrow \Symm(S_x)\:.
\eeq
A short consideration shows that
\[ {\mathcal{C}}_x = \dsf {\mathcal{A}}(T_xM)\:, \]
so that the tangent cone simplifies to a plane in~$\Symm(S_x)$.

We next assume that~$\Cl M$ is a tangential Clifford section
(see Definitions~\ref{defRnondeg} and~\ref{defLnondeg}).
Then at every point~$x \in M$ we can form the mapping
\beq \label{gxRiem}
\gamma_x = \pi_{\Cl_x} \circ \dsf {\mathcal{A}} \::\: T_xM \rightarrow \Cl_x
\eeq
(with~$\dsf {\mathcal{A}}$ according to~\eqref{kappaform} and~\eqref{kappadef}).
Then~$\gamma$ gives rise to a spin structure, provided that the mapping~$\gamma_x$
is bijective at every point~$x \in M$.
We point out that this construction is local, and in applications it is easy to verify whether~$\gamma_x$
is bijective. However, if this local condition is satisfies at every point~$x \in M$, this
gives rise to global properties of~$M$ (see Theorem~\ref{thmspin}).

In applications, it may well happen that the mapping~$\gamma_x$ is {\em{not}} bijective.
In particular, for causal fermion systems, this issue is closely related to the problem of distinguishing a
direction of time. We will analyze this problem in detail in Section~\ref{secssMin} in the example
of two-dimensional Minkowski space. Here we merely
explain the difficulty in the case that~${\mathcal{A}}$
depends only on the signature operator at~$y$.
For clarity, we first restate Lemma~\ref{lemmaac} in the differentiable setting and give a
different proof.
\begin{Lemma} \label{lemmasxk} Let~${\mathcal{A}}$ be chosen according to~\eqref{As}. Then for all~$u \in T_xM$,
\[ \big\{ s_x, \dsf_u {\mathcal{A}} \big\} = 0 \:. \]
\end{Lemma}
\Proof We denote the orthogonal projections onto the positive and negative spectral subspaces
of~$x$ by~$\pi_x^+$ and~$\pi_x^-$, respectively.
Differentiating the relation~$(\pi^+_{\gamma(t)})^2 = \pi^+_{\gamma(t)}$ yields
\[ \pi^+_x\, (\dsf_u \pi^+_x) + (\dsf_u \pi^+_x)\, \pi^+_x = (\dsf_u \pi^+_x)\:, \]
where~$\dsf_u \pi^+_x \equiv \partial_t \pi^+_{\gamma(t)}|_{t=0}$.
Multiplying from the left and right by~$\pi^+_x$, we obtain
\beq \label{pip}
\pi^+_x\, (\dsf_u \pi^+_x)\, \pi^+_x = 0 \:.
\eeq
Similarly,
\beq \label{pim}
\pi^-_x\, (\dsf_u \pi^-_x)\, \pi^-_x = 0 \:.
\eeq
Next, we differentiate the relation~$\pi^+_{\gamma(t)} \pi^-_{\gamma(t)} = 0$ to obtain
\[ \pi^+_x\, (\dsf_u \pi^-_x) + (\dsf_u \pi^+_x)\, \pi^-_x = 0 \:. \]
Multiplying from the left by~$\pi^-_x$, we conclude that
\beq \label{pipm1}
\pi^-_x\, (\dsf_u \pi^+_x)\, \pi^-_x = 0\:.
\eeq
Similarly, multiplying from the right by~$\pi^+_x$, we get
\beq \label{pipm2}
\pi^+_x\, (\dsf_u \pi^-_x)\, \pi^+_x = 0\:.
\eeq

Using the identities~\eqref{pip}--\eqref{pipm2} in the equations
\beq \label{pisrel}
\pi_x = \pi_x^+ + \pi_x^- \:,\qquad s_x = \pi_x^+ - \pi_x^- \:,
\eeq
we obtain
\[ \pi^+_x\, (\dsf_u s_x)\, \pi^+_x = 0 = \pi^-_x\, (\dsf_u s_x)\, \pi^-_x \:. \]
A straightforward calculation using again~\eqref{pisrel} gives the claim.
\QED
Now suppose that~$\Cl_x$ is a Clifford extension (as is always the case if
we use the construction leading to Definition~\ref{defLnondeg}).
Then the statement of Lemma~\ref{lemmasxk} implies that the image of~$\dsf {\mathcal{A}}$
is orthogonal to the vectors~$s_x \in \Cl_x$. As a consequence, the composition~\eqref{gxRiem}
is necessarily singular. We will come back to this problem in Section~\ref{secssMin}.

We finally point out that the above construction should be handled with care in the sense
that it gives topological, but no geometric information on the fermion system.
More specifically, a spin structure~$\gamma \::\: TM \rightarrow \Cl M$ induces
a Riemannian or Lorentzian metric on~$M$ by
\beq \label{Lorentz}
\la X, Y \ra := \frac{1}{2}\, \big\la \gamma(X), \gamma(Y) \big\ra \:.
\eeq
However, it would be too naive to interpret this metric as describing the geometry of space
or space-time. Namely, for a causal fermion system constructed on a globally hyperbolic space-time
(see~\cite{lqg}, \cite[Section~4]{finite} or the introduction~\cite{nrstg}), the Lorentzian metric 
and the spin connection are recovered
by considering the closed chain~$A_{xy}$ for pairs of points~$x,y$ whose distance is much larger than the
regularization length (see the constructions in~\cite{lqg}; note that in~\cite[Theorem~4.7]{lqg}
one first takes the limit~$\varepsilon \searrow 0$ and then the limit~$N \rightarrow \infty$). Moreover,
the Clifford extension at~$x$
must be chosen as a function of~$y$ (as is made precise by the synchronizations and splice maps
introduced in~\cite[Section~3]{lqg}). In view of these constructions, the linearization of~$s_x$
as captured by the mapping~$\dsf {\mathcal{A}}$ in~\eqref{kappaform} does not encode the macroscopic geometry of the causal fermion system. But it can be used to obtain topological information.

\section{The Topology of Discrete and Singular Fermion Systems} \label{secdiscrete}
In the previous Sections~\ref{sectopspin} and~\ref{sectangential}, it was essential that~$M$
was a topological manifold (or at least a finite cell complex, because otherwise the
cohomology groups cannot be introduced). We now explain how our methods and results can be
applied even in cases when~$M$ is so singular that it has no manifold structure or is discrete.
Our technique is to ``extend''~$M$ to a larger set~$\tilde{M} \subset \F$, and to analyze
the topology of the enlarged space. For technical simplicity, we restrict attention to the
case that the Hilbert space~$\H$ is finite-dimensional.

We begin with a simple method which allows us to associate to~$M$ a manifold.
For given~$r>0$, we first take an $r$-neighborhood of~$M$,
\beq \label{Mr}
M_r := B_r(M) \subset \F
\eeq
(where we work with the distance function induced by the $\sup$-norm on~$\Lin(\H)$).
Next, for any~$p, q$ with~$0 \leq p \leq \p$ and~$0 \leq q \leq \q$ we define the sets
\[ \F^{p,q} = \{ x \in \F \,|\, \text{$x$ has~$p$ positive and~$q$ negative eigenvalues} \}\:. \]
Obviously, these sets form a partition of~$\F$,
\[ \F = \bigcup_{p,q=0}^n \F^{p,q} \:, \qquad
\F^{p,q} \cap \F^{p',q'} = \varnothing \quad \text{if~$(p,q) \neq (p',q')$}\:. \]
Now we set
\[ M_r^{p,q} = M_r \cap \F^{p,q}\:. \]
If~$\H$ is finite-dimensional, the sets~$\F^{p,q}$ are manifolds.
Hence the sets~$M_r^{p,q}$ are open submanifolds of~$\F^{p,q}$.
Then~$S M_r^{p,q}$ is a bundle over a smooth manifold,
so that the methods of Sections~\ref{sectopspin} and~\ref{sectangential} apply.
Clearly, the construction depends on the choice of the parameter~$r$.

An alternative method is to use the universal measure~$\rho$ in the construction of~$\tilde{M}$:
Given a parameter~$\delta>0$, we define the function~$r_\delta : \F \rightarrow \R^+_0$ by
\beq \label{rdelta}
r_\delta(x) = \sup \big\{ r \in \R \:|\: \rho \big( B_r(x) \big) < \delta \big\} \:.
\eeq
Moreover, we set
\beq \label{Mdelta}
M_\delta = \bigcup_{x \in M} B_{r_\delta(x)}(x) \qquad \text{and} \qquad
M_\delta^{p,q} = M_\delta \cap \F^{p,q}\:.
\eeq
Again, the sets~$M_\delta^{p,q}$ are smooth submanifolds of~$\F^{p,q}$.

The above constructions give rise to sets~$M_\delta, M_r \subset \F$
(and similarly~$M^{p,q}_\delta, M^{p,q}_r$)
which carry topological information, but unfortunately these sets are no longer
the support of a measure, so that they cannot be regarded as the base spaces
of corresponding topological fermion systems. This disadvantage can be
removed with the following construction, provided that another measure~$\mu$
on~$\F$ is given. In examples when~$\F$ is finite-dimensional, one can choose~$\mu$
as the Lebesgue measure. The infinite-dimensional situation is definitely more difficult,
but one could choose~$\mu$ for example as a Gaussian measure.
Given~$\mu$, we can choose a smooth test function~$\eta_r : C^\infty(\F \times \F, \R^+_0)$
and define a measure~$\rho_r$ by
\beq \label{rhordef}
\rho_r(\Omega) := \int_M \left( \int_{\Omega} \eta_r(x,y)\, d\mu(y) \right) d\rho(x) \:.
\eeq
A typical example is to choose~$\eta_r(x,y) = \eta(\|x-y\|^2/r^2)$
with~$\eta \in C^\infty_0([0,1))$. The effect of this construction is that the universal measure
is ``smeared out'' in the sense that the support of~$\rho_r$ is an $r$-neighborhood
of~$\supp \rho$. Hence the effect on the base space is the same as in the construction~\eqref{Mr},
but with the advantage that~$(\H, \F, \rho_r)$ is again a topological fermion system.
Modifying~\eqref{rhordef} to
\beq \label{rhoddef}
\rho_\delta(\Omega) := \int_M \left( \int_\Omega \eta_{r_\delta(x)}(x,y)\, d\mu(y) \right) d\rho(x) \:,
\eeq
one obtains similarly a universal measure~$\rho_\delta$ whose support coincides with~$M_\delta$
as defined in~\eqref{Mdelta}.

The above constructions will be illustrated in Section~\ref{seclattice}
in the example of a lattice system with a non-trivial topology.

\section{Basic Examples} \label{secexamples}
In this section, we illustrate our abstract constructions in a set of different simple geometric situations and
indicate potential applications.

\subsection{The Euclidean Plane}  \label{sec61}
The Dirac operator on the Euclidean~$\R^2$ can be written as
\beq \label{dir2}
\Dir = i \sigma^1 \partial_{\zeta_1} + i \sigma^2 \partial_{\zeta_2} \:,
\eeq
where we denote the points of~$\R^2$ by~$\zeta=(\zeta_1, \zeta_2)$.
The spinor space~$(S_\zeta \R^2, \Sl .|. \Sr_\zeta)$ at a point~$\zeta$ can be
identified with~$\C^2$ with the canonical Euclidean scalar product. For clarity, we denote
this standard spinor space by~$(Y \simeq \C^2, \Sl .|. \Sr)$.
We shall consider eigensolutions of the Dirac operator corresponding to an eigenvalue~$m \geq 0$,
\beq \label{Direigen}
\Dir \psi = m \psi \:.
\eeq
Particular solutions can be written as plane waves
\beq \label{plane}
\e_{k}(\zeta) = (k_1 \sigma^1 + k_2 \sigma^2 + \1)\,\chi\: e^{-i k\, m \zeta} \:,
\eeq
where the momentum (which for convenience we rescaled by the mass)
lies on the unit sphere, $k := (k_1, k_2) \in S^1 \subset \R^2$,
and~$\chi$ is the fixed spinor~$\chi=(1,0)$ (note that the matrix~$k \!\cdot\! \sigma+1$
has rank one, and therefore we can choose~$\chi$ arbitrarily, provided that it does not lie
in the kernel of this matrix).
A general solution can be written as a linear combination of these plane waves,
\beq \label{psirep}
\psi(\zeta) = \int_{S^1} \hat{\psi}(k)\, \e_{k}(\zeta)\: d\nu(k)\:,
\eeq
where~$\nu$ is the normalized Lebesgue measure on the sphere.

We want to introduce~$\H$ as the solution space of the Dirac equation~\eqref{Direigen}.
However, since the solutions~\eqref{psirep} are in general not square integrable,
we cannot use the $L^2$-scalar product. Instead, we make use of the fact that, in view of~\eqref{psirep},
the solution space can be identified with the space of complex-valued functions on the unit sphere.
We can thus take the $L^2$-scalar product on the sphere,
\beq \label{sprodS1}
\la \psi | \phi \ra_\H = \int_{S^1} \overline{\hat{\psi}(k)} \hat{\phi}(k)\, d\nu(k)\:.
\eeq
Combining the estimate
\[ \big| \e_{k}(\zeta) - \e_{k}(\zeta') \big| \overset{\eqref{plane}}{\leq}
2 \, \big| e^{-i k\, m \zeta} - e^{-i k\, m \zeta'}\big|
\leq 4 \left| \sin \Big(-\frac{i k\, m}{2}\: (\zeta-\zeta') \Big) \right|
\leq 2 m\, |\zeta-\zeta'| \]
(where~$|\psi|$ denotes the $\C^2$-norm of a spinor) with the
H\"older inequality, we find that
\begin{align*}
\big| \psi(\zeta) - \psi(\zeta') \big| &\leq  \int_{S^1} \big| \hat{\psi}(k) \big|\, \big| \e_{k}(\zeta) - \e_k(\zeta')
\big|\: d\nu(k) \\
&\leq \| \hat{\psi}\|_{L^1(S^1, d\nu)} \sup_{k \in S^1} \big| \e_{k}(\zeta) - \e_k(\zeta') \big|
\leq 2 m\, |\zeta-\zeta'| \; \| \hat{\psi}\|_{L^2(S^1, d\nu)} \:.
\end{align*}
This shows that the functions in~$\H$ are
all continuous. Hence we can introduce the local correlation operator at a point~$\zeta$ again by~\eqref{loccorr}.

The local correlation operators can be described conveniently with the help of
the so-called evaluation map, as we now explain (see also~\cite[\S1.2.4]{cfs} or~\cite[Section~4.1]{lqg}
and~\cite[Section~4]{finite}).
For any~$\zeta \in \R^2$, the {\em{evaluation map}} $e_\zeta$ is defined by
\beq \label{evalmap}
e_\zeta \::\: \H \rightarrow S_\zeta \:,\qquad e_\zeta \,\psi = \psi(\zeta) \:.
\eeq
We denote its adjoint by~$\iota_\zeta$,
\[ \iota_\zeta := (e_\zeta)^* \::\: S_\zeta \rightarrow \H\:. \]
Combining the computation
\[ \la \psi \,|\, \iota_\zeta(u) \ra_\H = \Sl e_\zeta(\psi) \,|\, u \Sr = \Sl \psi(\zeta) \,|\, u \Sr
\overset{\eqref{psirep}}{=}  \int_{S^1} \overline{\hat{\psi}(k)} \,\Sl \e_{k}(\zeta) \,|\, u \Sr \: d\nu(k) \]
with~\eqref{sprodS1}, one sees that
\beq \label{iotak}
\widehat{\iota_\zeta(u)}(k) = \Sl \e_{k}(\zeta) | u \Sr \:.
\eeq
It then follows by definition that the local correlation operators take the form
\beq \label{Fzeta}
F(\zeta) = -\iota_\zeta \,e_\zeta \:.
\eeq
We also introduce the so-called {\em{kernel of the fermionic operator}}~$P(\zeta',\zeta) \in \Lin(Y)$ by
\beq \label{Pzeta}
P(\zeta',\zeta) = -e_{\zeta'} \, \iota_\zeta \:.
\eeq

\begin{Lemma} The kernel of the fermionic operator~\eqref{Pzeta} is given by
\begin{align}
P(\zeta',\zeta) &= - \left(\frac{1}{m}\:\D_{\zeta'} + \1 \right) \int_{S^1} e^{-ik m (\zeta'-\zeta)} \:d\nu(k) \label{Pform} \\
&= i m (\zeta'-\zeta) \!\cdot\!\sigma\: \frac{J_1(m |\zeta-\zeta'|)}{m |\zeta-\zeta'|} -
J_0 \big( m |\zeta-\zeta'| \big) \:. \label{Prep}
\end{align}
\end{Lemma}
\Proof A short computation using~\eqref{psirep} and~\eqref{iotak} gives
\[ P(\zeta',\zeta) = - \int_{S^1} \e_k(\zeta') \otimes \e_k(\zeta)^* \:d\nu(k)
= \int_{S^1} (k \!\cdot\! \sigma+\1) \:e^{-ik m (\zeta'-\zeta)} \:d\nu(k) \:, \]
where in the last step we applied~\eqref{plane} and simplified the Pauli matrices according to
\[ (k \!\cdot\! \sigma+\1) \begin{pmatrix} 1 & 0 \\ 0 & 0 \end{pmatrix} (k \!\cdot\! \sigma+\1)
= (k \!\cdot\! \sigma+\1) \:. \]
Rewriting the factors~$k$ as derivatives, we obtain~\eqref{Pform}.
Setting~$r=|\zeta-\zeta'|$ and
denoting the angle between~$k$ and~$\zeta-\zeta'$ by~$\varphi$, we
can compute the integral in terms of Bessel functions
\[ \int_{S^1} e^{-ik m (\zeta'-\zeta)} \:d\nu(k)
= \frac{1}{2 \pi} \int_0^{2 \pi} e^{i m r \cos \varphi}\:d\varphi = J_0(m r) \:. \]
Using this equation in~\eqref{Pform}, we can compute the derivative with
the help of~\cite[\S10.6.3]{DLMF}. This gives~\eqref{Prep}.
\QED
We remark that~$P$ can also be written as the distributional Fourier transform
\[ P(\zeta', \zeta) = -\int_{\R^2} \frac{d^2 k}{\pi}\: (k \!\cdot\! \sigma + \1)\:\delta(k^2-1)\, e^{-i k m (\zeta-\zeta')} \:. \]
In this form, it resembles closely the kernel of the so-called fermionic projector of the vacuum
in Minkowski space (see for example~\cite[Lemma~1.2.8]{cfs}, \cite[Lemma~1.1]{rrev} or~\cite[Section~5]{srev}).
To clarify the notions, we remark that the {\em{``fermionic projector''}} differs from the ``fermionic operator''
in that it involves additional normalization conditions (as is explained in more detail in~\cite[\S1.1.3]{cfs}).
Since we here disregard such normalization conditions, we prefer to use the term ``fermionic operator''
throughout.

Having introduced the local correlation operators (see~\eqref{loccorr} or~\eqref{Fzeta}),
we can introduce the universal measure by~$\rho = F_* \mu$, where~$d\mu=d^2\zeta$ is the Lebesgue measure.
We thus obtain a Riemannian fermion system~$(\H, \F, \rho)$ of spin dimension two.
Since for every~$\zeta$ in~$\R^2$, there are two functions~$\psi, \phi \in \H$ such that~$\psi(\zeta)$
and~$\phi(\zeta)$ are linearly independent, we conclude that the Riemannian fermion system is regular.
In what follows, we again identify~$\zeta$ with the corresponding local correlation operator~$F(\zeta)$.

\begin{Lemma} \label{lemmaid}
The above Riemannian fermion system~$(\H, \F, \rho)$ has the properties
\begin{align}
\zeta &= -\pi_{\zeta} \label{zetatwo} \\
\pi_\zeta \,\zeta' \,\zeta &= \big(
\big|J_1(m |\xi|) \big|^2 + \big| J_0(m |\xi|) \big|^2 \big)\: \pi_\zeta \:, \label{zzp}
\end{align}
where we used the abbreviation~$\xi = \zeta'-\zeta$.
\end{Lemma}
\Proof Let us compute the two non-trivial eigenvalues of~$F(\zeta)$.
Using~\eqref{plane}, we obtain
\[ \la \psi | F(\zeta) \,\phi \ra_\H = -\int_{S^1} d\nu(k') \,\overline{\hat{\psi}(k')} \int_{S^1} d\nu(k) \,\hat{\phi}(k)\:
e^{i (k'-k) m \zeta} \:\la (k'  \!\cdot\! \sigma + \1) \,\chi \,|\, (k \!\cdot\! \sigma + \1) \,\chi \ra_{\C^2} \:. \]
Comparing with~\eqref{sprodS1} and simplifying the Pauli matrices by
\[ \la (k' \!\cdot\! \sigma + \1) \,\chi \,|\, (k \!\cdot\! \sigma + \1) \,\chi \ra_{\C^2} = 1+e^{i \varphi(k',k)} \:, \]
where~$\varphi(k',k)$ denotes the angle between~$k'$ and~$k$ (measured from~$k'$ in counter-clockwise
direction), we conclude that
\begin{align}
F(\zeta) \,\psi &= - \int_{S^1} d\nu(k')  \int_{S^1} d\nu(k) \,
\hat{\psi}(k)\: e^{i (k'-k) m \zeta} \:\big(1+e^{i \varphi(k',k)} \big)\,\e_{k'} \label{Frel} \\
&= \int_{S^1} \hat{\psi}(k) \left[ - \int_{S^1} e^{i (k'-k) m \zeta} \:\big(1+e^{i \varphi(k',k)} \big)\,\e_{k'} 
\:  d\nu(k') \right] d\nu(k) \:. \notag
\end{align}
Comparing with~\eqref{psirep} and using linearity,
one sees that the square bracket coincides precisely with~$F(\zeta)$
applied to~$\e_k$,
\[ F(\zeta) \,\e_k = - \int_{S^1} e^{i (k'-k) m \zeta} \:\big( 1+e^{i \varphi(k',k)}  \big)\, \e_{k'}\: d\nu(k') \:. \]
Iterating this relation, we obtain
\begin{align*}
F(\zeta) \,F(\zeta) \,\e_k &= \int_{S^1} d\nu(k'')\,\e_{k''} \: e^{i (k''-k) m \zeta}  \int_{S^1} d\nu(k') \:
\big( 1+e^{i \varphi(k'',k')} \big) \big( 1+e^{i \varphi(k',k)} \big) \\
&= \int_{S^1} d\nu(k'')\,\e_{k''} \: e^{i (k''-k) m \zeta} \:\big( 1+e^{i \varphi(k'',k)} \big)
= - F(\zeta)\, \e_k \:,
\end{align*}
where in the last line we carried out the integral over~$k' \in S^1$.
We conclude that~$F(\zeta)$ has the eigenvalue~$-1$ with multiplicity two
(note that~$F(\zeta)$ has rank two because our Riemannian fermion system
is regular). Identifying~$\zeta$ with~$F(\zeta)$, we thus obtain~\eqref{zetatwo}.

In preparation for proving~\eqref{zzp}, we form the so-called
{\em{closed chain}} by taking the product of
the kernel of the fermionic operator with its adjoint,
\[ A_{\zeta \zeta'} := P(\zeta, \zeta') \, P(\zeta', \zeta) \:. \]
(for the motivation of the name ``closed chain'' we refer to~\cite[\S3.5]{PFP}).
Using the explicit formula~\eqref{Prep}, we obtain
\begin{align}
A_{\zeta \zeta'} 
&= \Big( i\,m \,\xi \!\cdot\!\sigma\: \frac{J_1(m |\xi|)}{m |\xi|} + J_0(m |\xi|) \Big)
\left( -i\,m \,\xi \!\cdot\!\sigma\: \frac{J_1(m |\xi|)}{m |\xi|} + J_0(m |\xi|) \right) \nonumber \\
&= \big| J_1(m |\xi|) \big|^2 +  \big| J_0(m |\xi|) \big|^2 \:. \label{chainid}
\end{align}
In particular, the closed chain is a multiple of the identity matrix. As a consequence,
\begin{align*}
\pi_\zeta \,\zeta' \zeta \;\;& \!\!\!\overset{\eqref{zetatwo}}{=} -\zeta\, \zeta'\, \zeta
\overset{\eqref{Fzeta}}{=} (\iota_\zeta \,e_\zeta)\: (\iota_{\zeta'} \,e_{\zeta'})\: (\iota_\zeta \,e_\zeta)
\overset{\eqref{Pzeta}}{=} \iota_\zeta \,P(\zeta, \zeta')\, P(\zeta', \zeta) \,e_\zeta \\
&= \big( |J_1|^2 + |J_0|^2 \big)\: \iota_\zeta \,e_\zeta
\overset{\eqref{Fzeta}}{=} -\big( |J_1|^2 + |J_0|^2 \big)\: \zeta
= \big( |J_1|^2 + |J_0|^2 \big)\: \pi_\zeta\:.
\end{align*}
This concludes the proof.
\QED
We remark that, combining~\eqref{Frel} and~\eqref{psirep}, we can write~$F(\zeta)$ as
the integral operator
\beq \label{S1ker}
(\widehat{F(\zeta)\, \psi})(k') = \int_{S^1} F_\zeta(k',k)\, \hat{\psi}(k)\: d\nu(k)
\eeq
with the kernel
\[ F_\zeta(k',k) = -e^{i (k'-k) m \zeta} \:\big(1+e^{i \varphi(k',k)} \big) \:. \]
This makes it possible to compute the trace of~$F(\zeta)$ by
\[ \tr \big( F(\zeta) \big) = \int_{S^1} F_\zeta(k,k)\: d\nu(k) = -2 \:, \]
in agreement with~\eqref{zetatwo} and the fact that~$\zeta$ has rank two.

The main conclusion from Lemma~\ref{lemmaid} is that the
operators~${\mathcal{A}}$ chosen according to~\eqref{cchain}, \eqref{As} or~\eqref{Ap}
are all multiples of the identity.
This implies that by analyzing the operator~$\zeta'$ in a neighborhood of~$\zeta$ (for example using
tangent cone measures corresponding to the mapping~\eqref{Ap}),
it is impossible to distinguish a Clifford subspace at a point~$\zeta$.
In more technical terms, the non-degeneracy property of Definition~\ref{defRnondeg}
is necessarily violated. A possible method to avoid this shortcoming is to use
the decomposition into left- and right-handed spinors.
Before introducing this method in Section~\ref{secmix} below, we proceed
by adapting the present example to Lorentzian signature.

\subsection{Two-Dimensional Minkowski Space} \label{secminkowski}
We now work out an example in two-dimensional Minkowski space (a similar example in four-dimensional Minkowski space is given in~\cite[Section~1.2]{cfs} and~\cite[Section~4]{lqg}).
We let~$(\scrM,g)$ be two-dimensional Minkowski space.
We work in the coordinates~$\zeta = (t,x)$ in which~$ds^2 = dt^2 - dx^2$. The Dirac operator
can be written as
\beq \label{dir3}
\Dir = i \gamma^0 \partial_t + i \gamma^1 \partial_x
\eeq
with the Dirac matrices given by
\[ \gamma^0 = \begin{pmatrix} 1 & 0 \\ 0 & -1 \end{pmatrix} \:,\qquad
\gamma^1 = \begin{pmatrix} 0 & 1 \\ -1 & 0 \end{pmatrix} \:. \]
The spinor space~$(S_\zeta \scrM, \Sl .,. \Sr_\zeta)$ at a point~$\zeta$ can
be identified with the inner product space~$(Y \simeq \C^2, \Sl .|. \Sr)$,
where the spin scalar product is defined by
\[ \Sl \psi | \phi \Sr = \la \psi | \gamma^0 \phi \ra_{\C^2}\:. \]
The Dirac matrices are obviously symmetric with respect to the spin scalar product.

We consider solutions of the Dirac equation
\[ (\Dir - m) \,\psi = 0 \:, \]
where~$m>0$ is a given mass parameter.
Since the Dirac equation is a linear hyperbolic equation, we know that the
initial value problem is well-posed, and that there is a finite speed of propagation.
This implies that if a solution has compact support at some time~$t_0$, it will also
have compact support at any other time. On such {\em{spatially compact solutions}}
one can introduce the scalar product
\beq \label{print}
( \psi | \phi)_{t_0} = \int_{-\infty}^\infty \Sl \psi | \gamma^0 \phi \Sr(t_0, x)\: dx\:.
\eeq
The integrand of~$(\psi | \psi)_{t_0}$ has the physical interpretation as the probability
density of a quantum mechanical particle to be at the position~$x$.
Due to current conservation\footnote{In our context, current conservation means that
for any two solutions~$\psi, \phi$ of the Dirac equation, the vector field~$\Sl \psi | \gamma^j \phi \Sr$
is divergence-free, $\partial_j \Sl \psi | \gamma^j \phi \Sr = 0$
(for details see for example~\cite[Section~1.2]{PFP}).
Integrating this equation over a space-time region $\{(t,x) \,|\, t_0 < t < \tilde{t}_0\}$
and using the Gauss divergence theorem, one finds that~\eqref{print} is indeed
independent of~$t_0$.}, this scalar product is independent of the time~$t_0$.
Therefore, we can simply denote it by~$(.|.)$.

Similar as explained in Section~\ref{sec61} in the Euclidean setting, the Dirac equation
can again be solved by plane wave solutions, which we write as
\beq \label{plane2}
\e_{k}(\zeta) = \frac{1}{\sqrt{|k_0|}}\: (k^0 \gamma^0 - k^1 \gamma^1 + m)\,\chi\: e^{-i k \zeta} \:,
\eeq
where the momentum lies on the mass shell
\beq \label{massshell}
k_0^2 - k_1^2 = m^2 \:,
\eeq
and~$\chi$ is the fixed spinor~$\chi=(1,i)/\sqrt{2}$ (here~$k \zeta = k_0 \zeta^0 - k_1 \zeta^1 $ is a
Minkowski inner product).
A general solution can be written as an integral over the mass shell, which is most conveniently
written with a $\delta$-distribution,
\beq \label{psifour}
\psi(\zeta) = \int_{\R^2} \frac{d^2k}{(2 \pi)^2}\: \hat{\psi}(k) \,\e_k(\zeta) \:\delta \big(k_0^2 - k_1^2 - m^2 \big)\:,
\eeq
where~$\hat{\psi}$ is a complex-valued function on the mass shell.
If the function~$\hat{\psi}$ satisfies suitable regularity and decay conditions, the wave function~$\psi$
will decay at spatial infinity, so that the scalar product~\eqref{print} is finite.
More specifically, the scalar product can be expressed by an integral over the mass shell:
\begin{Lemma} For solutions in the Fourier representation~\eqref{psifour}, the
scalar product~\eqref{print} becomes
\beq \label{printf}
( \psi | \phi ) = \frac{1}{(2 \pi)^3} \int_{\R^2} d^2k\: \delta \big(k_0^2 - k_1^2 - m^2 \big)
\: \overline{\hat{\psi}(k)} \,\hat{\phi}(q)\:.
\eeq
\end{Lemma}
\Proof A direct computation gives
\begin{align*}
( \psi | \phi ) &= \int_{-\infty}^\infty dx \; e^{-i (q_1-k_1) x}
\int_{\R^2} \frac{d^2k}{(2 \pi)^2}\:\delta \big(k_0^2 - k_1^2 - m^2 \big)
\int_{\R^2} \frac{d^2q}{(2 \pi)^2}\: \delta \big(q_0^2 - q_1^2 - m^2 \big) \\
&\qquad \times \frac{1}{\sqrt{|k_0|\, |q_0|}}\:  \overline{\hat{\psi}(k)} \,\hat{\phi}(q) \:
\Sl (k^0 \gamma^0 - k^1 \gamma^1 + m)\,\chi \,|\, \gamma^0\: (q^0 \gamma^0 + q^1 \gamma^1 + m)\,\chi \Sr \\
&= \frac{1}{(2 \pi)^3} \int_{\R^2} d^2k\: \delta \big(k_0^2 - k_1^2 - m^2 \big) \sum_{s=\pm1}
\frac{1}{2 \sqrt{m^2+k_1^2}} \\
&\qquad \times \frac{1}{|k_0|}\: \overline{\hat{\psi}(k)} \,\hat{\phi}(q) \:
\Sl (k^0 \gamma^0 - k^1 \gamma^1 + m)\,\chi \,|\, \gamma^0\: (s \,k^0 \gamma^0 - k^1
\gamma^1 + m)\,\chi \Sr \:.
\end{align*}
The Dirac matrices can be simplified by
\begin{align*}
(k^0 &\gamma^0 - k^1 \gamma^1 + m) \:\gamma^0\: (s \,k^0 \gamma^0 - k_1 \gamma^1 + m) \\
&= \gamma^0 (k^0 \gamma^0 + k^1 \gamma^1 + m) (s \,k^0 \gamma^0 - k^1 \gamma^1 + m) \\
&= \gamma^0 \left( s\, k_0^2 + k_1^2 + m^2 + (1+s) \,m \,k^0 \gamma^0
- (1+s) \,\gamma^0 \gamma^1\, k^0 k^1\right) .
\end{align*}
In the case~$s=-1$, this expression vanishes in view of~\eqref{massshell}. In the case~$s=1$, on the
other hand, we obtain
\beq \label{matmult}
\begin{split}
(k^0 &\gamma^0 - k^1 \gamma^1 + m) \:\gamma^0\: (k^0 \gamma^0 - k^1 \gamma^1 + m) \\
&= \gamma^0 \big( 2 k_0^2 + 2 m\, k^0 \gamma^0 - 2\,\gamma^0 \gamma^1\, k^0 k^1 \big) 
= 2 k^0 \big( k^0 \gamma^0 - k^1 \gamma^1 + m \big) \:.
\end{split}
\eeq
Next, the expectation values of the spinor~$\chi$ are computed by
\[ \Sl \chi | \gamma^0 \chi \Sr = \la \chi | \chi \ra_{\C^2} = 1 \:, \qquad
\Sl \chi | \chi \Sr = 0 = \Sl \chi | \gamma^1 \chi \Sr \:. \]
Finally, we again use~\eqref{massshell} to obtain the result.
\QED

With~\eqref{printf} we rewrote the scalar product~\eqref{print} coming from the probabilistic interpretation
of the Dirac equation in a way which is very similar to the scalar product~\eqref{sprodS1}.
The only difference is that, instead of integrating over the circle, we now integrate over
the hyperbolas of the mass shell.
Since these hyperbolas are non-compact, the estimate used after~\eqref{sprodS1} to prove the
continuity of the wave functions fails. This makes it necessary to introduce
an ultraviolet regularization, as we now explain.
First, we consider solutions of negative energy, obtained from~\eqref{psifour} by integrating
only over the lower mass shell,
\begin{align*}
\psi(\zeta) &= \int_{\R^2} \frac{d^2k}{(2 \pi)^2}\: \hat{\psi}(k) \,\e_k(\zeta)
\:\delta \big(k_0^2 - k_1^2 - m^2 \big)\: \Theta(-k^0) \\
&= \frac{1}{(2 \pi)^2} \int_{-\infty}^\infty \frac{dp}{2 \sqrt{p^2+m^2}}\: \hat{\psi}(k) \,\e_k(\zeta)
\Big|_{k= \big( -\sqrt{p^2+m^2}, p \big)} \:.
\end{align*}
We denote the scalar product~\eqref{print} on these wave functions by~$\la .|. \ra_\H$.
Taking the completion, we obtain the Hilbert space~$(\H, \la .|.\ra_\H)$.
For a wave function~$\psi \in \H$, the evaluation map~\eqref{evalmap} will in general be ill-defined,
because~$\psi$ need not be continuous. The simplest method to cure this problem is to
insert a convergence-generating factor into the Fourier integral. Thus for a given parameter~$\varepsilon>0$
we set
\beq \label{ezdef}
e_\zeta^\varepsilon(\psi) = \int_{\R^2} \frac{d^2k}{(2 \pi)^2} \: e^{\frac{\varepsilon k^0}{2}}\: \hat{\psi}(k) \,\e_k(\zeta)
\:\delta \big(k_0^2 - k_1^2 - m^2 \big)\: \Theta(-k^0) \::\: \H \rightarrow S_\zeta \:.
\eeq
We define~$\iota_\zeta^\varepsilon$ as its adjoint,
\[ \iota_\zeta^\varepsilon = \big( e_\zeta^\varepsilon \big)^* \::\: S_\zeta \rightarrow \H \:. \]
Comparing the computation
\[ \la \psi \,|\, \iota^\varepsilon_\zeta(u) \ra_\H = \Sl e^\varepsilon_\zeta(\psi) \,|\, u \Sr 
= \int_{\R^2} \frac{d^2k}{(2 \pi)^2} \: e^{\frac{\varepsilon k^0}{2}} \: \overline{\hat{\psi}(k)} \, \Sl \e_k(\zeta) | u \Sr
\:\delta \big(k_0^2 - k_1^2 - m^2 \big)\: \Theta(-k^0) \]
with~\eqref{printf}, we obtain
\beq \label{iotarep}
\widehat{ \iota^\varepsilon_\zeta(u) }(k) = e^{\frac{\varepsilon k^0}{2}}\:\Theta(-k^0) \: \Sl \e_k(\zeta) | u \Sr \:.
\eeq
Before going on, we remark that there are of course many other ways to introduce an ultraviolet
regularization. In generalization of the convergence-generating factor in~\eqref{ezdef},
one can work with so-called regularization operators as introduced in~\cite[Section~4]{finite}
(see also~\cite[\S1.2.2]{cfs}).

We next define the regularized local correlation operators in analogy to~\eqref{Fzeta} by
\beq \label{Fzeta2}
F(\zeta) = -\iota^\varepsilon_\zeta \,e^\varepsilon_\zeta \:.
\eeq
Moreover, the kernel of the fermionic operator is introduced similar to~\eqref{Pzeta} by
\beq \label{Pdef}
P(\zeta',\zeta) := -e^\varepsilon_{\zeta'} \, \iota^\varepsilon_\zeta \:.
\eeq
The next lemma represents~$P(\zeta, \zeta')$ as an integral over the lower mass shell,
which is Lorentz invariant except for the convergence-generating factor~$e^{\varepsilon k^0}$.
\begin{Lemma} \label{lemmaPkern}
The kernel of the fermionic operator is given by
\begin{align}
& P(\zeta',\zeta) = \int_{\R^2} \frac{d^2k}{(2 \pi)^2} \: e^{\varepsilon k^0}\:
(k^0 \gamma^0 - k^1 \gamma^1 + m) \:\delta \big(k_0^2 - k_1^2 - m^2 \big)\: \Theta(-k^0)
 \: e^{-ik(\zeta'-\zeta)} \nonumber \\
&= \frac{m}{4 \pi^2}\: K_0 \Big(-i m \sqrt{(t'-t+i \varepsilon)^2 - (x'-x)^2} \Big) \nonumber \\
&\quad + \frac{m}{4 \pi^2}\: \frac{K_1 \Big(-i m \sqrt{(t'-t+i \varepsilon)^2 - (x'-x)^2} \Big)}
{\sqrt{(t'-t+i \varepsilon)^2 - (x'-x)^2}}\: \begin{pmatrix} t'-t+i \varepsilon & -x'+x \\
x'-x & -t'+t-i \varepsilon \end{pmatrix} . \label{mat3}
\end{align}
\end{Lemma}
\Proof
Substituting~\eqref{iotarep} and~\eqref{ezdef} into~\eqref{Pdef}, we obtain
\begin{align*}
P(\zeta',\zeta) &= -\int_{\R^2} \frac{d^2k}{(2 \pi)^2} \: e^{\varepsilon k^0} \,\e_k(\zeta')
\otimes \e_k(\zeta)^* \:\delta \big(k_0^2 - k_1^2 - m^2 \big)\: \Theta(-k^0)
\end{align*}
A short calculation shows that~$\chi \otimes \chi^* = (\gamma^0 - \gamma^2)/2$ with
\[ \gamma^2 := -\begin{pmatrix} 0 & i \\ i & 0 \end{pmatrix}\:. \]
Hence can apply~\eqref{matmult} together with
\begin{align*}
(k^0 &\gamma^0 - k^1 \gamma^1 + m) \:\gamma^2\: (k^0 \gamma^0 - k^1 \gamma^1 + m) \\
&= \gamma^2 (-k^0 \gamma^0 + k^1 \gamma^1 + m) (k^0 \gamma^0 - k^1 \gamma^1 + m)
= \gamma^2 \left(-k_0^2 + k_1^2 + m^2 \right) \overset{\eqref{massshell}}{=} 0 
\end{align*}
to obtain
\begin{align*}
P(\zeta',\zeta) &= -\int_{\R^2} \frac{d^2k}{(2 \pi)^2} \: e^{\varepsilon k^0}\: \frac{k_0}{|k_0|}\,
 (k^0 \gamma^0 - k^1 \gamma^1 + m) \:\delta \big(k_0^2 - k_1^2 - m^2 \big)\: \Theta(-k^0)
 \: e^{-ik(\zeta'-\zeta)} \:.
\end{align*}
This gives the Fourier integral in the statement of the lemma.
It can be computed by
\begin{align*}
P(\zeta',\zeta)&=(i\slashed{\partial}_{\zeta'}+m) 
	\int_{\R^2} \frac{d^2k}{(2 \pi)^2} \: e^{\varepsilon k^0}\: \delta \big(k_0^2 - k_1^2 - m^2 \big)\: \Theta(-k^0)
 \: e^{-ik(\zeta'-\zeta)} \\ 
&= (i\slashed{\partial}_{\zeta'}+m) \int_{-\infty}^{-m}
\frac{d\omega}{(2 \pi)^2} \: \frac{e^{\varepsilon \omega}}{|k^1|} \: 
e^{-i \omega (t'-t)}\: \cos(k^1 \,(x'-x)) \Big|_{k^1 =
\sqrt{\omega^2-m^2}} \\
&= \frac{1}{4 \pi^2}\: (i\slashed{\partial}_{\zeta'}+m) K_0 \Big(-i m \sqrt{(t'-t+i \varepsilon)^2 - (x'-x)^2} \Big) \:,
\end{align*}
where~$K_0$ is the modified Bessel function (see~\cite[\S10.25]{DLMF}).
Using the formulas for derivatives of Bessel functions (see~\cite[\S10.29(ii)]{DLMF}), we obtain~\eqref{mat3}.
\QED

Having introduced the local correlation operators by~\eqref{Fzeta2},
we can introduce the universal measure again by~$d\rho = F_*(dt \,dx)$ to obtain a causal fermion
system~$(\H, \F, \rho)$ of spin dimension one. In what follows, we again identify
the space-time point~$\zeta$ with the corresponding local correlation operator~$F(\zeta) \in \F$.
For the detailed computations, it is convenient to choose pseudo-orthonormal bases $(\f_\alpha(\zeta))_{\alpha=1,2}$
of the spin spaces~$S_\zeta$. To this end, we first need to calculate the nontrivial
eigenvalues of~$F(\zeta)$. Comparing~\eqref{Fzeta2} and~\eqref{Pdef}, these eigenvalues
are the same as those of the matrix~$P(\zeta, \zeta)$. Applying Lemma~\ref{lemmaPkern}
and using the symmetry under the transformation~$k_1 \rightarrow -k_1$, we obtain
\begin{align*}
P(\zeta, \zeta) &= \int_{\R^2} \frac{d^2k}{(2 \pi)^2} \: e^{\varepsilon k^0}\:
(k^0 \gamma^0 + m) \:\delta \big(k_0^2 - k_1^2 - m^2 \big)\: \Theta(-k^0) \\
&= \int_{-\infty}^{-m} \frac{dk_0}{(2 \pi)^2} \: e^{\varepsilon k^0}\:\frac{k^0 \gamma^0 + m}{2 \sqrt{k_0^2-m^2}}\:.
\end{align*}
This matrix has the eigenvalues
\[ \nu_{1\!/\!2} := \int_{-\infty}^{-m} \frac{dk_0}{(2 \pi)^2} \: e^{\varepsilon k^0}\:\frac{\pm k^0 + m}
{2 \sqrt{k_0^2-m^2}} \:, \]
and the corresponding eigenvectors~$\e_1 = (1,0)$ and~$\e_2 = (0,1)$ are the canonical
pseudo-orthonormal basis of~$Y$ (note that~$\nu_1<0$ and~$\nu_2>0$).
A straightforward computation using again~\eqref{Fzeta2}
and~\eqref{Pdef} shows that corresponding eigenvectors of~$F(\zeta)$ are given by
\beq \label{falpha}
\f_\alpha(\zeta)
:= \frac{1}{\nu_\alpha}\: \iota_\zeta \,\e_\alpha \:,\qquad \alpha=1,2\:.
\eeq
Moreover, the computation
\begin{align*}
\Sl \f_\alpha|\f_\beta \Sr_x
&= - \la \f_\alpha | F^\varepsilon(p)\, \f_\beta \ra_\H 
= - \nu_\beta \: \la \f_\alpha | \f_\beta \ra_\H  \nonumber \\
&= -\frac{1}{\nu_\alpha} \:
\la \iota^\varepsilon_x \e_\alpha | \iota^\varepsilon_x \e_\beta \ra_\H
= -\frac{1}{\nu_\alpha} \:
\Sl \e_\alpha | e^\varepsilon_x\, \iota^\varepsilon_x \e_\beta \Sr \nonumber \\
&\!\!\!\overset{\eqref{Pdef}}{=} 
= \frac{1}{\nu_\alpha} \:
\Sl \e_\alpha | P(\zeta, \zeta)\, \e_\beta \Sr
= \frac{\nu_\beta}{\nu_\alpha} \:
\Sl \e_\alpha | \e_\beta \Sr = s_\alpha\: \delta_{\alpha \beta}
\end{align*}
(with~$s_{1\!/\!2} = \pm1$) shows that~$(\f_\alpha(\zeta))$ is indeed a pseudo-orthonormal basis
of~$(S_\zeta, \Sl .|. \Sr_\zeta)$. 

\begin{Lemma} \label{lemma75}
In the pseudo-orthonormal spinor bases~$\f_{1\!/\!2}(\zeta)$, we have
\begin{align}
\zeta &= \begin{pmatrix} \nu_1 & 0 \\ 0 & \nu_2 \end{pmatrix} \:,\qquad
\pi_\zeta = \begin{pmatrix} 1 & 0 \\ 0 & 1 \end{pmatrix} \:,\qquad
s_\zeta = \begin{pmatrix} 1 & 0 \\ 0 & -1 \end{pmatrix} \label{mat1} \\
\pi_{\zeta'} \,\zeta &= P(\zeta',\zeta) \quad \text{as defined by~\eqref{Pdef}
and given by~\eqref{mat3}} \label{mat2} \\
\pi_\zeta \,\zeta'\, \zeta &= P(\zeta, \zeta')\, P(\zeta', \zeta) \label{mat2a} \\
\pi_\zeta \,s_{\zeta'}\, \zeta &= -P(\zeta, \zeta') \begin{pmatrix} |\nu_1|^{-1} & 0 \\
0 & |\nu_2|^{-1}  \end{pmatrix} P(\zeta', \zeta) \label{mat4} \\
\pi_\zeta \,\pi_{\zeta'}\, \zeta &= P(\zeta, \zeta') \begin{pmatrix} \nu_1^{-1} & 0 \\
0 & \nu_2^{-1}  \end{pmatrix} P(\zeta', \zeta) . \label{mat5}
\end{align}
\end{Lemma}
\Proof The equations~\eqref{mat1} are obvious from the fact that the~$(\f_\alpha(\zeta))_{\alpha=1,2}$
are eigenvectors of~$\zeta$. In order to derive the other formulas, we first note that
the matrix representation of a general operator~$B : S_\zeta \rightarrow S_\zeta'$ is computed by
\[ B^\alpha_\beta = s_\alpha \:\Sl \f_\alpha(\zeta') \,|\, B\, \f_\beta(\zeta) \Sr_{\zeta'} 
= - s_\alpha \:\la  \f_\alpha(\zeta') \,|\, F(\zeta') \,B\, \f_\beta(\zeta) \ra_\H \:. \]
In particular,
\begin{align*}
(\pi_{\zeta'} \,\zeta)^\alpha_\beta &= -s_\alpha\, 
\la  \f_\alpha(\zeta') \,|\, F(\zeta')\, \pi_{\zeta'} \,\zeta \,\f_\beta(\zeta) \ra_\H \\
&= -s_\alpha\, \la  \f_\alpha(\zeta') \,|\, F(\zeta')\, \,\zeta \,\f_\beta(\zeta) \ra_\H 
= -s_\alpha \nu_\alpha \nu_\beta \: \la \f_\alpha(\zeta') \,|\, \f_\beta(\zeta) \ra_\H \\
&= -s_\alpha \: \la \iota^\varepsilon_{\zeta'} \e_\alpha \,|\, \iota^\varepsilon_\zeta \e_\beta \ra_\H 
= -s_\alpha \: \Sl \e_\alpha \,|\, (\iota^\varepsilon_{\zeta'})^* \iota^\varepsilon_\zeta\, \e_\beta \Sr 
\overset{\eqref{Pdef}}{=} s_\alpha \: \Sl \e_\alpha | P(\zeta', \zeta) \,\e_\beta \Sr  \:,
\end{align*}
proving~\eqref{mat2}. The relation~\eqref{mat2a} follows immediately from~\eqref{mat2}.

Next, comparing the matrices in~\eqref{mat1} and using that they are all diagonal, we obtain
\[ (\pi_{\zeta} \,s_{\zeta'})^\alpha_\beta = (\pi_{\zeta} \,\zeta')^\alpha_\beta\:
\Big( - \frac{1}{|\nu_\beta|} \Big) =  -P(\zeta, \zeta')^\alpha_\beta \: \frac{1}{|\nu_\beta|} \:. \]
Multiplying by~\eqref{mat2} gives~\eqref{mat4}. The identity~\eqref{mat5} is obtained similarly.
\QED
We remark that the relation~\eqref{mat2} can also be used to define the kernel of the
fermionic operator on general topological fermion systems by~$P(x,y) = \pi_x \,y|_{S_y} : S_y \rightarrow S_x$.
This is indeed the procedure in~\cite{cfs, rrev, lqg} (cf.~\cite[eq.~(1.1.13)]{cfs},
\cite[eq.~(1.15)]{rrev} and~\cite[eq.~(2.7)]{lqg}).
In order to present a somewhat different point of view, we here preferred to define the kernel
of the fermionic operator by~\eqref{Pdef}, which is also more convenient for doing computations.

\subsection{The Euclidean Plane with Chiral Asymmetry} \label{secmix}
We now return to the Euclidean plane as considered in Section~\ref{sec61}.
We modify this example as follows. On~$Y$ we introduce the linear
operator~$\pseudo =\sigma^3$. This operator anti-commutes with the Dirac operator~\eqref{dir2},
\[ \Dir \pseudo = -\pseudo \Dir \:. \]
The eigenspaces of~$\pseudo$ give a $\Z_2$-grading of~$S_x$. In analogy to the splitting
into right- and left-handed components in four space-time dimensions, we we refer
to this grading as the {\em{chiral grading}}. Next, we modify the evaluation map by
inserting the operator~$\pseudo$,
\beq \label{evalmaprho}
e_\zeta \::\: \H \rightarrow S_\zeta \:,\qquad e_\zeta \,\psi = (\1+\tau \pseudo) \psi(\zeta) \:,
\eeq
where~$\tau$ is a real parameter. We again denote the adjoint of the evaluation map
by~$\iota_\zeta$ and introduce the local correlation operators and the kernel of the fermionic
operator by~\eqref{Fzeta} and~\eqref{Pzeta}. More explicitly,
the kernel of the fermionic operator~$P(\zeta', \zeta)$ is obtained by multiplying~\eqref{Prep}
from the left and right by~$(\1+\tau \pseudo)$,
\beq \label{Pmix}
P(\zeta', \zeta) = (\1+\tau \pseudo) \left( 
i m(\zeta-\zeta') \!\cdot\!\sigma\: \frac{J_1(m |\zeta-\zeta'|)}{m |\zeta-\zeta'|} -
J_0 \big( m |\zeta-\zeta'| \big) \right)  (\1+\tau \pseudo) \:.
\eeq

Having introduced the local correlation operators (see~\eqref{loccorr} or~\eqref{Fzeta}),
we can introduce the universal measure again by~$\rho = F_* \mu$, where~$d\mu=d^2\zeta$ is the Lebesgue measure.
We thus obtain a Riemannian fermion system~$(\H, \F, \rho)$ of spin dimension two.
For the detailed computations, it is convenient to work with an orthonormal
basis~$(\f_\alpha(\zeta))_{\alpha=1,2}$ of the spin spaces~$S_\zeta$.
We again follow the procedure explained before~\eqref{falpha}.
Expanding the Bessel functions in~\eqref{Pmix} around zero,
one readily finds that the matrix~$P(\zeta, \zeta)$ has the eigenvectors~$\e_1$ and~$\e_2$
with corresponding eigenvalues
\[ \nu_1 = -(1+\tau)^2 \qquad \text{and} \qquad \nu_2 = -(1-\tau)^2 \:. \]
As a consequence, the vectors~$\f_{1\!/\!2}(\zeta)$ defined in analogy to~\eqref{falpha} by
\beq \label{falpha2}
\f_\alpha(\zeta)
:= \frac{1}{\nu_\alpha}\: \iota_\zeta \,\e_\alpha \:,\qquad \alpha=1,2\:,
\eeq
form an orthonormal basis of~$S_\zeta$. 

\begin{Lemma} \label{lemmamix} In the orthonormal spinor bases~$\f_{1\!/\!2}(\zeta)$, we have
\begin{align}
\zeta &= \begin{pmatrix} \nu_1 & 0 \\ 0 & \nu_2 \end{pmatrix} \:,\qquad
\pi_\zeta = s_\zeta = \begin{pmatrix} 1 & 0 \\ 0 & 1 \end{pmatrix} \label{mat1mix} \\
\pi_{\zeta'} \zeta &= P(\zeta',\zeta) \quad \text{as defined by~\eqref{Pdef}
and given by~\eqref{Pmix}} \label{mat2mix} \\
\pi_\zeta \,\zeta'\, \zeta &= P(\zeta, \zeta')\, P(\zeta', \zeta) \\
\pi_\zeta \,s_{\zeta'}\, \zeta &= \pi_\zeta \,\pi_{\zeta'}\, \zeta = -P(\zeta, \zeta') \begin{pmatrix} |\nu_1|^{-1} & 0 \\
0 & |\nu_2|^{-1}  \end{pmatrix} P(\zeta', \zeta) . \label{mat4mix}
\end{align}
\end{Lemma}
\Proof Follows exactly as in the proof of Lemma~\ref{lemma75}.
\QED

We finally explain the notion ``chiral asymmetry.'' As one sees best in~\eqref{Pmix},
the matrix~$(1+\tau \pseudo)$ inserted in~\eqref{evalmaprho} multiplies the left- and right-handed
components of the fermionic operator by certain prefactors. In the case~$\tau \neq 0$,
these prefactors are different for the left- and right-handed components.
Thus the chiral symmetry is broken.
Such a chiral asymmetry is present in a more realistic physical situation in the neutrino
sector of the fermionic projector in Minkowski space (for details see~\cite[Section~4.2]{cfs}).

\subsection{The Spin Structure of the Euclidean Plane with Chiral Asymmetry} \label{secssEuklid}
We shall now explore the constructions of Section~\ref{sectangential} in the example
of the Euclidean plane with chiral asymmetry.
This is a preparation for a similar analysis for the two-dimensional Minkowski space,
which is a bit more subtle and will be given in Section~\ref{secssMin} below.
Our first step is to construct the tangent cone measure, following the general construction
described in Section~\ref{sectcm}.
In preparation, we need to specify the functional~${\mathcal{A}}$ in~\eqref{Afunct}.
In all the following computations, we work in the basis~$(\f_{1\!/\!2}(\zeta))$ of the spin spaces
introduced in~\eqref{falpha2}.
Combining the result of Lemma~\ref{lemmamix} with the explicit form of the kernel of the
fermionic operator~\eqref{Pmix}, we obtain
\begin{align}
\pi_\zeta \,s_{\zeta'}\, \zeta |_{S_\zeta} &= -\big( J_0(m |\xi|)^2 + J_1(m |\xi|)^2 \big)
\begin{pmatrix} (1+\tau^2)^2 & 0 \\ 0 & (1-\tau^2)^2 \end{pmatrix} \label{Anogo} \\
\pi_\zeta \,\zeta'\, \zeta |_{S_\zeta}  &= (\1 + \tau \pseudo ) \Big( im \,\xi \!\cdot\!\sigma\: \frac{J_1(m |\xi|)}{m |\xi|}
-J_0(m |\xi|) \Big)  (\1 + \tau \pseudo )^2 \nonumber \\
&\qquad \times \Big( -im \,\xi \!\cdot\!\sigma\: \frac{J_1(m |\xi|)}{m |\xi|} - J_0(m |\xi|) \Big) 
(\1 + \tau \pseudo ) \nonumber \\
&= a(\xi)\, \1 + b(\xi)\, \pseudo + c(\xi)\, i \,\xi \!\cdot\!\sigma\, \pseudo \:, \label{Ago}
\end{align}
where we again set~$\xi = \zeta'-\zeta$ and
\begin{align*}
a(\xi) &= \left( (1+6 \tau^2 + \tau^4) \, J_0(m |\xi|)^2 - (1-\tau^2)^2\, J_1(m |\xi|)^2 \right) \\
b(\xi) &= 4 \tau^2 (1+\tau^2) \, J_0(m |\xi|)^2 \\
c(\xi) &= 4 \tau (1-\tau^2) \:\frac{J_0(m |\xi|) \,J_1(m |\xi|)}{|\xi|}\:.
\end{align*}
Note that~\eqref{Anogo} is a linear combination of the matrices~$\1$ and~$\pseudo$, whereas~\eqref{Ago}
involves an additional Clifford multiplication by the vector~$\xi$.
For the construction of the tangent cone measure, the contribution involving the Clifford multiplication
is crucial. Therefore, \eqref{Anogo} is of no use, leading us to dismiss~\eqref{As} and~\eqref{Ap}.
The functional~\eqref{cchain}, on the other hand, looks promising in view of~\eqref{Ago}.
The function~$a$, $b$ and~$c$ all are all smooth and non-zero at the origin. Namely,
\begin{align*}
a &= \big(1 + 6 \tau^2 + \tau^4 \big) + \O \big( |\xi|^2 \big) \\
b &= 4 \tau\, \big(1 + \tau^2 \big) + \O \big( |\xi|^2 \big) \\
c &= 2 m \tau \big( 1-\tau^2 \big) + \O \big( |\xi|^2 \big) \:.
\end{align*}
In view of the additional factor~$\xi$ in the last summand in~\eqref{Ago}, this means that
at~$\xi=0$, the contributions proportional to~$\1$ and~$\pseudo$ dominate.
It is preferable to remove these contributions, because we want to focus on the
Clifford multiplication part. Clearly, for a Riemannian fermion system, where~$s_\zeta=\pi_\zeta$,
the decomposition~\eqref{Adodef} into block diagonal and off-diagonal parts cannot be used.
But we can subtract off the contribution for~$\xi=0$ according to~\eqref{cchain}.
We thus obtain
\beq \label{Aexp}
{\mathcal{A}}(\zeta') = \pi_\zeta \,(\zeta'-\zeta)\, \zeta|_{S_\zeta} =
c_0\, i \,\xi \!\cdot\!\sigma\, \pseudo + \O \big( |\zeta|^2 \big)
\eeq
with~$c_0 = 2 m^3 \tau \big( 1-\tau^2 \big)$. Restricting attention to the case~$\tau \neq 1$,
the coefficient~$c_0$ is non-zero.

The resulting tangent cone measure of Lemma~\ref{lemmameas} is most conveniently described
as follows. We consider the embedding of~$S^1$ into~$\Symm(S_\zeta)$
\[ f : S^1 \rightarrow \Symm(S_\zeta) \:,\qquad e^{i \varphi} \mapsto \sigma^1 \cos \varphi
+ \sigma^2 \sin \varphi \:. \]
Moreover, we let~$\mu_{S^1}$ be the normalized Lebesgue measure on~$S^1$. Then
\beq \label{muS1}
\mu_\con = f_* \mu_{S^1}
\eeq
(meaning that~$\mu_\con(\Omega) = \mu_{S^1}(f^{-1}(\Omega))$ for
any conical set~$\Omega \in {\mathfrak{M}}_\con$).

We next analyze the functional~$L(U)$ introduced in~\eqref{Ldef}.
First, the two-dimensional Clifford subspaces~$\K_\zeta^{(2,0)}$
are the two-dimensional subspaces of the space spanned by the three Pauli matrices.
Describing them by a unit vector normal to this subspace, we find
\[ \K_\zeta^{(2,0)} = \{ K^{(\nu)} \:|\: \nu \in S^2 \subset \R^3 \}
\qquad \text{where} \qquad
K^{(\nu)} := \{ \vec{x} \!\cdot\! \vec{\sigma} \text{ with } \vec{x} \in \R^3, \:\vec{x} \perp \vec{\nu} \} \:. \]
Since~$K^{(\nu)}=K^{(-\nu)}$, one sees that~$\K_\zeta^{(2,0)}$ is homeomorphic to
the real projective plane~${\mathbb{P}}^2(\R).$ Now, using~\eqref{muS1},
\begin{align*}
L \big( K^{(\nu)} \big)
&= \frac{1}{2 \pi} \int_0^{2 \pi} \Tr_{\text{Symm}(S_x)} \big( \pi_{K^{(\nu)}} \,\pi_{\bra \sigma^1 \cos \varphi
+ \sigma^2 \sin \varphi \ket} \big) \: d\varphi \\
&= \frac{1}{\pi} \int_0^{2 \pi} \left( 1 - \nu_1^2\, \cos^2 \varphi - \nu_2^2 \sin^2 \varphi \right) d\varphi
= \frac{1}{2 \pi} \int_0^{2 \pi} \left( 2 - \nu_1^2 - \nu_2^2 \right) d\varphi \:.
\end{align*}
Obviously, this functional is maximal if~$\nu_1=\nu_2=0$. As a consequence, there is the
unique maximizer
\[ K = \{ x \sigma^1 + y \sigma^2 \text{ with } x,y \in \R \} \:. \]
Varying~$\zeta$, we obtain a tangential Clifford section,
\beq \label{Clmix}
\Cl_\zeta = K = \{ x \sigma^1 + y \sigma^2 \text{ with } x,y \in \R \} \:.
\eeq
We conclude that the tangent cone measure~$\mu_\con$ is non-degenerate
(see Definition~\ref{defRnondeg}).

It remains to construct a spin structure. Linearizing~\eqref{Aexp}, the mapping~$\dsf {\mathcal{A}}$
defined by~\eqref{kappaform} is readily computed by
\beq \label{kappamix}
\dsf {\mathcal{A}}(u) = c_0\, i \,u \!\cdot\!\sigma\, \pseudo \::\: T_\zeta M \simeq \R^2 \rightarrow \Symm(S_\zeta)\:.
\eeq
Comparing with~\eqref{Clmix}, one sees that the image of~$\dsf {\mathcal{A}}$ lies in~$\Cl_\zeta$.
Hence
\beq \label{gzdef}
\gamma_\zeta = \pi_{\Cl_\zeta} \circ \dsf {\mathcal{A}} = \dsf {\mathcal{A}}\:,
\eeq
where in the last step we used that $\dsf {\mathcal{A}}$ maps to~$\Cl_\zeta$
(as is obvious from~\eqref{kappamix} and~\eqref{Clmix}).
In this way, the spin structure gives back the usual Clifford multiplication.

We conclude that for the Euclidean plane with chiral asymmetry, the constructions
in Section~\ref{sectangential} yield a unique tangential Clifford section
and a unique spin structure, in agreement with the usual Clifford multiplication.
Apart from providing a consistency check, this result is important because it
shows that the constructions in Section~\ref{sectangential} also apply to
spaces which are sufficiently small perturbations of the Euclidean plane,
even if these perturbations destroy the structure of a two-dimensional spin manifold.
More precisely, suppose that we perturb the universal measure of the Euclidean plane.
Then the resulting topological fermion system will in general no longer have
a smooth manifold structure. Nevertheless, provided that the perturbation is so small that
the tangent cone measure remains regular in the sense of Definition~\ref{defRnondeg},
we still have a tangential Clifford section~$\Cl M$,
giving rise to a well-defined notion of Clifford multiplication.
If the resulting quantum space still has the structure of a two-dimensional
differentiable manifold, we can again use~\eqref{gzdef} to obtain a canonical spin structure.
In this way, many of our notions and constructions can be carried over to
non-smooth or singular spaces.

\subsection{The Spin Structure of Two-Dimensional Minkowski Space} \label{secssMin}
We now return to the example of two-dimensional Minkowski space
introduced in Section~\ref{secminkowski}.
Our first step is the analysis of the tangent cone measures. In preparation, we need
to choose the function~${\mathcal{A}}$ in~\eqref{Afunct}.
In all the following computations, we again work in the basis~$(\f_{1\!/\!2}(\zeta))$ of the spin spaces
introduced in~\eqref{falpha}.
For clarity, we write the result of Lemma~\ref{lemmaPkern} symbolically as
\beq \label{PPbex}
\begin{split}
P(\zeta, \zeta') &= \left( (i \xi^0+\varepsilon)\, \gamma^0 - i \xi^1 \gamma^1 \right) \alpha(\xi) + \beta(\xi) \\
P(\zeta', \zeta) &=  \left( (-i \xi^0+\varepsilon)\, \gamma^0 + i \xi^1 \gamma^1 \right) \alpha(-\xi) + \beta(-\xi) \:,
\end{split}
\eeq
where we again set~$\xi=\zeta'-\zeta$ and
\beq \label{abex}
\begin{split}
\alpha(\xi) &= \frac{i m}{4 \pi^2}\: \frac{K_1 \Big(-i m \sqrt{(-\xi^0+i \varepsilon)^2 - (\xi^1)^2} \Big)}
{\sqrt{(-\xi^0+i \varepsilon)^2 - (\xi^1)^2}} \\
\beta(\xi) &= \frac{m}{4 \pi^2}\: K_0 \Big(-i m \sqrt{(-\xi^0+i \varepsilon)^2 - (\xi^1)^2} \Big) \:.
\end{split}
\eeq
The symmetry of the kernel of the fermionic operator yields
\[ P(\zeta', \zeta) = P(\zeta, \zeta')^*
= \left( (-i \xi^0+\varepsilon)\, \gamma^0 + i \xi^1 \gamma^1 \right) \overline{\alpha(\xi)} + \overline{\beta(\xi)}\:, \]
implying that
\beq \label{abrel}
\overline{\alpha(\xi)} = \alpha(-\xi) \qquad \text{and} \qquad \overline{\beta(\xi)} = \beta(-\xi) \:.
\eeq
These relations can also be verified directly by taking the complex conjugates of the above Bessel functions
(and choosing the correct branch of the square root).
The composite expressions in~\eqref{mat2a}--\eqref{mat5} are all of the form that
a diagonal matrix is multiplied from the left
by~$P(\zeta, \zeta')$ and from the right by~$P(\zeta', \zeta)$.
In order to see the general structure, we now compute and expand such expressions in the case
that the diagonal matrix is the identity matrix or the matrix~$\gamma^0$.
The corresponding expressions in~\eqref{mat2a}--\eqref{mat5} are then readily obtained by
taking suitable linear combinations. A straightforward computation yields
\begin{align}
P(\zeta, \zeta')\,P(\zeta', \zeta) &= \begin{pmatrix} (\beta(0) + \varepsilon \alpha(0))^2
& 2 i \varepsilon \,\xi^1\, \alpha(0)^2 \\
2 i \varepsilon \,\xi^1\, \alpha(0)^2 &
(\beta(0) - \varepsilon \alpha(0))^2 \end{pmatrix} + o(\xi) \label{AMin} \\
P(\zeta, \zeta')\,\gamma^0\, P(\zeta', \zeta) &=
\begin{pmatrix} (\beta(0) + \varepsilon \alpha(0))^2
& 2 i \,\xi^1\, \alpha(0) \,\beta(0) \\
2 i \,\xi^1\, \alpha(0) \,\beta(0) &
-(\beta(0) - \varepsilon \alpha(0))^2 \end{pmatrix} + o(\xi) \label{gMin}
\end{align}
(note that~$\alpha(0)$ and~$\beta(0)$ are real by~\eqref{abrel}).
The off-diagonal matrix elements both have the desired dependence on~$\xi^1$.
In~\eqref{AMin}, this dependence drops out in the limit~$\varepsilon \searrow 0$.
Thus in order for our construction to be independent of the regularization, it is preferable
to work with~\eqref{gMin}, which corresponds to choosing~${\mathcal{A}}$ according to~\eqref{Ap}.
If one is not concerned about the dependence on~$\varepsilon$ (which is allowed as long as one does not
intend taking the limit~$\varepsilon \searrow 0$), one can just as well choose~${\mathcal{A}}$
according to~\eqref{cchain} or~\eqref{As}.

The resulting tangent cone measure~$\mu_\con$ 
(see Lemma~\ref{lemmameas} and~\eqref{defmus}) is most conveniently described
as follows. Since the linearization of~${\mathcal{A}}$ maps to the line~$\R \gamma^2 \subset \Symm(S_x)$,
the measure~$\mu_\con$ will also be supported only on this line.
Therefore, we introduce the mapping
\[ f : \{\pm 1\} \rightarrow \Symm(S_\zeta) \:,\qquad f(\pm 1)=\pm \gamma^2 \:, \]
where the Dirac matrix~$\gamma^2$ is defined by
\[ \gamma^2 := i \gamma^1 \gamma^0 = -\begin{pmatrix} 0 & i \\ i & 0 \end{pmatrix} . \]
Moreover, we let~$\mu$ be the normalized counting measure on~$\{\pm 1\}$. Then
\beq \label{mupt}
\mu_\con = f_* \mu\:.
\eeq
If one prefers the notation with Dirac measures, one can also write
\[  \mu_\con(B) = \frac{1}{2} \Big( \delta_{\gamma^2}(B) +  \delta_{-\gamma^2}(B) \Big) , \]
to be evaluated for conical sets.

We next analyze the variational principle~\eqref{Ldef2} and~\eqref{max2}.
Fist of all, the symmetric operators which anti-commute with~$s_\zeta$ 
as well as the Clifford extensions can be parametrized by an angle~$\varphi$,
\begin{align*}
\AC\{s_\zeta\} &= \left\{ \gamma^1 \cos \varphi + \gamma^2 \sin \varphi \text{ with } \varphi \in \R \right\} \\
\K_\zeta^{s_\zeta, 1} &= \left\{ K^{(\varphi)} \:|\: \varphi \in \R \right\} \qquad \text{where} \\
K^{(\varphi)} &\!:= \{ t \gamma^0 + x (\gamma^1 \cos \varphi + \gamma^2 \sin \varphi) \text{ with } 
t, x \in \R \}\:.
\end{align*}
Using~\eqref{mupt}, the functional~\eqref{Ldef2} becomes
\[ L(U) = \frac{1}{2} \sum_\pm
\Tr_{\text{Symm}(S_x)}(\pi_U \,\pi_{\la \pm \gamma^2 \ra}) =
\Tr_{\text{Symm}(S_x)}(\pi_U \,\pi_{\la \gamma^2 \ra}) \:, \]
and thus
\[ L\big( K^{(\varphi)} \big) = \sin^2 \varphi \:. \]
Maximizing~$L$, we obtain the unique Clifford extension
\[ K = \{ t \gamma^0 + x \gamma^2 \text{ with } t, x \in \R \}\:. \]
We conclude that the tangent cone measure~$\mu_\con$ is non-degenerate
(see Definition~\ref{defLnondeg}).

Linearizing~\eqref{AMin} and~\eqref{gMin}, the mapping~$\dsf {\mathcal{A}}$ defined by~\eqref{kappaform}
is readily computed by
\beq \label{dA1}
\dsf {\mathcal{A}}(u) = c_1\,u^1 \gamma^2 \::\: T_\zeta M \simeq \R^2 \rightarrow \Symm(S_\zeta)
\eeq
with a non-zero constant~$c_1$. Again, the image of~$\dsf {\mathcal{A}}$ lies in~$K$.
However, $\dsf {\mathcal{A}}$ is {\em{not}} bijective. In the case that~${\mathcal{A}}$ is chosen
according to~\eqref{As}, this is precisely the observation made after the proof of Lemma~\ref{lemmasxk}.
More generally, the shortcoming is that the functionals~${\mathcal{A}}$ considered so far
do not distinguish a time direction.

In~\cite[Section~3.5]{lqg} a time direction was distinguished in spin dimension two
by working in suitable ``synchronized'' bases of the spin spaces.
Here we give a somewhat different construction which does not involve a choice of bases
(for a similar construction in four-dimensional Minkowski space see~\cite[Section~\S1.2.5]{cfs}).
In preparation, we work out the general structure of the formulas~\eqref{PPbex}
and of resulting composite expressions. Since we are interested in the local behavior near~$\xi=0$,
it is preferable to expand~\eqref{PPbex} and~\eqref{abex} in a Taylor series in~$\xi^0$.
Noting that the functions in~\eqref{abex} are even in~$\xi^1$, we have
\beq \label{PPbex2}
\begin{split}
P(\zeta, \zeta') &= \sum_{p=0}^\infty \Big( i \xi^0 \gamma^0 \,v^0_p + i \xi^1 \gamma^1\, v^1_p
+ s_p \Big)  \,(i \xi^0)^p \\
P(\zeta', \zeta) &= \sum_{p=0}^\infty \Big( -i \xi^0 \gamma^0 \,v^0_p - i \xi^1 \gamma^1\, v^1_p
+ s_p \Big)  \,(-i \xi^0)^p \:,
\end{split}
\eeq
where the coefficient functions~$v^0_p$, $v^1_p$ and~$s_p$ depend only on~$(\xi^1)^2$.
The relations~\eqref{abrel} immediately yield that the functions~$v^0_p$, $v^1_p$ and~$s_p$
are real-valued. Hence
\beq \label{A1}
\begin{split}
P&(\zeta, \zeta')\, P(\zeta', \zeta) = \sum_{p,p'=0}^\infty 
(i \xi^0)^{p} \,(-i \xi^0)^{p'} \;\Big( (\xi^0)^2 \,v^0_p \,v^0_{p'} - (\xi^1)^2 \,v^1_p \,v^1_{p'}  + s_p\, s_{p'} \\
&+ \xi^0 \xi^1 \, \gamma^0 \gamma^1 \big( v^0_p \,v^1_{p'} - v^1_p \,v^0_{p'} \big)
+ i \xi^0 \gamma^0 \big( v^0_p s_{p'} - s_p v^0_{p'}\big)
+ i \xi^1 \gamma^1 \big( v^1_p s_{p'} - s_p v^1_{p'}\big) \Big) \:.
\end{split}
\eeq
We now analyze the symmetry if~$p$ and~$p'$ are exchanged.
Under this replacement, the factor~$(i \xi^0)^{p} \,(-i \xi^0)^{p'}$ changes sign if~$p+p'$ is odd, whereas it
does not change sign if~$p+p'$ is even.
As a consequence, the terms in the first line vanish if~$p+p'$ is odd,
whereas the second line vanishes if~$p+p'$ is even. Thus one sees that
the closed chain~$P(\zeta, \zeta')\, P(\zeta', \zeta)$ remains unchanged under the replacements
\beq \label{reps}
\xi \rightarrow -\xi \qquad \text{and} \qquad \gamma^1 \rightarrow -\gamma^1\:.
\eeq
Thus the closed chain does {\em{not}} distinguish a time direction. Similarly,
\beq \label{A2}
\begin{split}
P&(\zeta, \zeta')\,\gamma^0\, P(\zeta', \zeta) = \sum_{p,p'=0}^\infty 
(i \xi^0)^{p} \,(-i \xi^0)^{p'} \:\gamma^0\, \Big( (\xi^0)^2 \,v^0_p \,v^0_{p'}
+ (\xi^1)^2 \,v^1_p \,v^1_{p'}  + s_p\, s_{p'} \\
&+ \xi^0 \xi^1 \, \gamma^0 \gamma^1 \big( v^0_p \,v^1_{p'} + v^1_p \,v^0_{p'} \big)
+ i \xi^0 \gamma^0 \big( v^0_p s_{p'} - s_p v^0_{p'}\big)
- i \xi^1 \gamma^1 \big( v^1_p s_{p'} + s_p v^1_{p'}\big) \Big) \,.
\end{split}
\eeq
Again considering the symmetry when exchanging~$p$ with~$p'$, we conclude that
the composite expression~$P(\zeta, \zeta')\,\gamma^0\, P(\zeta', \zeta)$ remains unchanged
again under the replacements~\eqref{reps}. Thus again, no time direction is distinguished.

Another method for understanding this shortcoming is to take the trace of composite
expressions and use the invariance under cyclic commutations. For example,
\[ \Tr \big( P(\zeta, \zeta')\, P(\zeta', \zeta) \big) = \Tr \big( P(\zeta', \zeta)\, P(\zeta, \zeta') \big) \:, \]
showing that the scalar component of the closed chain is invariant under the replacement~$\xi \rightarrow -\xi$.
By inserting additional matrices into the trace, one can verify the symmetries of all
the contributions in~\eqref{A1} and~\eqref{A2}. Moreover, the following consideration shows that
it is impossible to distinguish a time direction if~$\zeta^1=0$: Consider the trace of a composite
expression of the form
\beq \label{trace}
\Tr \big( M_0\, P(\zeta, \zeta') \,M_1\, P(\zeta', \zeta) \,M_2\:\cdots\: M_{2k} \,P(\zeta, \zeta')
\,M_{2k+1}\, P(\zeta', \zeta) \big) \:,
\eeq
where the factors~$M_\ell$ are linear combinations of~$\1$ and~$\gamma^0$
(note that in view of~\eqref{mat1}, these are the only matrices which we can form locally
at the point~$\zeta$ or~$\zeta'$). Assuming that~$\zeta^1=0$, the matrices~$P(\zeta, \zeta')$
and~$P(\zeta', \zeta)$ are also multiples of~$\1$ and~$\gamma^0$
(see for example~\eqref{PPbex2}). As a consequence, all the matrices in~\eqref{trace}
commute with each other. Thus we can reorder the matrices to obtain~\eqref{trace}
with the arguments~$\zeta$ and~$\zeta'$ exchanged.
Hence~\eqref{trace} is symmetric under the replacement~$\zeta \rightarrow -\zeta$,
proving that no time direction is distinguished.

We now give a method for distinguishing a time direction. We first note that
taking the commutator of~$\P(\zeta, \zeta')$
with~$\gamma^0$, only the component involving~$v^1_\bullet$ remains, i.e.\ in view of~\eqref{PPbex2}
\[ \big[ \gamma^0, P(\zeta, \zeta') \big] = 2 \sum_{p=0}^\infty i \xi^1 \gamma^0 \gamma^1\, v^1_p \,(i \xi^0)^p \:.\]
We now take the trace of a product which involves two such commutators,
\[ \B(\zeta, \zeta') :=
\Tr \Big( \big[ P(\zeta, \zeta'), \gamma^0 \big]\, P(\zeta', \zeta)\, \big[ P(\zeta, \zeta'), \gamma^0 \big]\,
P(\zeta', \zeta) \Big) \:. \]
Then of both factors~$P(\zeta, \zeta')$ only the component~$v^1_\bullet$ enters, whereas
the factors~$P(\zeta', \zeta)$ also contribute via the components~$v^0_\bullet$ and~$s_\bullet$.
In this way, the symmetry $\zeta \leftrightarrow \zeta'$ is broken. In particular, we may
anti-symmetrize to obtain a real-valued functional which is anti-symmetric under the
replacement~$\xi \rightarrow -\xi$,
\beq \label{Cdef}
{\mathscr{C}}(\zeta, \zeta') := \frac{1}{2i} \big( \B(\zeta, \zeta') - \B(\zeta', \zeta) \big) \:.
\eeq
A straightforward computation shows that
\beq \label{Cform2}
{\mathscr{C}}(\zeta, \zeta') = 16 \,\xi^0\, (\xi^1)^2 \,\Big( i \alpha(0) \beta(0)  
\Big[ \beta(0)\: \frac{\partial \alpha(0)}{\partial \xi^0} - \alpha(0)\:\frac{\partial \beta(0)}{\partial \xi^0} \Big]
- \varepsilon \alpha(0)^4 \Big) + \O \big( \|\xi\|^4 \big)
\eeq
(where~$\alpha$ and~$\beta$ are again the functions~\eqref{abex}, and~$\| \xi \|^2
:= |\xi^0|^2 + |\xi^1|^2$). A direct calculation shows that the coefficient in the round brackets
in~\eqref{Cform2} is indeed non-zero.

Using the notation in Lemma~\ref{lemma75}, we can write the functional~${\mathscr{C}}(\zeta, \zeta')$
in a simpler form:
\begin{Lemma} The functional~${\mathscr{C}}(\zeta, \zeta')$ as defined by~\eqref{Cdef} is given by
\beq \label{Cform}
{\mathscr{C}}(\zeta, \zeta') = i c \Tr \big( \zeta \,\pi_{\zeta'}\,
\pi_{\zeta}\, \zeta' - \zeta\,\zeta'\,\pi_\zeta \,\pi_{\zeta'} \big) \:,
\eeq
where~$c$ is the positive constant
\[ c = \frac{4\, \nu_1^2 \nu_2^2}{(\nu_2-\nu_1)^2}\:. \]
\end{Lemma}
\Proof
Writing out the commutators and cyclically commuting the factors inside the trace, we obtain
\[ {\mathscr{C}}(\zeta, \zeta') = i \Tr \Big( P(\zeta, \zeta')\, P(\zeta', \zeta)\,\Big[ P(\zeta, \zeta')\, 
\gamma^0\,P(\zeta', \zeta)\,\gamma^0 - \gamma^0\, P(\zeta, \zeta')\,\gamma^0\, P(\zeta', \zeta) \Big]
\Big) \:. \]
Using~\eqref{mat2}, this simplifies to
\beq \label{Cform3}
{\mathscr{C}}(\zeta, \zeta') = i \Tr \Big( \zeta'\, \zeta \,\big( \zeta' \,s_{\zeta'}\,
\zeta \,s_{\zeta} - s_{\zeta} \,\zeta' \,s_{\zeta'} \,\zeta \big) \Big) \:.
\eeq
According to~\eqref{mat1}, we can write the matrix product~$\zeta s_{\zeta}$ as a linear combination
of~$\zeta$ and~$\pi_\zeta$,
\[ \zeta \,s_{\zeta} = \frac{\nu_1+\nu_2}{\nu_1 - \nu_2}\: \zeta - \frac{2 \nu_1 \nu_2}{\nu_1-\nu_2}\: \pi_\zeta\:. \]
Substituting this formula into~\eqref{Cform3}, multiplying out and cyclically commuting the factors inside the trace,
we obtain the result.
\QED
The trace in~\eqref{Cform} is obviously anti-symmetric under permutations of~$\zeta$ and~$\zeta'$.
It seems that this trace can be used on general causal fermion systems to distinguish the future from
the past. Indeed, when evaluating this trace in spin dimension two, one gets
(up to irrelevant prefactors) the same expression which was used in~\cite[Section~3.5]{lqg}
to distinguish a time direction.

In order to complete the construction of the spin structure, 
we need to modify~\eqref{Ap} by adding a term which gives a linear time dependence.
According to~\eqref{Cform2}, the function~${\mathscr{C}}$ has this desired linear dependence on~$\xi^0$,
but unfortunately it vanishes to second order in the spatial coordinate~$\xi^1$.
The last problem can be handled by integrating over a small ball~$B_\delta(\zeta')$
(where we again work with the metric in~$\F \subset \Lin(\H)$ induced by the sup-norm).
Thus for for given~$\delta>0$ we choose the functional~\eqref{Afunct} as
\[ {\mathcal{A}}(\zeta') = \pi_\zeta \,\pi_{\zeta'}\, \zeta|_{S_\zeta}
+ s_\zeta \int_{B_\delta(\zeta')} {\mathscr{C}}(\zeta, \theta)\: d\rho(\theta)
+ c \:, \]
where the constant~$c$ is chosen such that~${\mathcal{A}}(\zeta)=0$
(we remark that, just as explained after~\eqref{gMin}, we could also have modified~\eqref{cchain}
or~\eqref{As} instead of~\eqref{Ap}). Then the derivative of~${\mathcal{A}}$
becomes in modification of~\eqref{dA1}
\beq \label{dA2}
\dsf {\mathcal{A}}(u) = c_1\, u^1 \gamma^2 
+ c_0\, u^0 \gamma^0 \::\: T_\zeta M \simeq \R^2 \rightarrow \Symm(S_\zeta)
\eeq
with a non-zero constant~$c_0$ which depends on the coefficient in~\eqref{Cform2}
and on~$\delta$. Now~$\dsf {\mathcal{A}}$ is injective, and its image lies in~$\Cl_\zeta$.
Hence the composition
\beq \label{Cliff}
\gamma_\zeta = \pi_{\Cl_\zeta} \circ \dsf {\mathcal{A}} = \dsf {\mathcal{A}}
\eeq
gives the desired spin structure.

We finally point out that the Clifford multiplication induced by the spin structure~\eqref{Cliff}
differs from the expected Clifford multiplication by the fact that in~\eqref{dA2} the
component~$u^1$ is multiplied by~$\gamma^2$ instead of~$\gamma^1$.
This difference is of no relevance as long as topological questions are considered.
However, it shows that the Clifford multiplication induced by~\eqref{Cliff} 
does not have a geometric meaning, exactly as explained at the end of
Section~\ref{secconstruct} after~\eqref{Lorentz}.
The significance of these constructions lies in the fact that they are robust to general perturbations.
Thus we obtain a canonical spin structure for small perturbations of the two-dimensional
Minkowski space, even if the manifold structure ceases to exist.

\section{Spinors on Singular Spaces} \label{secsing}
We now turn attention to examples on curved space or in curved space-time.
Our main interest is to describe singularities of the geometry.
For the computation of the Dirac operator, we shall always use the following convenient method,
which makes it unnecessary to compute all the spin coefficients.
Instead, the zero order term in the Dirac operator is written as a  ``covariant divergence.''
Suppose that the metric of our manifold~$(\scrM^k, g)$ has signature~$(r,s)$, and that the
spin scalar product on the corresponding spinor bundle has signature~$(\mathfrak{q,p})$.
Our starting point are symmetric matrices~$(\sigma^i)_{i=1,\ldots, k}$ on~$\C^{\mathfrak{q,p}}$
which satisfy the canonical anti-commutation relations in the corresponding Euclidean or
Minkowski space,
\beq \label{anticomm}
\{ \sigma^i, \sigma^j \} = 2 s_i\: \delta^{ij} \:,\qquad
\text{with} \qquad s_1,\ldots, s_r=1,\; s_{r+1},\ldots, s_k=-1 \:.
\eeq
A typical choice are the Pauli matrices or the usual Dirac matrices.

\begin{Prp} \label{prpDOP} Let~$(\scrM^k,g)$ be a spin manifold with a metric
of signature~$(r,s)$ with~$r \leq 1$ and~$s \leq 3$. Let~$\Dir$ be the Dirac operator acting
on smooth sections~$\Gamma(\scrM, S\scrM)$ of the corresponding spinor bundle
(for an arbitrarily chosen spin structure).
Assume that there is a local chart~$(x^i, U)$ in which the metric is diagonal, i.e.\
\[ ds^2 = g_1\, dx_1^2 + \cdots + g_r\, dx_r^2
\:-\: g_{r+1} \,dx_{r+1}^2 - \cdots - g_k\, dx_k^2 \]
(where the coefficient functions~$g_i \in C^\infty(U, \R^+)$ may depend on all~$k$ variables).
Then around every point in~$U$ there is a local trivialization of the spinor bundle by pseudo-orthonormal
bases $(\e_\alpha)_{\alpha=1,\ldots, {\mathfrak{p+q}}}$ such that in this local chart and trivialization,
the Dirac operator takes the form
\beq \label{Dirac}
\Dir = i G^j \frac{\partial}{\partial x^j} + B \:,
\eeq
where
\begin{align}
G^j(x) &= g_j(x)^{-\frac{1}{2}}\: \sigma^j \label{Gchoice} \\
B(x) &= \frac{i}{2 \sqrt{|\det g|}}\: \partial_j \left( \sqrt{|\det g|}\, G^j \right) . \label{Bchoice}
\end{align}
More generally, if in a chart and local trivialization of the spinor bundle by pseudo-orthonormal
bases $(\e_\alpha)_{\alpha=1,\ldots, {\mathfrak{p+q}}}$ the matrices~$G^j$ are of the form
\beq \label{GOchoice}
G^j(x) = g_j(x)^{-\frac{1}{2}}\: O^j_k(x)\, \sigma^k \qquad \text{with~$O(x) \in \SO(r,s)$}\:,
\eeq
then the Dirac operator is of the form~\eqref{Dirac} with~$B$ according to~\eqref{Bchoice}.
\end{Prp}
\Proof By taking the Cartesian product with a Euclidean or Minkowski space,
we obtain a Lorentzian manifold of signature~$(1,3)$. Moreover, the spinor bundle of~$\scrM$
can be recovered as a sub-bundle of the spinor bundle on the Lorentzian manifold
(in particular, the spinors of~$\scrM$ may be recovered as the left- or right-handed
component of the four-component Dirac spinors). Then we can use the formalism developed in~\cite{u22}
(see also~\cite[\S1.5]{PFP}).
With~\eqref{Gchoice} we have satisfied the anti-commutation relations
\[ \{G^j, G^k \} = 2\, g^{jk}\, \1 \:. \]
Moreover, the choice~\eqref{Gchoice} ensures that the pseudoscalar matrix is constant,
and that all derivatives of the~$G^j$ are in the span of~$\sigma^1, \ldots, \sigma^k$.
Then the zero order term of the Dirac operator can be written as
(see~\cite[eqs~(41), (42) and~(51)]{u22})
\[ B = G^j E_j \qquad \text{with} \qquad
E_j = -\frac{i}{16}\: \Tr \big( G^m \,(\partial_j G^n + \Gamma^n_{jl} \,G^l) \big) \: G_m G_n\:, \]
where~$\Gamma^n_{jl}$ are the Christoffel symbols of the Levi-Civita connection
(and the partial derivatives simply act on each component). Hence
\beq \label{Bform}
B =  -\frac{i}{16}\: \Tr \big( G_m \,(\nabla_j G_n) \big) \: G^j G^m G^n \:,
\eeq
where~$\nabla_j G_n \equiv \partial_j G_n - \Gamma^k_{jn} G_k$ is the covariant derivative
acting on the components of the spinorial matrix.
Using the algebra of the Dirac matrices, one finds that~\eqref{Bform} has a vectorial component
(obtained by using the anti-commutation relations), and an axial component which 
is totally antisymmetric in the indices~$j$, $m$, and~$n$.
This totally antisymmetric term vanishes for the following reasons: First, since the Levi-Civita
connection is torsion-free, we may replace the covariant derivative by a partial derivative.
Second, it follows from~\eqref{Gchoice} that the matrix~$\partial_j G_n$ is a multiple of~$G_n$,
implying that the trace~$\Tr(G_m (\partial_j G_n))$ is symmetric in the indices~$m$ and~$n$.

It remains to compute the vectorial component of~\eqref{Bform}. A short computation yields
\[ B = \frac{i}{2}\, \nabla_j G^j\:, \]
and the usual formula for the covariant divergence of a vector field gives the result.
\QED

This proposition also gives a method for constructing the Dirac operator on a manifold.
To this end, one takes~\eqref{Dirac} as the definition of the Dirac operator in a local chart
and trivialization.
Identifying these so-defined Dirac operators in different charts by suitable transformations
of the coordinates and spinor frames, one obtains a globally defined Dirac operator.
In all the following examples, it will be straightforward to match the Dirac operators in the different charts.
However, we point out that in general, this is a non-trivial task, which amounts to verifying that
the manifold is spin and to choosing a specific spin structure.
In order to keep the following examples as simple as possible, we do not discuss the
freedom in choosing different spin structures.

\subsection{Singularities of the Conformal Factor} \label{secconformal}
We first consider curvature singularities which can be removed by
a conformal transformation of the metric. To this end, we assume that~$g$ is
a smooth metric, and that
\[ \tilde{g} = f^2\: g \]
with a smooth positive function~$f$. We denote the corresponding spinor
bundles and Dirac operators by~$S\scrM$, $S \:\tilde{\!\!\scrM}$ and~$\Dir$, $\tilde{\Dir}$.
According to~\cite{hitchin, hijazi} there is a fibrewise isometry~$\psi \mapsto \iota \psi$
of the spinor bundles  such that
\[ \tilde{\Dir} (\iota \psi)
= f^{-\frac{k+1}{2}}\,\iota \left( \Dir (f^{\frac{k-1}{2}}\psi ) \right) \]
(where~$k$ again denotes the dimension of~$\scrM$).
For ease of notation we usually omit the identification map~$\iota$ and simply write
\[ \tilde{\Dir} \psi  = f^{-\frac{k+1}{2}} \Dir \big( f^{\frac{k-1}{2}} \psi \big) \:. \]
In particular, solutions of the massless Dirac equation transform conformally,
\[ \Dir \psi = 0  \qquad \Longrightarrow \qquad \tilde{\Dir} \tilde{\psi} = 0 \quad \text{with} \quad
\tilde{\psi} = f^{-\frac{k-1}{2}}\, \psi \:. \]
For the massive Dirac equation~$(\Dir-m)\psi=0$, the situation is no longer so simple, because the
mass parameter brings in a length scale and thus destroys the conformal invariance.
Nevertheless, if we consider singularities of the conformal factor~$f$
where~$f \searrow 0$, then typically the mass parameter does not affect the
behavior of the spinor near the singularity. Hence we expect that the rescaled spinor
\beq \label{psires}
f^{\frac{k-1}{2}}\, \tilde{\psi}
\eeq
has a well-defined limit even if~$f \searrow 0$.
Since the topological spinor bundle is defined purely in terms of wave functions,
this will imply that the topological spinor bundle has a well-defined regular limit
even if the metric and curvature become singular.

Clearly, such singularities of the conformal factor can be treated just as well by
a conformal rescaling of the metric, as is a common procedure when constructing
for example conformal compactifications or Penrose diagrams.
Thus at this point, working with topological spinor bundles gives no major benefit.
However, the main advantage of working with topological spinor bundles becomes apparent
when considering curvature singularities which do {\em{not}} come from a conformal
transformation of the metric. Such singularities, which we refer to as {\em{genuine curvature
singularities}}, will be treated in the next section~\ref{secgenuine}.
Before, we illustrate conformal transformations by two simple examples.

\begin{Example} (A neck singularity of a 2-d Minkowski cylinder) \label{ex2dlorentz}
{\em{ 
On~$\scrM=\R \times S^1$ we choose coordinates~$(t, \varphi)$ with~$t \in \R$ and~$0 < \varphi < 2 \pi$
and consider the two-dimensional Lorentzian metric
\beq \label{2dl}
ds^2 = dt^2 - R(t)^2\, d\varphi^2 \:.
\eeq
In order to construct the Dirac operator, we use the method of Proposition~\ref{prpDOP}.
We satisfy the anti-commutation relations~\eqref{anticomm} with the ansatz
\[ G^t = \sigma^3 \:,\qquad G^\varphi = \frac{i \sigma^1}{R(t)} \:. \]
In order for these matrices to be symmetric, we need to consider the spin scalar product
\beq \label{spinsp}
\Sl .|. \Sr = \la .| \sigma^3 . \ra_{\C^2} \:,
\eeq
which is clearly indefinite of signature~$(1,1)$. 
The matrices~$G^t$ and~$G^\varphi$ are of the form~\eqref{Gchoice}.
Thus, using~\eqref{Dirac}, the Dirac operator becomes
\[ \Dir = i \sigma^3 \partial_t + \frac{i \dot{R}(t)}{2 R(t)} \:\sigma^3 - \frac{1}{R(t)}\: \sigma^1 \partial_\varphi \:. \]
We consider the Dirac equation~$(\Dir-m) \psi=0$. This equation can be separated by the ansatz
\beq \label{lor2dans}
\psi = \frac{e^{i k \varphi}}{\sqrt{R(t)}} \begin{pmatrix} \chi_1(t) \\
i \chi_2(t) \end{pmatrix} \quad \text{with} \quad k \in \Z
\eeq
to obtain the ODE in time
\beq \label{ODEt}
i \dot{\chi}(t) = 
\begin{pmatrix} m & -k/R(t) \\ -k/R(t) & -m \end{pmatrix} \chi(t) \:.
\eeq
Since the matrix on the right is Hermitian, one readily verifies that
\beq \label{chiconst}
\frac{d}{dt}\, \|\chi\| = 0 \:,
\eeq
implying that the norm of~$\chi$ is time independent
(this conservation law can be identified with current conservation).
Since the norm is constant, the equation~\eqref{ODEt} can be understood as describing
``oscillations'' of the spinor (for more details on this geometric picture we refer to the similar
equation~\cite[eq.~(5)]{friedmann} and its reformulation with Bloch vectors in~\cite[Section~2]{friedmann}).
If the function~$R(t)$ gets small, the frequency of the oscillations gets larger.
In order to control these oscillations, we estimate the matrix in~\eqref{ODEt} to obtain the inequality
\[ \left\| \frac{d}{dt} \chi \right\| \leq 2 \,\Big(m + \frac{|k|}{R(t)} \Big)\, \|\chi\| \:. \]
Integration gives the inequality
\beq \label{cont}
\chi \big|_{t_0}^{t_1} \leq 2 \,\|\chi\| \int_{t_0}^{t_1} \Big(m + \frac{|k|}{R(t)} \Big)\, dt \:.
\eeq
This shows that if the function~$1/R$ is integrable, then the function~$\chi$ is continuous. As
a specific example, one can consider the family of metrics
\beq \label{Reps}
R(t) = (t^2+\varepsilon^2)^{\frac{1}{4}}\:.
\eeq
In the limit~$\varepsilon \searrow 0$, the metric becomes singular at~$t=0$,
forming a neck singularity. Nevertheless, in this limit the functions~$\chi$ converge locally uniformly
to continuous functions.

In order to construct a causal fermion system, we choose
a (for simplicity finite-dimensional) space of Dirac wave functions~$\H$
and endow it with a scalar product~$\la .|. \ra$.
We again introduce the local correlation operators~$F$ by~\eqref{loccorr}.
Due to the factor~$1/\sqrt{R}$ in~\eqref{lor2dans}, the Dirac wave functions
diverge in the limit~$\varepsilon \rightarrow 0$.
As a consequence, the local correlation operators will also diverge. But we can cure this problem
simply by rescaling the local correlation operators according to
\beq \label{Fscal1}
\tilde{F}(t, \varphi) = R(t)\, F(t, \varphi) \:.
\eeq
This rescaling corresponds precisely to the conformal rescaling~\eqref{psires} needed to
remove the curvature singularity at the cusp of the cylinder.
Since the functions~$\chi$ converge uniformly as~$\varepsilon \searrow 0$,
we conclude that the rescaled correlation operators converge.
Introducing the universal measure by $\rho = \tilde{F}_*(dt \,d\varphi)$,
for any~$\varepsilon>0$ we obtain a causal fermion system~$(\H, \F, \rho)$ of spin dimension one.
This family of causal fermion systems has a regular limit as~$\varepsilon \searrow 0$, despite the fact that
a curvature singularity forms. }} \QEDrem
\end{Example}

\begin{Example} (A conical singularity of a Riemannian surface) \label{ex2driemann}
{\em{ 
We choose~$\scrM = \R^2 \setminus \{0\}$ and consider the metric in polar coordinates~$(r, \varphi)$
of the form
\beq \label{polar2}
ds^2 = dr^2 + R(r)^2\, d\varphi^2 \:,
\eeq
where~$0<r$ and~$0< \varphi < 2 \pi$.
We again construct the Dirac operator with the method of Proposition~\ref{prpDOP}.
In order to satisfy the anti-commutation relations~\eqref{anticomm},
we take the ansatz
\[ G^r = \sigma^r\:, \qquad G^\varphi = \frac{\sigma^\varphi}{R(r)} \:, \]
where~$\sigma^r$ and~$\sigma^\varphi$ are the linear combinations of Pauli matrices
\[ \sigma^r = \sigma^1 \cos \varphi + \sigma^2 \sin \varphi \:,\qquad
\sigma^\varphi =-\sigma^1 \sin \varphi + \sigma^2 \cos \varphi \:. \]
The matrices~$G^r$ and~$G^\varphi$ are of the form~\eqref{GOchoice}.
Using~\eqref{Dirac}, the Dirac operator is computed by
\beq \label{D2dR}
\Dir = i \sigma^r \partial_r + \frac{i}{R(r)}\: \sigma^\varphi \partial_\varphi + 
\frac{i}{2} \left( \frac{R'(r)}{R(r)} - 1 \right) \sigma^r \:.
\eeq
Note that in the special case~$R(r)=r$, the metric~\eqref{polar2} becomes flat,
and the Dirac operator reduces to the Dirac operator~\eqref{dir2} in Euclidean~$\R^2$,
written in polar coordinates.

We consider the Dirac equation~$\Dir \psi = \lambda \psi$. This equation can be separated by the ansatz
\beq \label{sep2d}
\psi = \frac{1}{\sqrt{R(r)}} \begin{pmatrix} e^{i (k-\frac{1}{2}) \varphi}\, \chi_1(r) \\
i e^{i (k+\frac{1}{2}) \varphi}\, \chi_2(r) \end{pmatrix} \quad \text{with} \quad k \in \Z+\frac{1}{2}
\eeq
to obtain the radial ODE
\[ \chi'(r) = \begin{pmatrix} k/R & \lambda \\ -\lambda & -k/R \end{pmatrix} \chi(r) \]
($k$ must be a half integer in order for~$\psi$ to extend to a single-valued section of the spinor bundle).

We now choose~$R(r)$ as
\[ R(r) = \frac{r}{2} \:. \]
The corresponding metric is conical. It cannot be extended to the cusp singularity at~$r=0$.
In order to construct a regular topological fermion system, we need to choose at least two wave functions.
For simplicity, we choose
\[ \lambda=1 \:,\qquad k = \pm \frac{1}{2} \]
and let~$\psi_\pm$ the solution of the Dirac equation which is bounded at the origin.
These solutions can be computed explicitly by
\begin{align*}
\psi_+(r, \varphi) &= \frac{1}{r^\frac{3}{2}} \begin{pmatrix} r \sin(r) \\ -i e^{i \varphi} \, (\sin(r) - r \cos(r)) \end{pmatrix} \\
\psi_-(r, \varphi) &= \frac{1}{r^\frac{3}{2}} \begin{pmatrix} e^{-i \varphi} \, (\sin(r) - r \cos(r))  \\ i r \sin(r) \end{pmatrix} \:.
\end{align*}
We let~$(\H, \la .|. \ra_\H)$ be the vector space spanned by~$\psi_+$ and~$\psi_-$
with the scalar product such that~$\psi_\pm$ are orthonormal.
Then the local correlation operators (see Definition~\ref{defloccor}) are computed by
\beq \label{Fdef2}
F(r, \varphi) = -\left( \frac{\sin^2(r)}{r} + \frac{(\sin(r) - r \cos(r))^2}{r^3} \right) \begin{pmatrix} 1 & 0 \\ 0 & 1 \end{pmatrix} \:.
\eeq
Obviously, these local correlation operators tend to zero as~$r \searrow 0$.
In order to cure this problem, we rescale the local correlation operators similar to~\eqref{Fscal1} by setting
\[ \tilde{F}(r, \varphi) = \frac{1}{r}\, F(r, \varphi) \:. \]
The function~$\tilde{F}$ is continuous and has a non-zero limit at the origin. Thus we
can extend it continuously to the origin,
\[ \tilde{F} \::\: \scrM \cup \{0\} \simeq \R^2 \rightarrow \F \subset \Lin(\H) \:. \]
Taking the push-forward measure~$\rho = \tilde{F}_*(d^2x)$, we obtain a
Riemannian fermion system of spin dimension one.
We point out that this Riemannian fermion system is regular at the origin, although the Riemannian
metric of the underlying manifold is ill-defined.

For the just-constructed Riemannian fermion system,
the local correlation operators~\eqref{Fdef2} do not depend on the angular variable,
so that the mapping~$\tilde{F}$ is not injective. As a consequence,
the space~$M:= \supp \rho$ is homeomorphic to~$\R^+ \cup \{0\}$,
meaning that every circle~$r=\text{const}$ has been identified with a point.
Such identifications are useful in applications in which some degrees of freedom of~$\scrM$
are irrelevant or should be suppressed. Likewise, such identifications can be arranged to occur at the boundary
of~$\scrM$ or in an asymptotic end, making it possible to describe different
types of compactifications.

In order to avoid identifications, one simply extends~$\H$
by another wave function, like for example the eigensolution for~$\lambda=1$ and~$k=3/2$
being regular at the origin,
\[ \psi = \frac{1}{r^\frac{5}{2}} \begin{pmatrix} e^{i \varphi}\: \big(3 r \cos(r) - (3-r^2) \sin(r) \big) \\[0.1em] 
i e^{2 i \varphi} \, \big( 3 (5 - 2 r^2) \sin(r) - (15 r-r^3) \cos(r) \big) \end{pmatrix} \:. \]
A direct computation shows that on the resulting three-dimensional Hilbert space~$\H$,
the rescaled local correlation operators~$\tilde{F}$ are continuous at the origin,
and the mapping~$\tilde{F} : \R^2 \rightarrow \F$ is indeed injective.
}} \QEDrem
\end{Example}

\subsection{Genuine Singularities of the Curvature Tensor} \label{secgenuine}
We now consider curvature singularities which are genuine in the sense that the
singularity cannot be removed by a conformal transformation.
The next two examples illustrate that even in such cases, the topological fermion system
may be regular and well-behaved.

\begin{Example} (A genuine singularity on the Lorentzian torus times~$S^1$) {\em{ 
On~$\scrM = \R \times S^1 \times S^1$ we choose coordinates~$(t, \varphi, \alpha)$
with~$t \in \R$ and~$0 < \varphi, \alpha < 2 \pi$. As the Lorentzian metric on~$\scrM$
we take the warped product of~\eqref{2dl} with a metric on~$S^1$,
\[ ds^2 = dt^2 - R(t)^2\, d\varphi^2 - S(t)^2\, d\alpha^2 \:. \]
We satisfy the anti-commutation relations~\eqref{anticomm} with the ansatz
\[ G^t = \sigma^3 \:,\qquad G^\varphi = \frac{i \sigma^1}{R(t)} \:,\qquad
G^\alpha = \frac{i \sigma^2}{S(t)} \:. \]
In order for these matrices to be symmetric, we again consider
the spin scalar product~\eqref{spinsp} of signature~$(1,1)$. Again using~\eqref{Dirac},
the Dirac operator becomes
\[ G = i \sigma^3 \partial_t + \frac{i}{2} \left( \frac{\dot{R}(t)}{R(t)} 
+  \frac{\dot{S}(t)}{S(t)} \right) \sigma^3 - \frac{1}{R(t)}\: \sigma^1 \partial_\varphi 
-  \frac{1}{S(t)}\: \sigma^2 \partial_\alpha  \:. \]
The Dirac equation~$(\Dir-m) \psi=0$ can be separated by the ansatz
\beq \label{lor3dans}
\psi = \frac{e^{i k \varphi + i l \alpha}}{\sqrt{R(t) S(t)}} \begin{pmatrix} \chi_1(t) \\
i \chi_2(t) \end{pmatrix} \quad \text{with} \quad k, l \in \Z \:.
\eeq
We thus obtain the ODE in time
\[ i \dot{\chi}(t) = 
\begin{pmatrix} m & \displaystyle -\frac{k}{R(t)} + \frac{i l}{S(t)} \\[-.3em]
\displaystyle -\frac{k}{R(t)} - \frac{i l}{S(t)} & -m \end{pmatrix} \chi(t) \:. \]
Since the matrix on the right is Hermitian, the relation~\eqref{chiconst}
again holds, whereas equation~\eqref{cont} is modified to
\[ \chi \big|_{t_0}^{t_1} \leq 2 \,\|\chi\| \int_{t_0}^{t_1} \bigg(m + \frac{|k|}{R(t)} + \frac{|l|}{S(t)}
\bigg)\, dt \:. \]
If the functions~$1/R$ and~$1/T$ are integrable, we infer that~$\chi$ is continuous.

We let~$\H$ be a finite-dimensional space of Dirac wave functions
endowed with a scalar product~$\la .|. \ra$.
We again introduce the local correlation operators~$F$ by~\eqref{loccorr}.
For given~$\varepsilon>0$, we again choose the function~$R(t)$ according to~\eqref{Reps}.
In order for the factor~$(R(t) S(t))^\frac{1}{2}$ in~\eqref{lor3dans} to be regular in the
limit~$\varepsilon \searrow 0$, we choose
\[ S(t) = \frac{1}{R(t)} = (t^2+\varepsilon^2)^{-\frac{1}{4}}\:. \]
Then the functions~$\chi$ converge uniformly as~$\varepsilon \searrow 0$.
Since the prefactors in~\eqref{lor3dans} are also regular in this limit, we
conclude that the local correlation operators defined by~\eqref{loccorr}
also converge. Introducing the universal measure by $d\rho = F_*(dt \,d\varphi)$,
for any~$\varepsilon>0$ we obtain a causal fermion system~$(\H, \F, \rho)$ of spin dimension one.
This family of causal fermion systems has a regular limit as~$\varepsilon \searrow 0$, despite the fact that
a curvature singularity forms.

We point out that in this example, the curvature singularity is genuine (in the sense that it cannot
be removed by a conformal transformation). Nevertheless, the corresponding causal fermion systems have
a regular limit, even without rescaling the local correlation operators.
}} \QEDrem
\end{Example}

\begin{Example} (A genuine singularity on a cone times $S^1$) {\em{ 
We consider~$\tilde{M} = (\R^2 \setminus \{0\}) \times S^1$, choose polar coordinates~$(r, \varphi)$
in~$\R^2$ and the angular coordinate~$\alpha \in (0, 2 \pi)$ on the factor~$S^1$.
We take the warped product metric
\[ ds^2 = dr^2 + R(r)^2\, d\varphi^2 + S(r)^2\, d\alpha^2 \:. \]
The Dirac operator is computed in analogy to~\eqref{D2dR} by
\[ \Dir = i \sigma^r \partial_r + \frac{i}{R(r)}\: \sigma^\varphi \partial_\varphi
+ \frac{i}{S(r)}\: \sigma^3 \partial_\alpha
+ \frac{i}{2} \left( \frac{R'(r)}{R(r)} + \frac{S'(r)}{S(r)}- 1 \right) \sigma^r \:. \]
Employing similar to~\eqref{sep2d} the ansatz
\[ \psi = \frac{e^{i l \alpha}}{\sqrt{R(r)\, S(r)}} \begin{pmatrix} e^{i (k-\frac{1}{2}) \varphi}\, \chi_1(r) \\
i e^{i (k+\frac{1}{2}) \varphi}\, \chi_2(r) \end{pmatrix} \quad \text{with} \quad k \in \Z+\frac{1}{2},\; l \in \Z \:, \]
we obtain the radial ODE
\[ \chi'(r) = \begin{pmatrix} k/R & \lambda-l/S \\ -\lambda-l/S & -k/R \end{pmatrix} \chi(r) \:. \]

We now choose the parameters in such a way that we obtain solutions of the Dirac
equation in closed form which are continuous and non-zero at the origin.
For the metric functions we take
\[ R(r) = \frac{5}{6}\:r \qquad \text{and} \qquad S(r) = \frac{5}{4}\: r\:. \]
Thus the metric on~$\R^2 \setminus \{0\}$ is conical (similar to Example~\ref{ex2driemann}),
and the size of the factor~$S^1$ shrinks to zero as~$r \searrow 0$.
A direct computation shows that the wave functions
\begin{align*}
\psi_+(r, \varphi, \alpha) &= \frac{e^{i \alpha}}{r^2}
\begin{pmatrix} (1-2r) \sin(r) - r \cos(r) \\[0.2em] -i e^{i \varphi} \,\big((2-r) \sin(r) - 2 r \cos(r) \big) \end{pmatrix} \\
\psi_-(r, \varphi, \alpha) &= \frac{e^{i \alpha}}{r^2} \begin{pmatrix} e^{-i \varphi} \, \,\big((2-r) \sin(r)
- 2 r \cos(r) \big) \\[0.2em]
i \big (1-2r) \sin(r) - r \cos(r) \big) \end{pmatrix} .
\end{align*}
are solutions of the Dirac equation corresponding to the angular quantum numbers~$k= \pm \frac{1}{2}$
and~$l=1$. Note that the spinors stay bounded as~$r \searrow 0$ and do not converge to zero in this limit.

We let~$(\H, \la .|. \ra_\H)$ be the vector space spanned by~$\psi_+$ and~$\psi_-$
with the scalar product such that~$\psi_\pm$ are orthonormal.
Then the local correlation operators (see Definition~\ref{defloccor}) have the following
expansion near~$r=0$,
\[ F(r, \varphi, \alpha) = \begin{pmatrix} 5 & 4 e^{-i \varphi} \\ 4 e^{-i \varphi} & 5 \end{pmatrix} + \O(r^2) \:. \]
For any fixed~$\varphi$, these local correlation operators have a well-defined limit as~$r \searrow 0$.
This allows us to extend~$F$ by continuity to a mapping
\[ \tilde{F} \::\: (\scrM \cup S^1) \times S^1 \simeq [0, \infty) \times S^1 \times S^1 \rightarrow \F \subset \Lin(\H) \:. \]
Taking the push-forward measure~$\rho = \tilde{F}_*(d^2x)$, we obtain a
Riemannian fermion system of spin dimension two.
With this construction, we have compactified the manifold~$\scrM$ by glueing an~$S^1 \times S^1$
to the singularity point.

We remark that the mapping~$F$ is not injective. In order to cure this shortcoming, one
simply extends~$\H$ by wave functions with other quantum numbers~$k$ and~$l$
(similar as explained in Example~\ref{ex2driemann}).
Then the compactified manifold can be identified with the support of~$\rho$.
}} \QEDrem
\end{Example}

\subsection{The Curvature Singularity of Schwarzschild Space-Time} \label{secschwarzschild}
We now explain how causal fermion systems can be used to extend
the Schwarzschild geometry by a boundary describing a blow-up of the curvature singularity.
In polar coordinates~$(t,r,\vartheta, \varphi)$, the line element of the
Schwarzschild metric is given by
\beq \label{schmetric}
ds^2 = g_{jk}\:dx^j x^k =
\frac{\Delta}{r^2} \: dt^2
\:-\: \frac{r^2}{\Delta}\:dr^2 - r^2\: d \vartheta^2 - r^2\: \sin^2 \vartheta\:
d\varphi^2\:,
\eeq
where
\beq \label{Deldef}
\Delta = r^2 - 2M r \:,
\eeq
and~$M$ is the mass of the black hole.
The variables take values in the range~$t \in \R$, $r \in \R^+ \setminus \{r_1\}$, 
$\vartheta \in (0, \pi)$ and~$\varphi \in [0, 2 \pi)$, where the zero~$r_1:=2M$ of~$\Delta$ defines
the event horizon of the black hole. The metric has coordinate singularities at~$r=r_1$
and~$\vartheta = \{0, \pi\}$. Moreover, at~$r=0$ there is the curvature singularity at
the center of the black hole. For details on the Schwarzschild geometry we refer for example
to~\cite{hawking+ellis} or~\cite{wald}.

The Dirac operator in the Schwarzschild geometry can be computed just as explained
at the beginning of Section~\ref{secsing}.
Since we are interested in the curvature singularity at~$r=0$, we restrict attention to the
region~$r<r_1$ in the {\em{interior of the black hole}}.
We work in the spinor frame used in~\cite{kerr} in the Kerr-Newman geometry,
so that all our formulas can be obtained from those in~\cite{kerr} simply by setting~$a=Q=0$.
We let~$\gamma^j$ be the Dirac matrices in Minkowski space in the Weyl representation,
\[ \gamma^0 = \begin{pmatrix} 0 & \1 \\ \1 & 0 \end{pmatrix} \:,\qquad
\gamma^\alpha = \begin{pmatrix} 0 & \sigma^\alpha \\ -\sigma^\alpha & 0 \end{pmatrix} \]
(where~$\alpha=1,2,3$, and~$\sigma^\alpha$ are again the Pauli matrices~\eqref{Pauli}).
These matrices satisfy the anti-commutation relations~\eqref{anticomm}
(with~$\sigma^j$ replaced by~$\gamma^j$).
Moreover, the metric~\eqref{schmetric} is obviously diagonal. Hence Proposition~\ref{prpDOP}
applies. We choose the Dirac operator as
\[ \Dir = i G^j \partial_j + B \:, \]
where the Dirac matrices take the form
\[ G^0 = \frac{r}{\sqrt{|\Delta}|}\: \gamma^1 \:,\qquad
G^1 = -\frac{\sqrt{|\Delta|}}{r}\: \gamma^0 \:,\qquad
G^2 = \frac{\gamma^2}{r}\:,\qquad
G^3 = \frac{\gamma^3}{r \sin \vartheta}\:, \]
and~$B$ is the multiplication operator~\eqref{Bchoice}.
The Dirac equation is
\[ \Dir \Psi = m \Psi \:, \]
where~$\Psi$ is a section in the spinor bundle~$S \scrM$. The inner product on the fibre~$S_x \scrM$
takes the form
\[ \Sl \Psi | \Phi \Sr_x = \big\la \Psi \,\big|\, \begin{pmatrix} 0 & \1 \\ \1 & 0 \end{pmatrix} \Phi \big\ra_{\C^4}\:. \]

The Dirac equation in the Schwarzschild geometry can be
completely separated into ordinary differential equations.
We again use a method which immediately generalizes to the Kerr-Newman geometry
to give the formulas in~\cite{kerr}. Employing the ansatz
\beq \label{separate}
\Psi(t,r,\vartheta,\varphi) = \frac{e^{-i \omega t} \:e^{-i (k+\frac{1}{2}) \varphi}}{\:\sqrt{r}\, |\Delta|^{\frac{1}{4}}}
\: \left( \begin{array}{c} X_-(r) \:Y_-(\vartheta) \\
X_+(r) \:Y_+(\vartheta) \\
X_+(r) \:Y_-(\vartheta) \\
X_-(r) \:Y_+(\vartheta) \end{array} \right)
\eeq
with~$\omega \in \R$ and~$k \in \Z$
gives two ordinary differential equations for~$X$ and~$Y$. The angular equation for~$Y$
can be solved explicitly in terms of spin-weighted spherical harmonics. This determines the
separation constant~$\lambda$ to take one of the values
\[ \lambda=\pm 1, \pm 2, \pm 3, \ldots \:, \]
and the separation constant~$k$ must lie in the range
\beq \label{krange}
-|\lambda|+\frac{1}{2} \:\leq\: k \:\leq\: |\lambda|-\frac{1}{2}
\eeq
(see~\cite{goldberg} or the detailed computations for the operator~$K$ in~\cite[Appendix~A]{moritz},
noting that the separation constants~$\lambda$ and the functions~$Y$ coincide with the eigenvalues and
eigenfunctions of the operator~$K$ in suitable spinor bases).
The radial equation becomes
\[ \begin{pmatrix} \sqrt{|\Delta|} \:{\mathcal{D}}_+ & imr - \lambda \\
-imr - \lambda & -\sqrt{|\Delta|}
\:{\mathcal{D}}_- \end{pmatrix} \begin{pmatrix} X_+ \\ X_- \end{pmatrix} =  0 \:, \]
where
\[ {\mathcal{D}}_\pm = \frac{\partial}{\partial r} \:\pm\:i \omega\: \frac{r^2}{\Delta} \:. \]
This equation can be written in the more convenient form
\beq \label{radial}
\partial_r X = i \omega\: \frac{r^2}{|\Delta|} \begin{pmatrix} 1 & 0 \\ 0 & -1 \end{pmatrix} X
+\frac{1}{\sqrt{|\Delta|}} \begin{pmatrix} 0 & \lambda - imr \\
-\lambda - imr & 0 \end{pmatrix} X \:.
\eeq
In order to understand the separation ansatz~\eqref{separate},
one should keep in mind that
we restrict attention to the region~$r<r_1$ in the interior of the black hole.
Then the variable~$t$ is spatial, whereas~$r$ is the time coordinate. Therefore, the
plane-wave~$e^{-i \omega t}$ can be used to form
the Fourier decomposition of initial data given at some initial time $r_0<r_1$.
The radial equation~\eqref{radial} describes the time evolution of each Fourier mode.
Since the right of this equation is anti-Hermitian, one readily sees that
\beq \label{sepcurr}
\partial_r |X| = 0 \:.
\eeq
This corresponds to current conservation for each separated mode.

Suppose that~$\Psi$ is a solution of the form~\eqref{separate}.
Then the angular eigenfunction~$Y$ is smooth. Moreover, near the curvature singularity at~$r=0$, the
function~$\Delta$, \eqref{Deldef}, is smooth and tends linearly to zero.
As a consequence, near~$r=0$ the radial equation has the asymptotic form
\beq \label{asy0}
\partial_r X = \frac{\lambda}{\sqrt{2 M r}} \begin{pmatrix} 0 & 1 \\ -1 & 0 \end{pmatrix} X
+ \O \big( \sqrt{r} \big) \,X \:.
\eeq
Since the singularity of the coefficients at~$r=0$ is integrable, a Gr\"onwall estimate similar to~\eqref{cont}
shows that~$X$ can be extended continuously to~$r=0$, and that its norm~$|X|$
is bounded away from zero. We conclude that the only singular contribution at the origin
is the factor~$|\Delta|^{-\frac{1}{4}} \,r^{-1/2}$ in~\eqref{separate}. Therefore, we can remove the singularity
simply by rescaling the local correlation operators similar to our procedure in Section~\ref{secconformal}.
More precisely, we introduce the local correlation operators in modification of~\eqref{loccorr} by
\beq \label{loccorrSch}
-r^\frac{3}{2}\: \Sl \Psi | \Phi \Sr_{(t,r, \vartheta, \varphi)}
=  \la \Psi | F(t,r, \vartheta, \varphi)\, \Phi \ra_\H \qquad \text{for all~$\Psi, \Phi \in \H$} \:.
\eeq

It remains to decide of which solutions the space~$\H$ should be composed
and how to choose the Hilbert space scalar product~$\la .|. \ra_\H$. The only subtle point is that we
want the mapping~$F$ to be injective, making it necessary to choose ``sufficiently many'' wave
functions. We take the span of all wave functions of the form~\eqref{separate} for~$\omega \in \R$,
and~$|\lambda| \leq 2$, i.e.
\[ \Psi = (\Psi^{k \omega \lambda}) \qquad \text{with} \qquad
\omega \in \R\:, \lambda \in \{ \pm 1, \pm 2 \} \]
(and~$k$ in the range~\eqref{krange}).
For the scalar product we simply choose
\beq \label{sprod}
\la \Psi | \Psi \ra_\H = \int_{-\infty}^\infty e^{-\varepsilon^2 \omega^2} \:d\omega
\sum_{\lambda = -2}^2 \;\:\sum_{k =-|\lambda|+1/2}^{|\lambda|-1/2} |X^{k \omega \lambda}|^2 \:,
\eeq
where~$\varepsilon>0$ (we always assume the angular eigenfunctions~$Y$ to be normalized;
note that by~\eqref{sepcurr} the scalar product is independent of~$r$). Polarizing and taking the completion,
we obtain a Hilbert space~$(\H, \la .|. \ra_\H)$.
The factor~$e^{-\varepsilon^2 \omega^2}$ in~\eqref{sprod} can be regarded as a convergence-generating
factor describing an ultraviolet regularization on the length scale~$\varepsilon$.
It ensures that the functions in~$\H$ are all continuous,
so that the local correlation operators are well-defined by~\eqref{loccorrSch} for all~$r>0$.
Moreover, the continuity of our fundamental solutions at~$r=0$
makes it possible to extend the local correlation operators to~$r=0$,
\[ F(t,0, \vartheta, \varphi) := \lim_{r \searrow 0} F(t,r, \vartheta, \varphi)\:. \]
We thus obtain a mapping
\beq \label{FSch}
F \::\: \R \times [0, r_1) \times S^2 \rightarrow \F\:.
\eeq
Again defining the universal measure as the push-forward measure~$\rho = F_* \mu$,
we obtain a causal fermion system~$(\H, \F, \rho)$ of spin dimension two.

\begin{Lemma} The mapping~$F$ in~\eqref{FSch} is injective.
\end{Lemma}
\Proof In preparation, we need to construct approximate solutions of the ODE~\eqref{radial}.
Introducing the Regge-Wheeler coordinate~$u$ by
\[ \frac{du}{dr} = -\frac{r^2}{|\Delta|} \:,\qquad {\mbox{so}} \qquad
u = r + 2M\, \log |r-2M| \:, \]
the radial equation can be written as
\[ \partial_u X = -i \omega \begin{pmatrix} 1 & 0 \\ 0 & -1 \end{pmatrix} X
-\frac{\sqrt{|\Delta|}}{r^2} \begin{pmatrix} 0 & \lambda - imr \\
-\lambda - imr & 0 \end{pmatrix} X \:. \]
In order to describe the asymptotics for large~$\omega$, we employ the ansatz
\[ X = \begin{pmatrix} e^{-i \omega u} & 0 \\ 0 & e^{i \omega u} \end{pmatrix} Z \:. \]
For~$Z$ we obtain the equation
\beq \label{ODEkerr}
\partial_u Z = 
-\frac{\sqrt{|\Delta|}}{r^2} \begin{pmatrix} 0 & (\lambda - imr)\, e^{2 i \omega u} \\
(-\lambda - imr)\, e^{-2 i \omega u} & 0 \end{pmatrix} Z \:.
\eeq
Due to the oscillatory phase factors~$e^{\pm 2 i \omega u}$, the right side has no influence
on the solution if~$\omega$ gets large, provided that~$r$ stays away from zero
(this can be made precise for example by using that~$Z(u) - Z(u')
= \int_{u'}^u \partial_u Z$, inserting the differential equation~\eqref{ODEkerr},
writing the exponentials in the potential
as~$e^{\pm 2 i \omega u} = \mp \frac{i}{2 \omega}\: \partial_u e^{\pm 2 i \omega u}$,
integrating by parts and inserting again~\eqref{ODEkerr}).
Combining this fact with the observation made after~\eqref{asy0} that~$X$ is
continuous at~$r=0$ (and this argument is even locally uniform in~$\omega$), we conclude that
there are solutions with the asymptotics
\[ X(u) = \begin{pmatrix} c_1 \,e^{-i \omega u} \\ c_2 \,e^{i \omega u} \end{pmatrix}
+ \O \big( \omega^{-1} \big) \:. \]
In view of the factor~$e^{-i \omega t}$ in~\eqref{separate}, we thus obtain
solutions which depend on~$t+u$ and~$t-u$, respectively.
Taking superpositions of such solutions for~$\omega$ in a small neighborhood
of some fixed frequency~$\omega_0$, we can build up ``wave packet solutions,''
where the first component of~$X$ propagates along the curves~$t+u=\text{const}$, whereas the
second component propagates along the curves~$t-u=\text{const}$,
\beq \label{packet}
X(t,u) = \begin{pmatrix} X_1(t+u) \\ X_2(t-u) \end{pmatrix} + \O \big( \omega_0^{-1} \big) \:.
\eeq
We remark for clarity that this estimate is locally uniform in~$t$ and~$u$, meaning that~\eqref{packet}
holds with a fixed error term for all~$t$ and~$u$ in a compact set. 
Moreover, the error term clearly depends on the angular momentum mode. 
But this is of no relevance to us because the Hilbert space~$\H$ only involves the finite number of angular
momentum modes~$\lambda \in \{ \pm 1, \pm 2 \}$.

Let~$(t,r,\vartheta, \varphi) \neq (\tilde{t}, \tilde{r}, \tilde{\vartheta}, \tilde{\varphi})$
be two distinct space-time points.
Then either~$(t,r) \neq (\tilde{t}, \tilde{r})$ or~$(\vartheta, \varphi)
\neq (\tilde{\vartheta}, \tilde{\varphi})$.
In order to treat the first case~$(t,r) \neq (\tilde{t}, \tilde{r})$, we know that in
Regge-Wheeler coordinates either~$t+u \neq \tilde{t}+\tilde{u}$
or~$t-u \neq \tilde{t}-\tilde{u}$. Thus we can choose a wave packet of the form~\eqref{packet}
which goes through the point~$(t,r)$ but not through the point~$(\tilde{t}, \tilde{r})$.
This shows that the local correlation operators at the two points are necessarily different.

In the remaining case~$(t,r)=(\tilde{t}, \tilde{r})$ but~$(\vartheta, \varphi) \neq
(\tilde{\vartheta}, \tilde{\varphi})$, we know from
the explicit form of the angular eigenfunctions as worked out in~\cite{goldberg}
or in~\cite[Lemma~A.3]{moritz} that the span of these functions for
eigenvalues in the range~$-2 \leq \lambda \leq 2$ contains the constant and linear functions
in the Cartesian coordinates~$(x,y,z)$ restricted to the sphere~$S^2 \subset \R^3$. In particular,
the span contains the four functions
\[ \begin{pmatrix} 1 \\ 0 \end{pmatrix}, \begin{pmatrix} x \\ 0 \end{pmatrix},
\begin{pmatrix} y \\ 0 \end{pmatrix}, \begin{pmatrix} z \\ 0 \end{pmatrix} . \]
Forming suitable linear combinations, we can construct a spinor which vanishes at~$(\vartheta, \varphi)$
but is non-zero at~$(\tilde{\vartheta}, \tilde{\varphi})$. Taking the spin scalar product with the
constant spinor, one concludes that the local correlation operators at the two points are different.
\QED \noindent
This lemma allows us to identify the extended space-time~$\R \times [0, r_1) \times S^2$
with a subset of~$\F$. Thus the causal fermion system describes the whole interior
Schwarzschild geometry. Moreover, it includes the 
singularity at~$r=0$ as a boundary of space-time which is diffeomorphic to~$\R \times S^2$.

We finally remark that by going over to the Kruskal extension, our construction could readily be extended
to the exterior region of the Schwarzschild black hole. Moreover, using the formulas in~\cite{kerr},
the constructions immediately extend to the non-extreme Reissner-Nordstr\"om, Kerr and Kerr-Newman
geometries.

\subsection{A Lattice System with Non-Trivial Topology} \label{seclattice}
We now illustrate the constructions in Section~\ref{secdiscrete} by a simple example
of a lattice system. Before beginning, we remark that 
our constructions bear some similarity to ideas by M.~L\"uscher~\cite{luscher}, who
considers a lattice on the four-dimensional torus and shows that one can introduce a
non-trivial topological charge provided that the field strength of a lattice gauge field is small on the
lattice scale (see also the discussion in~\cite{woit}).
However, our construction is different and much more general because we do not need
the nearest-neighbor relation. Moreover, we do not assume a connection on the bundle, nor that the
corresponding field strength be small. Instead, we need to assume that the distance of the lattice
points is small on the ``macroscopic length scale'' on which the topology of the torus is visible.

We consider the two-dimensional torus~$T^2 = \R^2 / (2 \pi \Z)^2$ with the metric induced
from the Euclidean metric of~$\R^2$.
Moreover, for a given parameter~$\kappa>0$ with~$2 \pi/\kappa \in \N$, we consider the lattice
\[ \scrM = (\kappa \Z)^2 / (2 \pi \Z)^2 \:. \]
Thus~$\scrM$ is a lattice on~$T^2$ with lattice spacing~$\kappa$, consisting of~$(2\pi/\kappa)^2$
lattice points. We let~$\mu$ be the normalized counting measure on~$\scrM$,
\[ \mu(\Omega) = \frac{\kappa^2}{(2 \pi)^2}\: \# \Omega \:. \]

On the two-dimensional torus there are different spin structures with corresponding Dirac operators.
For simplicity, we take the Dirac operator~$\Dir$ obtained from the Dirac operator on~$\R^2$ \eqref{dir2}
by taking the quotient with~$(2 \pi \Z)^2$. Then an eigenvector basis of this Dirac operator is given
similar to~\eqref{plane} in terms of the plane wave solutions
\begin{align}
\e_0^+(\zeta) &= \begin{pmatrix} 1 \\ 0 \end{pmatrix} \:, \qquad
\e_0^-(\zeta) = \begin{pmatrix} 0 \\ 1 \end{pmatrix} \label{plane0} \\
\e_{k}^\pm(\zeta) &= \frac{1}{|k|}\,
(k_1 \sigma^1 + k_2 \sigma^2 \pm |k| \1)\,\begin{pmatrix} 1 \\ 0 \end{pmatrix}
\: e^{-i k \zeta} \qquad
\text{if~$k \in \Z^2 \setminus \{ 0 \}$}\:, \label{planenz}
\end{align}
where now~$k$ lies on the dual lattice~$\Z \times \Z$.
By direct computation one verifies that the eigenvalues of the wave functions~$\e_k^\pm$ are~$\pm |k|$.

We choose~$\H$ as the vector space spanned by a finite number of plane-wave solutions.
The scalar product~$\la .|. \ra_\H$ is defined by imposing that the plane-wave solutions~$\e_k^\pm$
are orthonormal. For computational simplicity, we choose the three-dimensional space
\beq \label{Hexp}
\H = \text{span} \big( \e_0^+, \e_{(1, 0)}^+, \e_{(0, 1)}^+ \big) \:,
\eeq
but any choice of~$\H$ which contains these three vectors would work just as well.
For any~$p \in T^2$, we introduce the local correlation operator again by~\eqref{loccorr}.
We now define the universal measure as the push-forward of the counting measure on the lattice,
\[ \rho = \big( F|_{\scrM} \big)_* \mu \:. \]
We thus obtain a Riemannian fermion system~$(\H, \F, \mu)$ of spin dimension two.

In order to analyze this Riemannian fermion system, it is useful to represent the local correlation operators
in the orthonormal basis of the vectors in~\eqref{Hexp}.
A short computation gives
\beq \label{Frep}
F(\zeta) = -\begin{pmatrix} 1 & e^{-i x} & e^{-i y} \\ e^{ix} & 2 & (1-i) \,e^{ix-iy} \\ 
e^{i y} & (1+i) \,e^{-ix+iy} & 2 \end{pmatrix} ,
\eeq
where we denote the components of~$\zeta \in T^2$ by~$(x,y)$.
Moreover, a short computation gives
\[ \|F(\zeta) - F(\zeta')\|^2 = 16 - 4 \cos(x-x') - 4 \cos(y-y') - 8 \cos(x-x'+y-y') \]
(where for convenience we work with the Hilbert-Schmidt norm on~$\Lin(\H)$).
Using the sum rules and the inequality~$|\cos \varphi| \leq 1$, we obtain
\begin{align*}
\|F(\zeta) - F(\zeta') \| &= 8 \sin^2 \Big( \frac{x-x'}{2} \Big) +8 \sin^2 \Big( \frac{y-y'}{2} \Big)
+ 16 \sin^2 \Big( \frac{x-x'}{2} + \frac{y-y'}{2} \Big) \\
&\leq 24 \sin^2 \Big| \frac{x-x'}{2} \Big| + 32 \sin \Big| \frac{x-x'}{2} \Big| \:\sin \Big| \frac{y-y'}{2} \Big|
+ 24 \sin^2 \Big| \frac{y-y'}{2} \Big| .
\end{align*}
Applying the Schwarz inequality, we conclude that
\beq \label{Fdes}
\|F(\zeta) - F(\zeta') \| \leq \sqrt{24} \;\bigg( \sin \Big| \frac{x-x'}{2} \Big| + \sin \Big| \frac{y-y'}{2} \Big| \bigg) \:.
\eeq
Moreover, the distance of antipodal points on the torus is computed by
\begin{align*}
\|F(0,0) - F(\pi,0)\| &= \|F(0,0) - F(0,\pi)\| = \sqrt{24} \\
\|F(0,0) - F(\pi,\pi)\| &= 4\:.
\end{align*}

After these preparations, we can discuss the constructions in Section~\ref{secdiscrete}.
The matrix representation~\eqref{Frep} shows in particular that the mapping
\[ F : T^2 \rightarrow \F \quad \text{is injective}\:. \]
Hence the image~$F(T^2)$ is topologically a torus.
Taking the image of the lattice~$\scrM$, we obtain a set of~$(2\pi/\kappa)^2$ of points in~$\F$.
The universal measure of our Riemannian fermion system is the normalized counting measure
of these points. In particular, its support are these finite number of points,
\[ M := \supp \rho = F(\scrM) \subset \F \:. \]
We conclude that the topology of the Riemannian fermion system is trivial.

The situation becomes more interesting when we consider the sets~$M_r$
defined by~\eqref{Mr}. For small~$r$, the balls around the points in~$M$ do not intersect,
so that the topology remains trivial (in view of~\eqref{Fdes}, this is the case if~$r < \sqrt{24}\: \sin (\kappa/4)$).
If, on the other hand, the parameter~$r$ is chosen larger than~$\sqrt{24}$, then each of these balls
contains all~$M$.
Then~$M_r$ has the trivial topology of a ball. In the intermediate range
\beq \label{k1}
\sqrt{24}\: \sin \Big( \frac{\kappa}{2} \Big) < r < 2 \:,
\eeq
the ball around a point in~$M$ intersects the neighboring balls, but not the balls around the antipodal
points. As a consequence, $M_r$ has the topology of a torus.
Even more, $M_r$ can be continuously deformed to the set~$F(T^2)$
(more precisely, $F(T^2)$ is a deformation retract of~$M_r$).
This implies that~$M_r$ has the same bundle topology as~$F(T^2)$.
In particular, $M_r$ encodes the topological data of the torus.
These considerations are illustrated in Figure~\ref{figlattice}.
\begin{figure}
\begin{center}
\includegraphics[width=4.5cm]{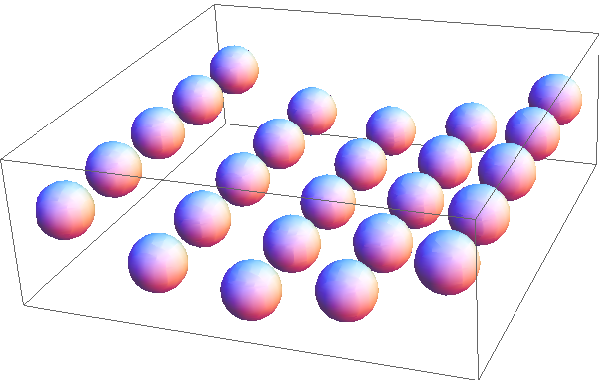} 
\includegraphics[width=4.5cm]{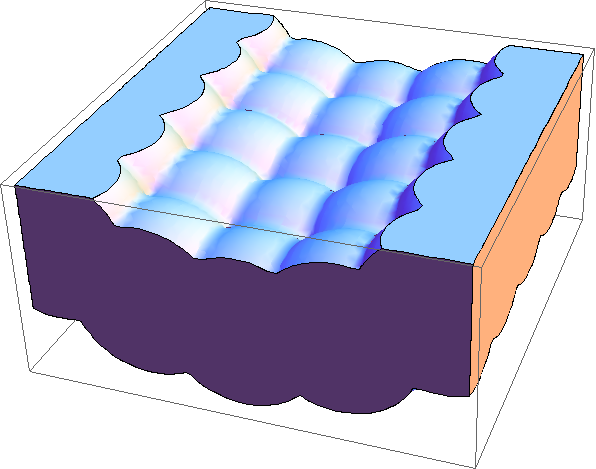} 
\includegraphics[width=4.5cm]{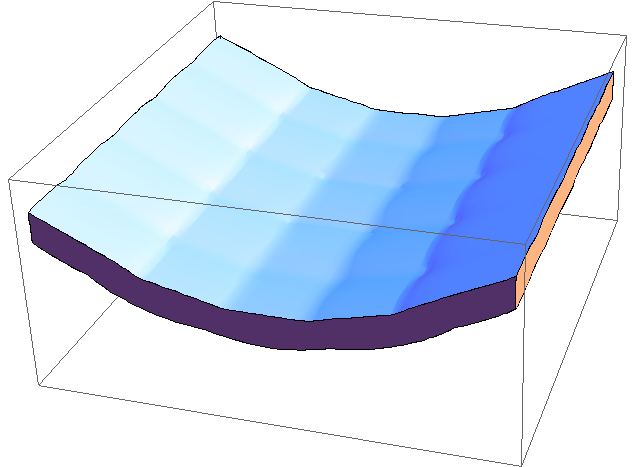}
\caption{The geometry for small~$r$ (left), larger~$r$ (middle),
and when deforming to~$F(T^2)$ (right).}
\label{figlattice}
\end{center}
\end{figure}

For the set~$M_\delta$ defined in~\eqref{Mdelta} the situation is similar,
except that we now need to specify the range of~$\delta$.
In order for the balls to include the nearest neighbors, we need to choose~$\delta$
larger than the volume of five points. In order to exclude the antipodal points,
we need to choose~$\delta<1/2$. Thus in order to recover the topology of the torus, we must
choose~$\delta$ in the range
\beq \label{k2}
\frac{5 \kappa^2}{4 \pi^2} < \delta < \frac{1}{2} \:.
\eeq

In order to implement the construction~\eqref{rhordef} or~\eqref{rhoddef}, we must choose
a measure~$\rho$ on~$\F$. A simple method is to choose a basis~$F_1, \ldots, F_9$
of the vector space of Hermitian~$3 \times 3$-matrices (for example an orthonormal basis with
respect to the Hilbert-Schmidt norm) and to represent~$F \in \Symm(\C^3)$ by
\[ F = \sum_{\alpha=1}^9 f_\alpha \,F_\alpha \:. \]
Let~$d\mu(F)$ be the Lebesgue measure~$df_1 \cdots df_9$ multiplied
by the Dirac distribution supported on the set~$\{ \det F = 0\}$. Then~$\mu$
is a measure supported on the~$3 \times 3$-matrices of rank at most two.
Since~$\F$ is a subset of these matrices of positive $\mu$-measure, restricting~$\mu$ to~$\F$
gives a non-trivial measure on~$\F$. Choosing~$\eta_r(x,y) = \eta(\|x-y\|^2/r^2)$
with~$\eta \in C^\infty_0([0,1))$, we can then introduce the measures~$\rho_r$ and~$\rho_\delta$
by~\eqref{rhordef} and~\eqref{rhoddef}.

To summarize, the constructions of Section~\ref{secdiscrete} make it possible to recover
topological information on a lattice, provided that the lattice is sufficiently fine and the 
parameters~$r$ respectively~$\delta$ are chosen such that
the microscopic discrete structure is ``smeared out'' without affecting the global topology.
In situations when the lattice spacing~$\kappa$ is very small, the inequalities~\eqref{k1}
and~\eqref{k2} leave a lot of freedom to choose~$r$ respectively~$\delta$.
Thus thinking of a discrete structure on the Planck scale, there is no problem in recovering
the global topology of space-time.

\Thanks {{\em{Acknowledgments:}}
We would like to thank Steven Boyer and Tristan Collins for helpful discussions. Moreover, we are grateful to
Michaela Fischer, Christina Pauly, Saeed Zafari and the referee for useful comments on the manuscript.
We would like to thank the Center of Mathematical Sciences and Applications at
Harvard University for hospitality and support.
We are grateful to the ``Regensburger Universit\"atsstiftung Hans Vielberth'' for generous support.
N.K.'s research was also supported by the NSERC grant RGPIN 105490-2011.

\providecommand{\bysame}{\leavevmode\hbox to3em{\hrulefill}\thinspace}
\providecommand{\MR}{\relax\ifhmode\unskip\space\fi MR }
\providecommand{\MRhref}[2]{%
  \href{http://www.ams.org/mathscinet-getitem?mr=#1}{#2}
}
\providecommand{\href}[2]{#2}

\end{document}